\documentclass[preprint]{iacrtrans}
\usepackage[utf8]{inputenc}
\usepackage{lipsum}
\usepackage{graphicx} % Required for inserting images
\usepackage{caption}
\usepackage{subcaption}
\usepackage[english]{babel}
\usepackage{xcolor}
\usepackage{tabularx}
\usepackage{tabularray}
\usepackage{amsmath,amsfonts,amsthm}
\usepackage{physics}
\usepackage{enumitem}
\usepackage[colorlinks=true, allcolors=blue]{hyperref}
\UseTblrLibrary{diagbox}
\usepackage{thmtools}
\usepackage{thm-restate}
\usepackage{algorithm}

\usepackage{subfiles} % Best loaded last in the preamble

\newcounter{protocol}
\makeatletter
\newenvironment{protocol}[1][htb]{
    \let\c@algorithm\c@protocol
    \renewcommand{\ALG@name}{Protocol}
    \begin{algorithm}[#1]
    }{\end{algorithm}
}

\newcommand{\neigh}[0]{\mathcal{N}}

\title{Quantum-Safe Identity Verification using Relativistic Zero-Knowledge Proof Systems}
\author{Yao Ma \and Wen Yu Kon \and Jefferson Chu \and Kevin Han Yong Loh \and Kaushik Chakraborty \and Charles Lim}
\institute{Global Technology Applied Research, JPMorganChase}
\date{}

\begin{document}

\maketitle

% \keywords{}

\begin{abstract}
Identity verification is the process of confirming an individual's claimed identity, which is essential in sectors like finance, healthcare, and online services to ensure security and prevent fraud. However, current password/PIN-based identity solutions are susceptible to phishing or skimming attacks, where malicious intermediaries attempt to steal credentials using fake identification portals. Alikhani et al. \cite{Alikhani_2021} began exploring identity verification through graph coloring-based relativistic zero-knowledge proofs (RZKPs), a key cryptographic primitive that enables a prover to demonstrate knowledge of secret credentials to a verifier without disclosing any information about the secret. Our work advances this field and addresses unresolved issues: From an engineering perspective, we relax further the relativistic constraints from 60m to 30m, and significantly enhance the stability and scalability of the experimental demonstration of the 2-prover graph coloring-based RZKP protocol for near-term use cases. 
% \wy{State specific improvements (e.g. distance -- from 60m to 30m?)}
% \wy{What's ``other performances?"}
At the same time, for long-term security against entangled malicious provers, we propose a modified protocol with comparable computation and communication costs, we establish an upper bound on the soundness parameter for this modified protocol. On the other hand, we extend the two-prover, two-verifier setup to a one-verifier, three-prover configuration, demonstrating the security of such relativistic protocols against entangled malicious provers. 
%\wy{This does not seem to show scalability?}
\end{abstract}

\section{Introduction}
\label{sec:introduction}
Identity verification is the process of confirming that a person is who they claim to be. This process is crucial to ensure security and prevent fraud in various sectors, including finance, healthcare, and online services. The current day password/PIN-based identity solutions are vulnerable to phishing or skimming attacks, where a malicious intermediate party tries to steal the credentials by using for example a fake identification portal. To address these vulnerabilities, Zero-Knowledge Proofs (ZKPs), first introduced by Goldwasser, Micali, Rackoff \cite{GMR85}, offer a compelling solution. It enables a prover to prove its knowledge about some secret credentials to a verifier without revealing any information about the secret. In modern day cryptography, ZKP plays a crucial role in providing privacy in domains like cryptocurrencies and blockchain, electronic voting etc. This paper focuses on the identity verification application of ZKP. Intuitively, ZKP can offer a perfect solution to make identity verification protocols phishing attack-proof. 

It is widely believed that perfect zero-knowledge proofs for any arbitrary NP language is impossible without any extra computational hardness assumption. In a multi-prover setting with non-communicating provers, it is possible to achieve a perfect zero-knowledge scheme \cite{Kil90}. However, in practice, it is difficult to satisfy the non-communication assumption. In \cite{Kent99}, Kent proposed a way to implement such a stringent requirement using the theory of special relativity, and introduced a statistically secure hiding and binding bit commitment protocol. It opened a new direction of research called \emph{relativistic cryptography}. Later in \cite{CL17}, Chailloux and Leverrier proposed a perfect zero-knowledge proof protocol for Hamiltonian cycle in a 2-prover relativistic setting by adopting Blum's ZKP~\cite{Blum86} and combining it with the relativistic bit commitment protocol from Ref.~\cite{CSST11}. However, the protocol is limited by its high communication costs. Following that, there are many proposals that try to circumvent the efficiency issues \cite{Alikhani_2021,Crepeau_2019,CB21,CS23}. Among all those attempts, the protocol proposed by Crepeau et al. in ~\cite{Crepeau_2019} is the most lightweight relativistic ZKP protocol for graph 3-coloring. It uses a approach called the \emph{unveil-via-commit principle} which is inspired by the double-spending detection mechanism of the untraceable electronic cash of Chaum, Fiat and Naor~\cite{CFN90}. 
The practicality of this protocol was demonstrated in a follow-up experiment with a real-time implementation and the provers being $60$ meters apart~\cite{Alikhani_2021}.
This can be useful in applications such as ATM-card authentication, as described by Brassard~\cite{brassard2021relativity}. All these results pave the way to bring this protocol from theory to practice. 

However, the protocol suffers from many issues: On the one hand, the two-prover protocol proposed in \cite{Crepeau_2019, Alikhani_2021} was proven to be sound only against unentangled adversaries. Except for modifying the protocol into a 3-prover scenario with adaptions, the authors left their protocol's soundness against entangled provers open. 
Note that, this protocol has lots of similarity with the \emph{graph coloring game} \cite{Gardner70}. For such non-local games there are some examples for which entangled provers can convince (with certainty) a verifier a graph is $k$-colorable even if its chromatic number is $k' > k$ \cite{MR18}. Therefore, it is not clear if the 2-prover relativistic zero-knowledge proof protocol for graph 3-coloring is sound against entangled provers. 
On the other hand, when transitioning \cite{Alikhani_2021} to real-world application, a key limitation is the lack of comprehensive performance data necessary for benchmarking. This gap in information can hinder the ability to accurately assess the system's efficiency, scalability, and reliability under practical conditions. Without detailed performance metrics, it becomes challenging to predict how the system will behave in diverse environments, potentially affecting its adoption and effectiveness in real-world scenarios. 

Even though the analysis of entangled malicious provers can lead to a significant increase in the complexity of the protocol, 
e.g., the cost associated with the protocol scales up sharply in terms of the number of required rounds, 
it is still feasible where additional setup and practical assumptions are considered. % here: 
While we consider quantum provers in general, it is in practice difficult for malicious provers to maintain quantum entanglement in the near-term.
In particular, the client's token, which is expected to be small, would not be capable 
of long-term storage of multiple quantum systems that are entangled with the private computing system of client bank server. 
This is a demanding requirement due to the lack of good quantum memories that do not easily decohere, the development of which is still in its nascent stages~\cite{Heshami2016}.
% It is widely known that maintaining quantum entanglement requires sophisticated quantum memory, which is currently a major technological hurdle. Quantum memory must be capable of storing quantum states without significant decoherence, a process where the quantum information is lost due to interactions with the environment. This requirement is particularly demanding because even slight disturbances can disrupt the entangled state. Furthermore, the infrastructure needed to support quantum communication and storage is still in its nascent stages. The complexity and cost of developing and maintaining such systems make the use of quantum provers in RZKP identification highly unrealistic with current technology. The need for stable quantum memory and reliable quantum communication channels adds layers of complexity that are not yet feasible for widespread implementation. 
As such, in the near-term, it is safe to assume that the provers are not entangled, i.e., the security reduces to one of quantum but not entangled provers, which requires a much lower number of rounds to ensure soundness. 

\begin{figure}[!h]
    \centering
    \includegraphics[width=0.9\linewidth]{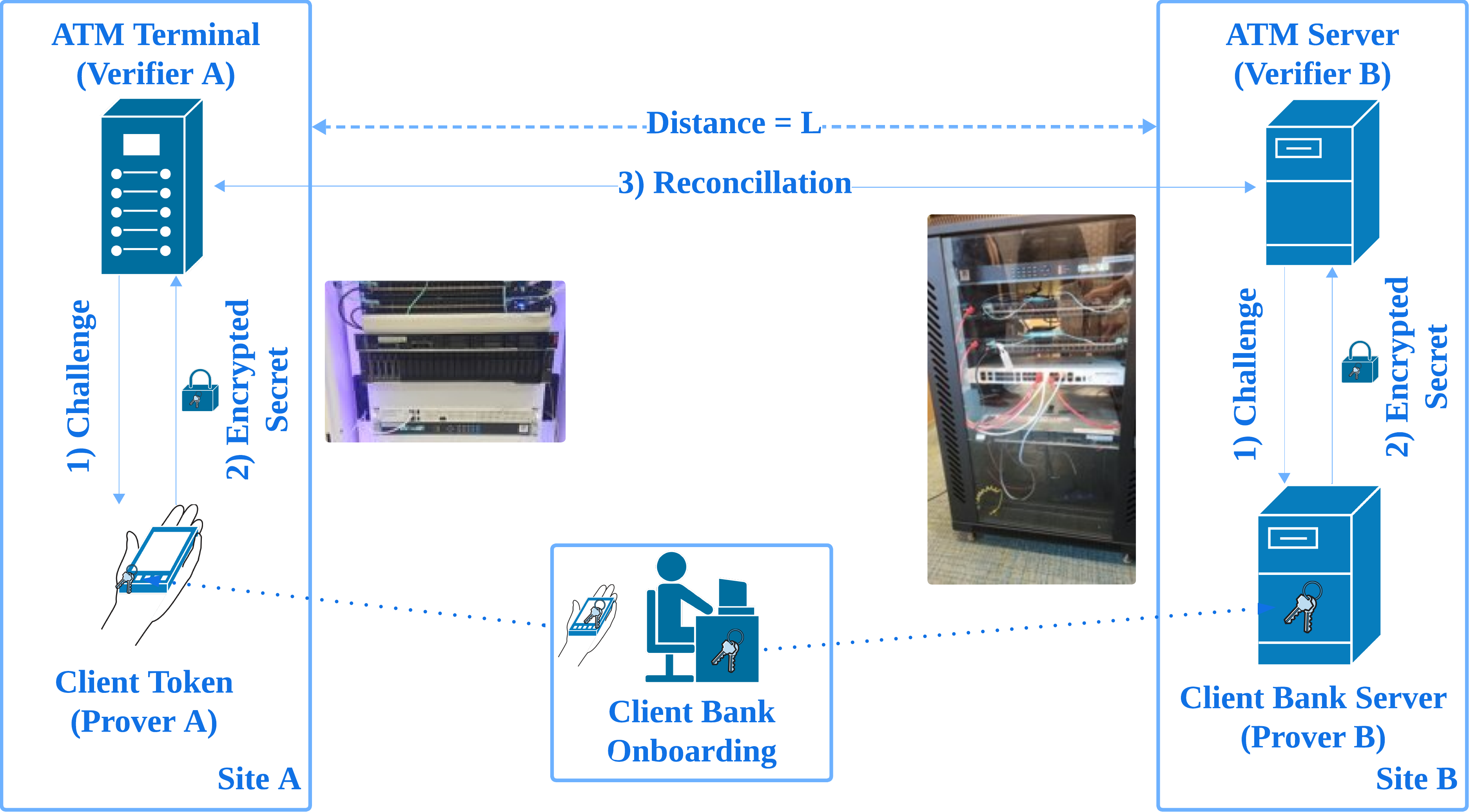}
    \caption{A RZKP-based Financial Identification System}
    \label{fig:identification}
\end{figure}

In the first part of our work, we demonstrated a more comprehensive graph coloring-based two-prover RZKP implementation~\cite{Alikhani_2021} in the context of a financial identification system application as a near-term use case.
Figure~\ref{fig:identification} showcases the architecture of the implementation.
Similar to current identification systems, the client is required to have a client token, such as an ATM card, and approach an identification terminal, such as an ATM terminal, for authentication.
The client token acts as Prover A, and the identification terminal serves as Verifier A, allowing the client to prove its identity to the identification terminal via ZKP.
To implement RZKP, we introduce a second verifier-prover pair, with Verifier B being a dedicated identification server outsourced to cloud or data center providers that host multiple entities, and Prover B being the private computing system of the financial institution that issued the client token.
The RZKP protocol can then run following the protocol in \cite{Crepeau_2019}.
Once a successful identification of the client is achieved, the terminal will allow for further operations, e.g., performing financial transactions.

% For RZKP protocol for graph 3-coloring, each client of the bank would receive a unique random 3-colorable graph as his user ID, and the corresponding coloring would be stored in the client's token as his credential. Meanwhile, the necessary shared randomness required by the protocol are also pre-loaded onto the client's token.

% When the client initiates the identification process on the terminal, all parties in the system are notified and starts preparing the required information. When all parties are prepared, the verifiers can synchronously start the transmission and receive responses from the provers independently. When the execution of transmission is completed, the verifiers can jointly verify the responses by consolidation. Once a successful identification of the client is achieved, the terminal will allow for further operations, e.g., performing financial transactions. 

Our implementation provides improvements over the RZKP protocol implemented in \cite{Alikhani_2021}.
From an engineering point of view, we modularize our system into an easy-to-deploy \emph{Preparation-Transmission-Verification} sequential process and provide a more measurable set of results. Furthermore, with sophisticated hardware/software design, we drastically improve the performance of the experimental demonstration of the 2-prover graph coloring protocol. For example, our new demonstration reduces the spatial separation of the provers from $60$ meters to $30$ meters. Note that, our engineering technique behind this improvement is independent of this 3-coloring protocol. It can be used to improve the performance of other relativistic zero-knowledge protocols as well. Last but not least, we adapt our RZKP implementation to the architecture of identity verification. Here, each client of the bank would receive a random 3-colorable graph as his unique user ID, and the corresponding coloring would be stored in the client's token as his credential. Additional data, such as random strings necessary for RZKP-3-COL (to generate permutations and trit vectors), are also pre-loaded onto the client's token. When the client initiates the identification process on the terminal, all parties in the system are notified and starts preparing the required information, following the preparation phase in our experiment. When all parties are prepared, the verifiers can synchronously start the transmission and receive responses from the provers independently. When all required rounds are completed for a batch, the verifiers can jointly verify the responses by consolidation. The procedure follows the specification in our design. Once a successful identification of the client is achieved, the terminal will allow for further operations, e.g., performing financial transactions.

The second part of our work evaluates the long-term security of graph coloring-based RZKP against entangled malicious provers from various dimensions: Experimentally, we extend the two-prover, two-verifier set-up to a one verifier three prover set-up, and demonstrate the security of this variant of relativistic protocols. On the other hand, we theoretically propose a 2-prover modified protocol with a similar computation and communication cost. We refer to our new protocol as Alt-RZKP-3-COL protocol. We prove an upper bound on the soundness parameter against entangled provers for this modified protocol.

In a recent study, Weng et al. \cite{WC25} proposed and experimentally implemented an unconditionally secure zero-knowledge proof (ZKP) for the graph three-coloring problem by combining subset relativistic bit commitments with a quantum non-local game. Unlike symmetric relativistic zero-knowledge protocols, their method requires one prover to commit to all vertices using modular arithmetic operations and to disclose the commitment of an arbitrary edge with the help of another prover. However, the soundness against entangled provers depends on the \(\mathbb{F}_Q\) commitment and the size of the prime \(Q\) used. As a result, the communication and computation costs per round are considerably higher, which may present challenges for practical applications.

\section{Results}
\label{sec:results}

\subsection{Two-Prover RZKP Implementation with Sophisticated Design}
\label{ssec:2pexp}
Guided by \cite{Alikhani_2021} on deploying the RZKP-3-COL protocol in practice, our implementation involves demonstration on multiple stages that integrate both hardware and software. Establishing benchmarks for each stage is beneficial for evaluating their practical applicability. To minimize resource requirements, we explore various methods to optimize the system's architectural design, balancing performance and stability. Our work aims to enhance the implementation of the RKZP-3-COL protocol for potential near-term applications through two main improvements. Firstly, we modularize our system into an easy-to-deploy \emph{Preparation-Transmission-Verification} sequential process and provide a more comprehensive set of results. In each stage, we detail the performance, which can allow for further analysis and development such as the identification of bottlenecks. Secondly, we optimize the interaction between the verifier and the prover, which allows us to halve the minimum distance (from 57.6m to 30m) between the prover-verifier pair compared to \cite{Alikhani_2021}.

\begin{figure}[!h]
    \centering
    \includegraphics[width=\linewidth]{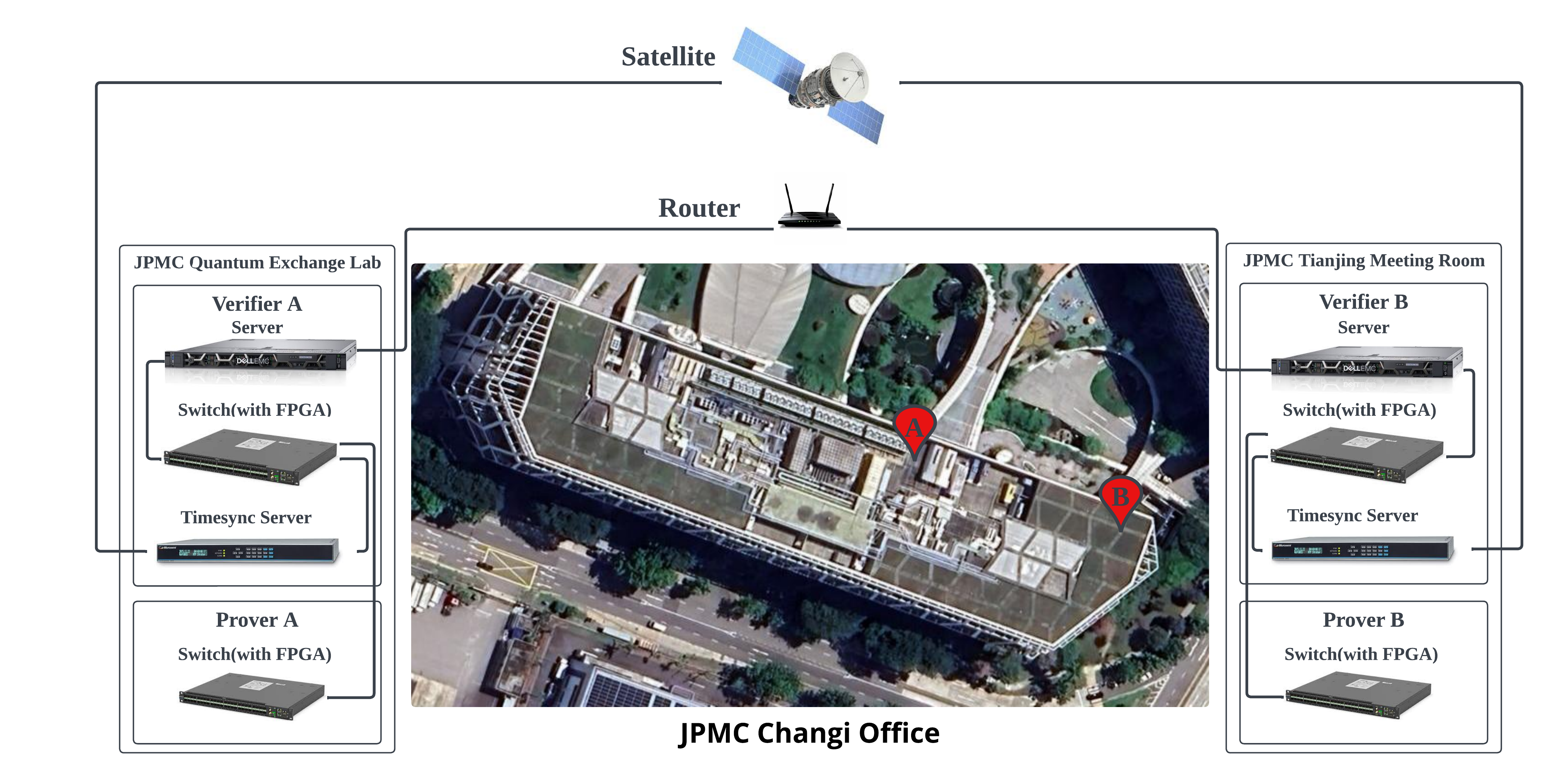}
    \caption{Schematic of the RZKP-3-COL implementation at the J.P. Morgan Singapore Corporate Center. The two verifier-prover pairs are located at locations A and B in the building, which are 40 meters apart, which is further than the 30 meters separation required. At each location, we use a single rack to house both the verifier and prover devices. The verifier has a server and FPGA for data processing, and a timesync server, while the prover has an FPGA for data processing. The servers at both locations are connected via a router link for communication, and the two verifiers achieve synchronization using their timesync servers via satellite.}
    \label{fig:2poverview}
\end{figure}

We implemented an RZKP-3-COL protocol (protocol summarized in Protocol~\ref{prot:2prover}) across two locations in the J.P. Morgan Singapore Corporate Center, as shown in Figure~\ref{fig:2poverview}. The two locations, separated by a distance of 40 meters, each houses a verifier-prover pair. They are labeled $\mathcal{V}_{A}-\mathcal{P}_{A}$ for the verifier-prover pair at location A and $\mathcal{V}_{B}-\mathcal{P}_{B}$ for the verifier-prover pair at location B. At each location, the verifier is equipped with a server and \emph{field-programmable gate array} (FPGA) for data processing, and a timesync server for synchronization, and the prover is equipped with an FPGA for data processing. Both prover and verifier devices at each location are mounted onto a single server rack. The servers of both verifiers are connected via a router to allow communication for the joint verification process and the timesync servers achieve time synchronization across locations A and B via satellite.

Due to hardware and software limitations, such as the limited FPGA memory for pre-loading the shared randomness required by provers and the volume of verification data, there is a limit to the number of rounds we can implement. As such, we design our implementation to take place in batches, with a choice of 600,000 rounds per batch for practical and efficiency reasons. Importantly, 600,000 rounds are sufficient to guarantee soundness against classical adversaries, using graph instances of size $|V_G|\sim600$ and $|E_G|\sim10^3$, with a soundness parameter of $e^{-100}$, i.e. with security parameter $k=100$~\cite{Alikhani_2021}. We perform a complete \emph{Preparation-Transmission-Verification} process within each batch. The process is re-initiated for the next batch after the end of the verification step, where the responses are consolidated and performed at location A. We note that the verifiers and provers can in principle pre-share the randomness required for all rounds, but due to memory limitations, we reload the shared randomness on the FPGA at the start of each batch. The batching of rounds is beneficial as it allows verifiers to abort the protocol whenever any errors are observed in a batch, rather than having to complete all rounds.

Once these messages are pre-loaded onto the FPGAs for each batch, we commence the process with a synchronous start message using the \emph{Network Time Protocol} (NTP) obtained from the GPS timesync server on $\mathcal{V}_A$ and $\mathcal{V}_B$. Additionally, we utilize the \emph{Pulse-per-Second} (PPS) signal from the same GPS server to initiate execution on the verifiers' FPGAs. This ensures that any network-induced latency and jitter are compensated by the PPS pulse. Moreover, we enhance our hardware system by deploying 10 PPS (i.e., a generation of ten pulses every second from the timesync server) instead of 1 PPS, which provides more frequent timing updates, and reduces the waiting time between batches. In our implementation, the start message sent to the FPGA includes the start and end round indices (up to 600,000 rounds) and the interval between each pair of rounds in a batch, ensuring that the main interactions, particularly the transmission between each verifier and prover within a batch, are automatically aligned by the FPGA once triggered. If more rounds of execution (i.e., more than one batch) is needed in practice, we reset the FPGA and resynchronize the system between every two batches. 
\begin{figure}[!h]
    \centering
    \includegraphics[width=0.9\linewidth]{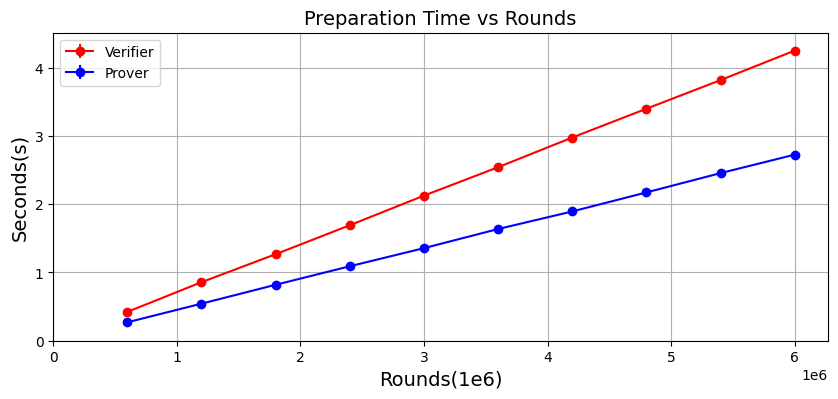}
    \caption{Preparation Time $t_{\text{prep}}$ of Verifier and Prover}
    \label{fig:2p_preparation}
\end{figure}

To demonstrate the scalability and stability of our system, we perform a series of 2-party experiments %completely 
from 600,000 to 6,000,000 rounds (10 batches): Each experimental set is with 1000 times of repetition using a generated graph instance $G$ with $|V_G|\sim600$ and $|E_G|\sim10^3$ (See Section \ref{ssec:graph_gen} for further details).
We record the time consumption on every step. For the preparation phase, it involves the preparation of the verifiers' challenges, and data and randomness required for the provers' responses, and loading them into the respective FPGAs. In Figure~\ref{fig:2p_preparation}, it shows the preparation time required for 1 to 10 batches of rounds, which are repeated for 1000 times for graphs with $\abs{V_G}\sim 600$ and $\abs{E_G}\sim 10^3$. This demonstrates that the preparation is stable across different batches, allowing us to scale the implementation by simply increasing the number of batches that leads to a linear increase in preparation time. The preparation time of the verifier is longer, with the additional time stemming from the sampling of edges. In our implementation, an initial edge is sampled from AES counter mode \emph{deterministic random bit generator} (AES-DRBG) \cite{BK15} by both verifiers using a common seed, followed by additional sampling on $\mathcal{V}_A$ to decide between the edge verification or well-definition tests, with probabilities of $\frac{1}{5}$ and $\frac{4}{5}$ respectively. To achieve low preparation time demonstrated here, we designed a highly efficient randomness generation and pre-loading process. More details of the preparation phase can be found in Section~\ref{sssec:resp_pre}. We note that there is potential for further reduction of preparation times by optimizing the method for sampling the verifiers' challenges.

\begin{figure}[!h]
    \centering
    \includegraphics[width=\linewidth]{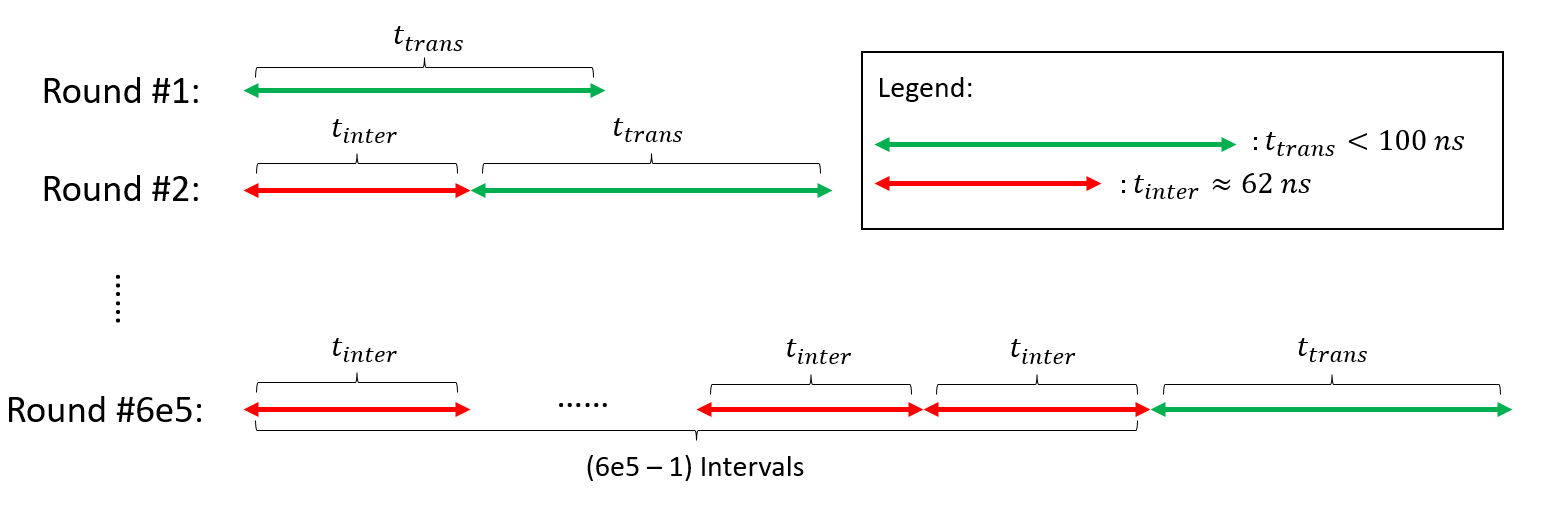}
    \caption{Illustration on Transmission Phase with a FPGA Timer}
    \label{fig:trans}
\end{figure}

During the transmission phase, we implement a timer within each verifier's FPGA to log the timestamp when a verifier receives each response. The transmission primarily occurs on FPGAs and adheres to relativistic constraints (see Section~\ref{sssec:trans} for details). Given that the relativistic setting parameter $t_{\text{trans}}$ is capped at 100 ns in our design, we further test the round interval $t_{\text{inter}}$, which is the time between sending two consecutive challenge messages by each verifier, using different parameters. Our system remains robust without any failures even when the round interval is down to 62 ns (20 clock cycles on FPGA). In Figure~\ref{fig:trans}, we illustrate how the transmission works within one batch. Since the FPGA is re-initialized for each batch, we calculate the total transmission time by summing the timestamps returned from the FPGA for each batch. Experimental results show that the transmission time for each batch consistently matches the sum of the round intervals plus a constant number of clock cycles, which consists with the duration of the last round and a fixed latency of front panels and transceivers on verifier's FPGA (see Section \ref{sssec:trans} for details). 

\begin{figure}[!h]
    \centering
    \includegraphics[width=0.9\linewidth]{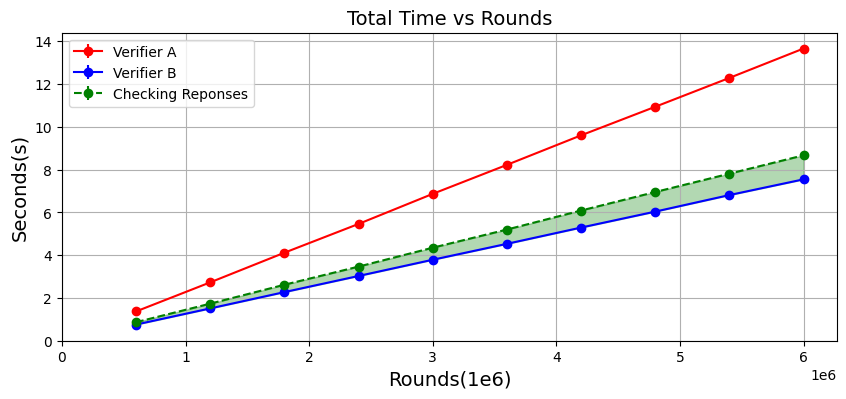}
    \caption{Total Execution Time $t_{\text{total}}$ of Verifier A ($\mathcal{V}_A$) and B ($\mathcal{V}_B$)}
    \label{fig:2p_total}
\end{figure}

Once all responses within a batch are received by each verifier, the verification phase enables verifiers $\mathcal{V}_A$ and $\mathcal{V}_B$ to consolidate their responses on $\mathcal{V}_A$ and validate them. However, we need to consider additionally various aspects of time consumption and uncertainties during the verification phase in the implementation, including network packet processing at various steps (see Section \ref{sssec:verf} for details). This explains exactly why the total time recorded by $\mathcal{V}_A$ exceeds that of $\mathcal{V}_B$, as illustrated in Figure~\ref{fig:2p_total}. Further analysis of the extra time recorded by $\mathcal{V}_A$ reveals that the time spent verifying the validity of the provers' responses and timings contributes only a small portion to the difference between $\mathcal{V}_A$ and $\mathcal{V}_B$, as indicated by the green dashed line in Figure~\ref{fig:2p_total}. We conclude that the primary source of delay is the preparation and processing of network packets, which acts as a significant bottleneck. This is corroborated by comparing the 3-prover implementation in Section \ref{ssec:3pexp}, where only a single verifier is involved, eliminating the need for network communication. Such a comparison provides us a more straightforward benchmarking result for performance evaluation.

\begin{table}[!h]
\centering
\begin{tblr}{
  width = \linewidth,
  hlines,
  vlines,
}
Graph Size $|E_G|$  & Required Rounds & $t_{prep}$(s) & $t_{chall}$(s) & $t_{vrfy}$(s) & $t_{total}$(s) \\
$\sim10^3$ & 600,000     &  0.4221  &  0.0373  &  0.1108  & 1.3702 \\
$\sim10^4$ & 6,000,000   &  4.2489  &  0.3726  &  1.1247  & 13.6652                 
\end{tblr}
\caption{Comparison of RZKP Implementation with different graph sizes (averaged over 1000 repetitions)}
\label{table: graph_rounds}
\end{table}

The experimental results in \cite{MN07} demonstrate that generated 3-colorable graphs can serve as challenging instances, with computational costs that grow exponentially with the number of vertices. Consequently, increasing the graph size can significantly enhance the security of the RZKP scheme. To accommodate potential future increases in graph sizes for improved security, our design offers the flexibility to easily adapt to larger graph sizes up to $|V_G|\sim8000$ and $|E_G|\sim10^4$, while maintaining consistent transmission times (see Section~\ref{sssec:trans} for details). However, this comes with a trade-off: the increase in graph size results in a proportional increase in the number of rounds. This is because the probability of detecting malicious provers in a single round by challenging an edge decreases as the graph instances grow larger. In this scenario, using graph instances with $|E_G|\sim10^4$ requires 10 times more rounds compared to graph instances with $|E_G|\sim10^3$, given the same security parameter. In Table~\ref{table: graph_rounds}, we compare the experimental performance of different graph sizes with the same security parameter $k=100$, showing that the data processing overhead in our hardware and software implementation is negligible concerning the performance for single round duration.

\subsection{Extension on Three-Prover RZKP Implementation}
\label{ssec:3pexp}
The two-prover RZKP implementation demonstrates its feasibility as a financial identification system in the near-term, where long-term quantum storage is unavailable.
To prepare for threats in the longer term, we seek solutions that is secure against entangled provers.
In \cite{Crepeau_2019}, a solution was proposed where a third party is introduced, and the resulting 3-prover RZKP-3-COL protocol is shown to be secure against entangled adversaries.
Here, we demonstrate the 3-prover RZKP-3-COL protocol in practice with a deployment in a laboratory environment.
Recall that the 3-prover protocol (See Protocol~\ref{prot:3prover}) is identical to the 2-prover one, except that the additional prover is asked the same challenge as either the first or second prover, with a probability of 1/2 each. That is, the additional prover is instructed to emulate the behavior of either the first or second prover. 
The verifier accepts the response of all three provers if the responses of the first two provers are accepted as in the 2-prover protocol, and the response from the additional prover repeats perfectly the response of either one of the first two provers according to the challenge.

\begin{figure}[!h]
    \centering
    \includegraphics[width=0.8\linewidth]{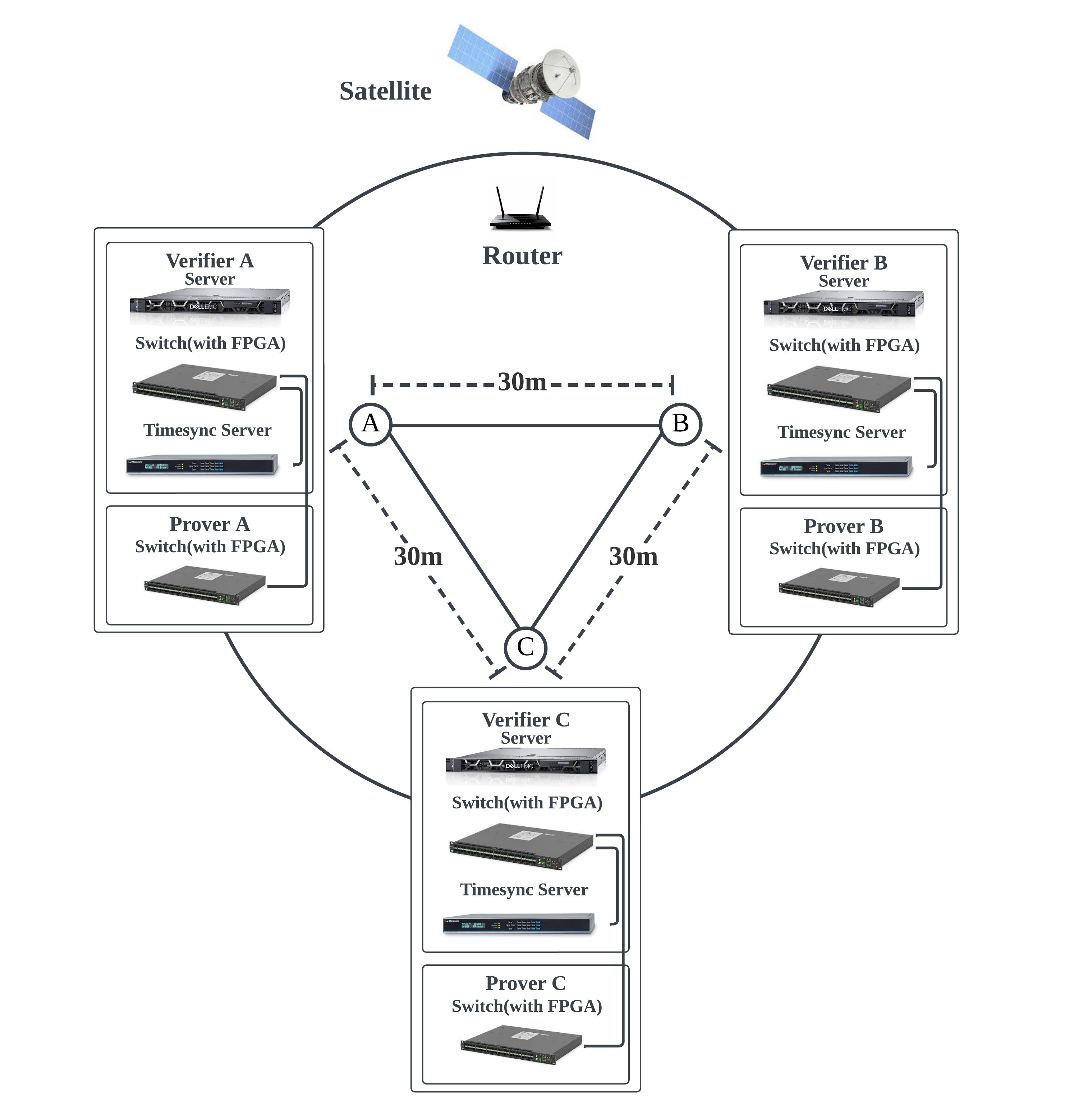}
    \caption{Intuitive extension of the RZKP-3-COL implementation to a 3-prover RZKP-3-COL implementation, with the addition of a third verifier-prover pair.}%3-Prover Intuitive Implementation by Replication
    \label{fig:3prover_ideal}
\end{figure}

Intuitively, a 3-prover protocol can be designed like Figure~\ref{fig:3prover_ideal}, where the three pairs of verifier-prover are topologically fixed at the vertices of a equilateral triangle with a length of 30 meters. 
This could be in principle implemented using the same techniques and optimizations as the 2-prover protocol. As the number of verifiers does not affect the proof system in theory, here, we instead choose to implement the 3-prover protocol in a setup where a single verifier that is connected to all three provers. The purpose of the switch is twofold. Firstly, it allows us to study the performance of a single verifier setup relative to a multi-verifier setup in Section~\ref{ssec:2pexp}, where the effect of network communication in the multi-verifier setup can be isolated. Secondly, it allows us to study the behavior and performance of the single verifier setup, which may be the natural scenario for some applications. For instance, provers may be connected to fixed network ports of an internal network that are connected to a central server that acts as a single verifier. 

\begin{figure}[!h]
    \centering
    \includegraphics[width=0.8\linewidth]{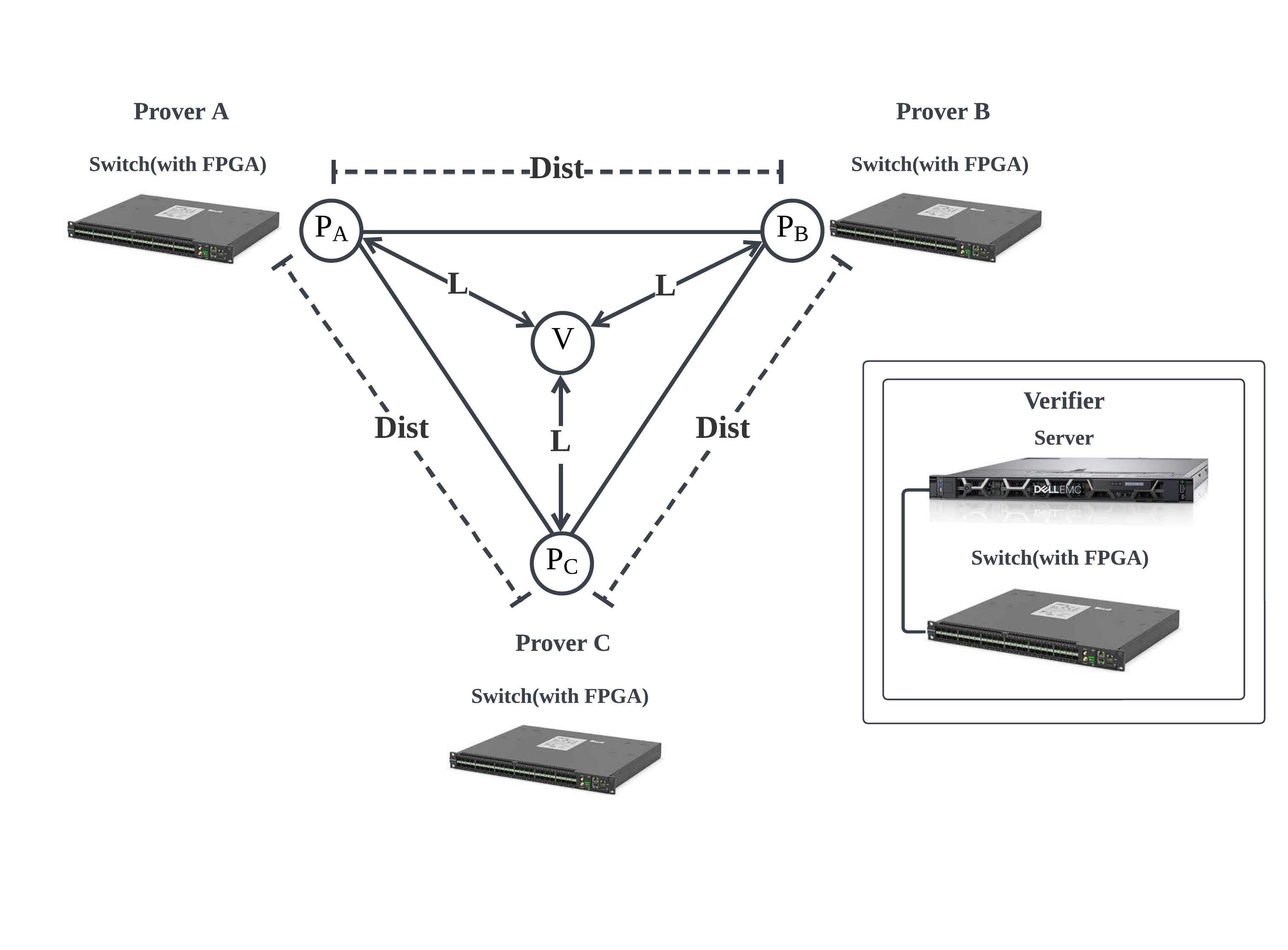}
    \caption{3-Prover Proof-of-Concept Implementation}
    \label{fig:3prover_real}
\end{figure}

Figure~\ref{fig:3prover_real} shows the single verifier setup, where we position the verifier $\mathcal{V}$ at the circumcenter of the equilateral triangle, and the three provers positioned at the vertices. In this setup, it is necessary to trust that the communication between the verifier and provers cannot be intercepted, even at long distance. As a result, an adversary has to wait for the challenges to arrive at each prover, before communicating the challenges between the provers to provide suitable responses that can pass the verification. This requires additional communication time between the provers relative to the honest case, and we can thus impose the relativistic constraint on the distance between the provers.

This change from a multi-verifier setting to a single verifier setting requires some changes to the implementation. In our implementation, we deployed three 30-meter long-range single mode fibers as the communication channels between the verifier and the provers. This results in a distance of
\begin{align*}
    \text{Dist} = 2\times\cos 30^{\circ} L=\sqrt{3}L\approx 52\text{m}
\end{align*}
between each pair of provers, satisfying the minimum separation for the relativistic constraint with our implementations.

Having a single verifier also removes the need for synchronization between separate verifiers, thereby simplifying the verifier's implementation. During the transmission phase, the verifier interfaces with a host server via a Network Interfacing Controller (NIC), allowing the verifier to send out challenges and receive responses from all three provers simultaneously. In the verification phase, the verifier performs the verification on its host server directly without additional communication. After testing, we enlarge the configuration of round interval from 20 to 50 clock cycles on FPGA to make sure that each round the communication between verifier and every prover can be completed properly. The remaining software and hardware designs are kept unchanged.

\begin{figure}[!h]
    \centering
    \includegraphics[width=0.8\linewidth]{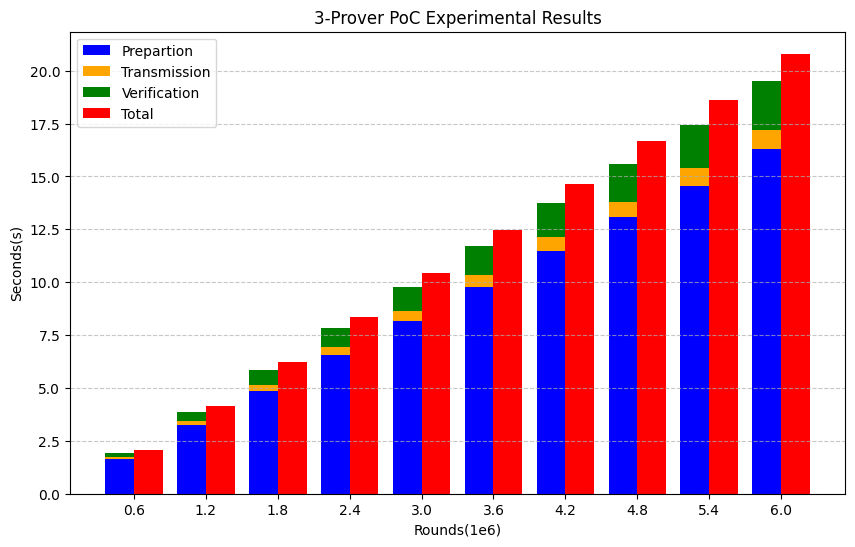}
    \caption{Time taken for the preparation, transmission and verification phases of the 3-prover RZKP-3-COL implementation with a single verifier.}
    \label{fig:3p_total}
\end{figure}
 
In Figure~\ref{fig:3p_total}, we show our 3-prover experimental results in terms of the time consumption at every phase. We first observe that the preparation takes more time than the two-prover experiment.
This is because the challenges to all three provers are now prepared by a single verifier, and more time is required to pre-load the necessary information on a single FPGA. Another crucial observation is that the relative gap between the cumulative time of the three phases and the total time is much smaller than that observed for the two-party case in Figure~\ref{fig:2p_total} by circumventing the preparation/processing of network packets and delays in verification phase. This highlights the fact that the network processing in the verification phase is a crucial bottleneck that should be optimized. Since we are focusing on the critical processes that is in the scope of the protocol specification, we leave the optimization on the network processing in the verification phase for future work.

\begin{table}[!h]
\centering
\begin{tblr}{
  width = \linewidth,
  hlines,
  vlines,
}
Reference & Rounds    & $t_{prep}$(s)          & $t_{chall}$(s)& $t_{vrfy}$(s) & $t_{total}$(s) \\
Experiment & 600,000 (6e5)   & 1.6253        &  0.0931        &  0.2212       &  2.0688    \\
Experiment & 6,000,000 (6e6) & 16.2881       &  0.9309        &  2.2791       &  20.8028   \\
% \cite{Crepeau_2019}  & $k(21|E_G|)^4$ ($\sim3\text{e}19$)  &  \textbf{---}             &  \textbf{---}              &   \textbf{---}            & $\sim1\text{e}14$ \\
\cite{Alikhani_2021} & $k(11|E_G|)^4$ ($\sim2\text{e}18$)  & \textbf{---}              &  \textbf{---}              &  \textbf{---}             & $\sim7\text{e}12$ \\
\cite{Shi_2024}      & $k\frac{2401}{16}(|E_G|)^4$ ($\sim1.5\text{e}16$)  &  \textbf{---}             & \textbf{---}               &   \textbf{---}            &  $\sim5.25\text{e}10$\\
\end{tblr}
\caption{3-Prover RZKP Performance}
\label{table: quantum_prover}
\end{table}

While the three-prover RZKP-3-COL protocol is %in the vanilla setting is proven to be 
sound against entangled provers, it remains impractical due to the large number of rounds required. This is illustrated by Table~\ref{table: quantum_prover}, which contains an estimation of time required for our implementation to reach the security bound from different existing works (with graph size $|E_G|\sim10^3$ and $k=100$).

\subsection{Two-Prover RZKP Security against Quantum Provers}
\label{ssec:2proveralt}
Recall that the security of RZKP can be defined by two properties: (1) soundness -- dishonest provers are only able to pass RZKP with low probability $\varepsilon_{sou}$, (2) zero-knowledge -- dishonest verifiers are unable to learn the secret (in this case, the coloring of graph $G$). 
It was proposed in \cite{Alikhani_2021} that the use of a commutative gadget presented in \cite{Ji_2013} can shed light on the security of RZKP-3-COL (see Methods) with two entangled provers.
However, it remains difficult to prove the security.
Here, we instead prove security for a similar RZKP protocol, termed Alt-RZKP-3-COL, where prover A provides responses corresponding to both bit values $b=0$ and $b=1$ instead of a chosen bit value $b$.
This modified protocol is presented below in Protocol~\ref{prot:Alt-ZK-3-COL}, where prover A provides four responses corresponding to both $b$, and additional checks are performed by the verifiers.

\begin{protocol}[!h]
\caption{Alt-RZKP-3-COL Game }
    Provers $\mathcal{P}_{A}$ and $\mathcal{P}_{B}$ pre-agree on a random 3-coloring of $G$: $\left\{ (i,c_{i}) | c_i \in \mathbb{F}_3\right\}_{i\in V_G}$ 
    such that $({i},{j}) \in E_G \implies c_{{j}} \neq c_{{i}}$, and two labellings $w_i^0,w_i^1\in_R \mathbb{F}_3$ per vertex that should sum up to a three-coloring, namely, ${w_i^0+w_i^1= c_i}$.
    
    {\bf Commit phase:}
    \begin{itemize}
        \item Verifier $\mathcal{V}$ picks $((i,j), ((i'\!,j'), b))\in_{\mathcal{D}_{iji'j'b}} \left(E_G^2\times \mathbb{F}_2\right)$, where $\mathcal{D}_{iji'j'b}=\frac{1}{2\abs{E_G}}(\frac{\delta_{ii'}+\delta_{ij'}}{2\abs{\neigh_{G}(i)}}+\frac{\delta_{jj'}+\delta_{ji'}}{2\abs{\neigh_{G}(j)}})$, where $\abs{\neigh_{G}(i)}$ refers to the number of neighbors of vertex $i$.% and $I[(i,j)\in E_G]$ is an indicator function requiring $(i,j)\in E_G$. 
        $\mathcal{V}$ then sends $(i,j)$ to $\mathcal{P}_A$ and $((i'\!,j'),b)$ to $\mathcal{P}_B$.
        \item If $(i,j) \in E_G$ then $\mathcal{P}_A$ replies $w_i^0$, $w_i^1$, $w_j^0$ and $w_j^1$.
        \item If $(i'\!,j') \in E_G$ then $\mathcal{P}_B$ replies $\tilde{w}_{i'}^{b}$ and $\tilde{w}_{j'}^{b}$.
    \end{itemize}
    {\bf Check phase:}
    \begin{itemize}
        \item[] {\bf Edge-Verification Test:}
        \item $\mathcal{V}$ checks if $w_i^0+w_i^{1}\neq w_j^0+w_j^1$.
        \item[] {\bf Well-Definition Test:}
        \item If $(i,j) = (i',j')$  then $\mathcal{V}$
        checks if $w_i^b=\tilde{w}_{i'}^b$, and $w_j^b=\tilde{w}_{j'}^b$.
        \item If $(i,j)\cap (i',j')=i$ then $\mathcal{V}$ checks if $w_i^b=\tilde{w}_i^b$.
        \item If $(i,j)\cap (i',j')=j$ then $\mathcal{V}$ checks if $w_j^b=\tilde{w}_j^b$.
        \item[] {\bf Accept Decision:}
        \item $\mathcal{V}$ accepts if all relevant checks passes.
    \end{itemize}
\label{prot:Alt-ZK-3-COL}
\end{protocol}

The soundness relies on the argument that if there exists an $\varepsilon$-perfect strategy for the Alt-RZKP-3-COL for an \emph{extended graph} $G'$ formed from graph $G$ using commutative gadgets, then one can construct a classical strategy for vertex-3-COL for graph $G$ that can pass the edge verification test with a probability of at least $1-O(\varepsilon^{-4})$.
Since the classical bound for vertex-3-COL is known, we can recover a bound on the range of $\varepsilon$-perfect strategies that are possible.
The overall soundness thus relies on the construction of the classical strategy, and can be split into three phases, each corresponding to a different observation:
\begin{enumerate}
    \item Reduction of a strategy for Alt-RZKP-3-COL game to a strategy for the binary constraint system (BCS) game, for which a construction of an almost satisfying assignment of $G'$~\cite{Slofstra2018,Paddock2024} is known.
    \item Lifting the almost satisfying assignment of $G'$ to an almost fully commuting satisfying assignment of $G$, which relies on observations that a gadget used to form $G'$ from $G$ ensures that a quantum satisfying assignment of $G'$ also implies that the assignment is fully commuting~\cite{Ji_2013}.
    \item Developing a classical strategy for vertex-3-COL based on the almost fully commuting satisfying assignment, which involves repeated measurements that has minimal disturbance on the pre-shared state $\ket{\psi}$ due to the almost commuting nature of the assignment.
\end{enumerate}

An almost satisfying assignment for a graph $G$ is given by a set of projectors $B^i_{\alpha}$ corresponding to measurement operators of vertex $i$ and color $\alpha$.
Importantly, they satisfy three conditions:
\begin{enumerate}
    \item Tracial property: Projectors almost commute with $\rho^{\frac{1}{2}}$, where $\rho=\Tr_A[\dyad{\psi}]$ is the partial trace of prover A's part of the pre-shared state. %\ym{What A here stands for? is it a partial trace? It's a bit confusing.}
    \item Neighboring vertices almost commute: Projectors corresponding to neighboring vertices almost commute, i.e. the trace of $[B^i_{\alpha},B^j_{\beta}]\ket{\psi}$ is small.
    \item Edge coloring: Projectors almost satisfy the edge coloring requirements, i.e., the trace of $B^i_{\alpha}B^j_{\alpha}\ket{\psi}$ is small.
\end{enumerate}
These assignments can be constructed from prover B's measurement operators in the BCS game, using results from \cite{Slofstra2018,Paddock2024}.
The construction of a BCS game strategy from the $\varepsilon$-perfect Alt-RZKP-3-COL strategy is straightforward by performing consecutive measurements, noting that the measurements do not alter the state much, by gentle measurement lemma~\cite{Wilde2017}.

\begin{figure}[!h]
    \centering
    \includegraphics[width=0.4\linewidth]{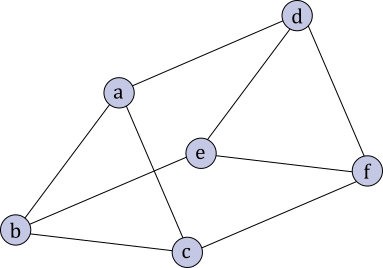}
    \caption{Triangular prism-based commutative gadget~\cite{Ji_2013}.}
    \label{fig:commutative_gadget}
\end{figure}

Having an almost satisfying assignment remains insufficient to construct a classical strategy.
Instead, we require that all projectors almost commute, since commuting projectors can form classical strategies -- they share the same eigenbasis that are unaltered by the projective measurements.
To achieve commutation, we use the commutative gadget introduced in \cite{Ji_2013}, depicted in Figure~\ref{fig:commutative_gadget}.
An extended graph $G'$ can be formed from graph $G$ by attaching a gadget (vertices a and e are a vertex pair of the original graph) to all vertex pairs that do not form an edge, $(i,j)\notin E_G$.
Note that the addition of the gadget does not change whether a graph is 3-colorable.
With this extension, we adapt a proof in \cite{Ji_2013} that shows perfect quantum satisfying assignment of $G'$ implies that the assignment operators are commuting to demonstrate that almost satisfying assignment operators of $G'$ implies that the assignment operators $B^i_{\alpha}$ corresponding to vertices of the original graph, $i\in G$, are almost commuting.
This, in turn, allows a classical strategy to be formed from the almost commuting projectors $B^i_{\alpha}$ for $i\in G$ (which are also an almost satisfying assignment of $G$) for a 3-coloring game with graph $G$, thereby completing the reduction.

Combining the results, we are able to demonstrate soundness by computing a quantum value $\omega_q$ (winning probability for any entangled adversary) in Thm.~\ref{thm:soundness_scaling} below, with the exact form of $\varepsilon$ provided in Thm.~\ref{thm:soundness_main_thm}.
\begin{theorem}
\label{thm:soundness_scaling}
    Consider a graph $G$ that is not 3-colorable. Then, the Alt-RZKP-3-COL game with extended graph $G'$ is sound against entangled provers with quantum value $\omega_q\sim 1-\frac{1}{\abs{E_G}^8}$, where $\abs{E_G}$ is the number of edges in graph $G$.
\end{theorem}
The modified protocol remains perfect zero-knowledge as it reduces directly to the perfect zero-knowledge property demonstrated in 3-prover-RZKP-3-COL in \cite{Crepeau_2019}.
Details of the proof is provided in Sec.~\ref{sec:Alt-RZKP-3-COL_ZK_Proof}.

\begin{theorem}
\label{thm:Alt-RZKP-3-COL_ZK}
    The Alt-RZKP-3-COL protocol for graph $G'$ is perfect zero-knowledge. 
\end{theorem}

It is straightforward to extend our implementation to Protocol~\ref{prot:Alt-ZK-3-COL} using the same design deployed here.
While we are able to demonstrate that Alt-RZKP-3-COL is sound against entangled adversaries, the bound obtained has room for improvement.
The tightening of the bound can be explored in future work, for instance, by examining the possibility of directly mapping the Alt-RZKP-3-COL operators to an almost fully commuting satisfying assignment without going through a reduction to Alt-Edge-3-COL, which incurs a $\sqrt{\varepsilon}$ penalty. Experimentally, the significant increase to graph size due to the introduction of gadgets and the significant larger number of rounds (see Appendix~\ref{app:perf}) would render the protocol impractical.

% \subsection{Improved Bound on Three-Prover RZKP}

\section{Discussion}

\subsection{Message Authentication}
\label{ssec:msg_auth}
While identification can be useful in many applications, there can sometimes be a need for the communication between a client and server to be authenticated.
These may include data such as financial transaction information, bank account details and account balances.

By adapting the RZKP as a sub-protocol, we propose a method of sending authenticated messages inspired from quantum position verification-based message authentication~\cite{Buhrman2014}. 
However, instead of utilizing valid and empty responses to represent bits, here we propose encode codewords via the timing of the responses to RZKP, allowing the responses to act as the identifier for the message -- if the response is correct (i.e. from honest provers), then the codeword encoded in the timing of that response would be from the honest prover.
To encode bit 0 and 1, we allow prover A to respond with at some delay time $t_0$ and $t_1$ respectively with $t_1>t_0$. 
The use of timing naturally fits into the RZKP scheme since these timings are already measured in the implementation to enforce the relativistic constraints.
Moreover, due to the sensitivity of the clock required for RZKP, small timing differences between $t_0$ and $t_1$ is sufficient for distinguishing the two delays.
% Here, we encode codewords via the timing of the responses to RZKP, allowing the responses to act as the identifier for the message -- if the response is correct (i.e. from honest provers), then the codeword encoded in the timing of that response would be from the honest prover.
% To encode bit 0 and 1, we allow prover A to respond with at some delay time $t_0$ and $t_1$ respectively with $t_1>t_0$. %note distance between prover-verifier pairs would increase due to additional delays).
% We note that the chosen delays have difference that is sufficiently large such that it is physically impossible for the adversary to force the response to arrive at the verifier with a shorter delay.
Since any adversary (man-in-the-middle) can only delay responses to change messages from an earlier to a later response (0 to 1), but not from a later response to an earlier response (1 to 0), we can encode single bit messages as 01 and 10 codewords, making it difficult for adversaries to switch between the two codewords.

The goal is to allow prover A to send a message to verifier A in an authentic manner -- i.e. verifier A is able to verify that the messages are sent from prover A.
We assume that provers hold functions $Rsp_1$ and $Rsp_2$ respectively to compute appropriate responses to pass the RZKP, and may share random string $s$ (including secrets, such as coloring). 
The protocol is presented in detail in Protocol 2.
Note that the timing $\tilde{\delta}$ is chosen at a value smaller than $\frac{t_1-t_0}{2}$.
We note that in general, we can also include additional steps where the actual message $M$ is sent, and the received message is checked against the decoded message.

\begin{protocol}[!h]
\caption{RZKP-based Message Authentication}
\begin{enumerate}[leftmargin=*,topsep=0pt]
    \item \textbf{Prover Encoding}: Prover A $\mathcal{P}_{A}$ encodes the message $M\in\{0,1\}$ as the codewords $C=01$ for $M=0$ and $C=10$ for $M=1$.
    \item \textbf{RZKP Sub-protocol}: For $i=1,2$, and $k=1,\cdots,n$
    \begin{enumerate}
        \item \textbf{Challenge}: Verifiers $\mathcal{V}_{A}$ and $\mathcal{V}_{B}$ sends challenges $x_{i,k}\in\mathcal{X}$ and $y_{i,k}\in\mathcal{Y}$ to provers $\mathcal{P}_{A}$ and $\mathcal{P}_{B}$ respectively such that the challenges arrive at the provers at some $t_{st,i,k}$.
        \item \textbf{Prover Computation}: $\mathcal{P}_{A}$ computes a response $a_{i,k}=Rsp_1(x_{i,k},s)$ and $\mathcal{P}_{B}$ computes a response $b_{i,k}=Rsp_2(y_{i,k},s)$.
        \item \textbf{Prover Response}: $\mathcal{P}_{B}$ sends the response $b$ to $\mathcal{V}_{A}$ immediately, while $\mathcal{P}_{A}$ sends the response $a$ to $\mathcal{V}_{A}$ at time $t_{st,i,k}+t_{C_i}$.
        \item \textbf{Verifier Receipt}: The verifiers records the time the messages are received $t_{rec,i,k,1}$ and $t_{rec,i,k,2}$.
        \item \textbf{Timing Check}: The verifiers checks if the received messages are within some tolerated time $t_{rec,i,k,j}-t_{start,i,k}<\delta_j$. If the timings are outside the tolerated range, the protocol aborts. $\mathcal{V}_{A}$ decodes the timing of the message $\hat{t}_{C_i,k}=t_{rec,i,k,A}-t_{start,i,k}-t_{resp,A}$, where $t_{resp,A}$ is the time it takes for the message sent from $\mathcal{P}_{A}$ to arrive at $\mathcal{V}_{A}$.
    \end{enumerate}
    \item \textbf{Message and Timing Checks}: The verifiers jointly check if $V(x_{i,k},y_{i,k},a_{i,k},b_{i,k})=1$ for all $i,k$. For each $i=1,2$, verifier 1 checks if $\abs{\hat{t}_{C_i,k}-\hat{t}_{C_i,k'}}\leq \delta'$ for all $k=k'$, i.e. whether the timing selected by $\mathcal{P}_{A}$ across $k=1,\cdots,n$ rounds are identical. If either check fails, the protocol aborts.
    \item \textbf{Message Decoding}: $\mathcal{V}_{A}$ decodes $\hat{C}_i=\begin{cases} 0 & \abs{\frac{1}{n}\sum_{k=1,\cdots,n}\hat{t}_{C_i,k}-t_0}<\tilde{\delta} \\ 1 & \abs{\frac{1}{n}\sum_{k=1,\cdots,n}\hat{t}_{C_i,k}-t_1}<\tilde{\delta} \\ \perp & otherwise \end{cases}$. If the received codeword is $\hat{C}=01$, verifier 1 decodes the message $\hat{M}=0$ and if the received codeword is $\hat{C}=10$, verifier 1 decodes the message $\hat{M}=1$. Otherwise, the protocol aborts.
\end{enumerate}
\label{prot:RZKP_AuthMsg}
\end{protocol}

The security for a message authentication protocol can be defined by its soundness, which is the probability that the authentication passes but the message sent by the prover does not match the message received by the verifier.
Here, we consider an adversary that is present at the two prover-verifier pair locations, and are able to intercept any messages between the prover and verifier, but has no knowledge of the secrets held by the provers.
Suppose that the RZKP sub-protocol utilized has a winning probability of $\varepsilon_{RZKP}$ for any adversary with no knowledge of the secret (e.g. the colorings for a 3-COL problem).
% By assumption, the adversary has little chance of impersonating the valid prover and send an authenticated message to the verifiers.
When a $t_1$ delay is chosen, the adversary is unable to force the response to arrive at verifier A at an earlier time $t_0$.
As such, the adversary is forced to provide a response on its own, i.e. it has to pass the RZKP with its own responses, which it can do with probability at most $\varepsilon_{RZKP}$.
Since the conversion for any message $M$ to $\bar{M}$ requires the flipping of at least one codeword bit from 1 to 0, the probability that he can succeed in altering the message without detection is $\varepsilon_{RZKP}$, i.e. $\Pr[M\neq\hat{M},acc]\leq\varepsilon_{RZKP}$.

\subsection{Distance Bounding}
\label{ssec:pos_veri}
In our demonstrated RZKP implementation, such as the 2-prover setup, communication between each verifier and its corresponding prover is facilitated through a 1-meter DAC fiber. This configuration not only enhances the permissible distance between verifiers but also reduces the requirements for relativistic constraints. Additionally, the setup closely resembles a real-world ATM authentication scenario. Despite this, as long as the non-signaling principle remains intact and prevents communication between provers within the protocol, we propose a natural extension of RZKP for identity verification use cases: a distance bounding protocol. This protocol enables a verifier to authenticate a prover using ZKP and evaluate whether the prover is located in his vicinity, based on the time taken for the prover to respond to a challenge.

Distance bounding protocols \cite{Lopez10,Rasmussen10,Capkun03,Hancke05} have become increasingly significant in contactless systems, such as electronic payment and access control systems \cite{Avoine18}, as a defense against relay attacks. These protocols are largely inspired by the first distance bounding protocol proposed in \cite{BC93}. In this setup, a verifier-prover pair shares a public/private key for generating signature and is separated by a distance \(d\). Assuming the communication between verifier/prover is with a propagation speed closed to speed of light, the protocol begins with the prover committing to a random string and sending the commitment to the verifier. The verifier then samples a challenge of the same size as the random string and transmits the challenge bit by bit to the prover. Upon receiving each challenge bit, the prover performs a bitwise XOR operation between the random string and the challenge. The verifier measures the time from sending each challenge to receiving the response, known as the round-trip time (RTT). Finally, the prover signs a message containing the received challenges, sent responses, and the opening of the commitment. The responses must be consistent, and the RTT for each round must be upper-bounded by \(2d/c + t_p\), where \(c\) is the speed of light and \(t_p\) is the estimated time the prover takes to compute the responses. By outlining the main concept of the distance bounding protocol, we recognize that the RZKP setup facilitates verification of provers in a similar manner, without sharing public/private key pairs for identity verification.

\begin{figure}[!ht]
    \centering
    \includegraphics[width=\linewidth]{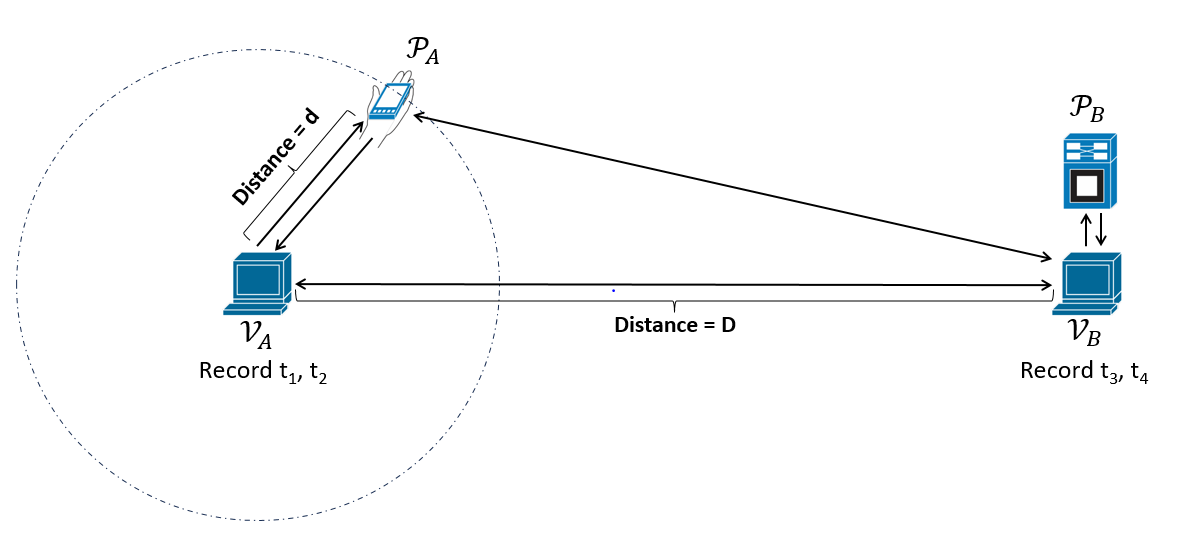}
    \caption{Distance Bounding on $\mathcal{P}_A$ based on RZKP}
    \label{fig:distance_bounding}
\end{figure}

Without loss of generality, we consider the same setup as 2-prover RZKP with two verifiers ($\mathcal{V}_A, \mathcal{V}_B$) and two provers ($\mathcal{P}_A, \mathcal{P}_B$). $\mathcal{V}_{A}, \mathcal{P}_{A}$ are at a certain location and $\mathcal{V}_{B}, \mathcal{P}_{B}$ are at a different location. Let $D$ denote the distance between the two locations. In terms of bounding the distance of $\mathcal{P}_A$ from its corresponding verifier $\mathcal{V}_A$, we further denote the allowed distance between $\mathcal{V}_A$ and $\mathcal{P}_A$ can maximally reach to $d$. On the other hand, $\mathcal{V}_B$ and $\mathcal{P}_B$ are close to each other by default. For a single round, let $t_1$, $t_2$ (resp. $t_3$, $t_4$) denote the time when $\mathcal{V}_A$ (resp. $\mathcal{V}_B$) sends the challenge and receives a response to/from $\mathcal{P}_A$ (resp. $\mathcal{P}_B$). To make sure that $\mathcal{P}_A$ and $\mathcal{P}_B$ do not communicate during execution, in addition to the consistency of responses, the following conditions should hold:
\begin{align*}
    \big|t_1-t_4\big|&<\frac{D}{c}\\
    \big|t_2-t_3\big|&<\frac{D}{c}\\
    \big|t_1-t_2\big|&<\frac{2d}{c}+t_p
\end{align*}
, where $c$ denotes the speed of light. In this case, $\mathcal{V}_A$ can effectively define a legitimate range of area where $\mathcal{P}_A$ is residing in, as shown in Figure~\ref{fig:distance_bounding}.

\section{Methods}

\subsection{Protocols}
\label{ssec:protocols}
In this section, we describe various relativistic 3-coloring protocols that we are studying and implementing. Note that in the protocol descriptions, the verifier acts as a single entity, but in practice, we divide the verifier into the same number of entities as there are provers. By spatially separating the provers and ensuring that each verifier interacts with only one prover, the system prevents the provers from communicating with each other during the interaction. This spatial separation is a practical consideration for implementing the relativistic setting by utilizing the distance between the verifiers themselves. Theoretically, splitting the verifier into multiple entities does not alter the computational power or the class of problems that the system can verify. The theoretical framework remains unchanged because the core principles of interactive proofs are preserved.

As demonstrated in Protocol~\ref{prot:2prover} with a 2-prover setting \cite{Crepeau_2019}, let $\mathcal{D}_{\text{2RZKP}}$ represent the distribution of challenges from the verifier to the two provers. In each round, the verifier randomly samples edges from $E_G$ and performs either an \emph{edge-verification test} or a \emph{well-definition test}. To align with the implementation, the probabilities of conducting these two tests are set at $\frac{1}{5}$ and $\frac{4}{5}$, respectively. Additionally, the well-definition test on two different vertices is conducted with a probability of $\frac{1}{2}$ for each vertex.

\begin{protocol}[!h]
\caption{RZKP-3-COL (from \cite{Alikhani_2021, Crepeau_2019})}
    Provers $\mathcal{P}_{A}$ and $\mathcal{P}_{B}$ pre-agree on a random 3-coloring of $G$: $\left\{ (i,c_{i}) | c_i \in \mathbb{F}_3\right\}_{i\in V_G}$ 
    such that $({i},{j}) \in E_G \implies c_{{j}} \neq c_{{i}}$, and two labellings $w_i^0,w_i^1\in_R \mathbb{F}_3$ per vertex that should sum up to a three-coloring, namely, ${w_i^0+w_i^1= c_i}$.
    
    {\bf Commit phase:}
    \begin{itemize}
        \item Verifier $\mathcal{V}$ picks $(((i,j),b), ((i'\!,j'), b'))\in_{\mathcal{D}_{\text{2RZKP}}} \left(E_G\times \mathbb{F}_2\right)^2$, where $\mathcal{D}_{\text{2RZKP}}=\frac{\delta_{ii'}\delta_{jj'}\delta_{b,b'\oplus 1}\varepsilon}{\abs{E_G}}+\frac{\delta_{bb'}(1-\varepsilon)}{2\abs{E_G}}(\frac{\delta_{ii'}+\delta_{ij'}}{\abs{\neigh_{G}(i)}}+\frac{\delta_{jj'}+\delta_{ji'}}{\abs{\neigh_{G}(j)}})$
        %$\mathcal{D}_{\text{2RZKP}}=\frac{\varepsilon}{\abs{E}}+\frac{1-\varepsilon}{2\abs{E}}(\frac{1}{\abs{\neigh_{G}(i)}}+\frac{1}{\abs{\neigh_{G}(j)}})$
        , where $\abs{\neigh_{G}(i)}$ refers to the number of neighbors of vertex $i$ and $\varepsilon=\frac{1}{5}$.
        $\mathcal{V}$ then sends $((i,j),b)$ to $\mathcal{P}_A$ and $((i'\!,j'),b')$ to $\mathcal{P}_B$.
        \item If $(i,j) \in E_G$ then $\mathcal{P}_A$ replies $w_i^b$ and $w_{j}^b$.
        \item If $(i'\!,j') \in E_G$ then $\mathcal{P}_B$ replies $\tilde{w}_{i'}^{b'}$ and $\tilde{w}_{j'}^{b'}$.
    \end{itemize}
    
    {\bf Check phase:}
    \begin{itemize}
        \item[] {\bf Edge-Verification Test:}
        \item If $(i,j) = (i'\!,j')$ and $b \neq b'$ then $\mathcal{V}$
         accepts iff $w_{i}^b+\tilde{w}_{i'}^{b'}= w_{j}^b+\tilde{w}_{j'}^{b'}$.
        \item[] {\bf Well-Definition Test:}
        \item If $(i,j)= (i'\!,j')$ and $b=b'$ then $\mathcal{V}$ accepts iff $(w_{i}^b = \tilde{w}_{i}^{b'})	\land (w_j^b = \tilde{w}_j^{b'})$.
        \item If $(i,j)\cap (i'\!,j')=i$ and $b=b'$ then $\mathcal{V}$ accepts iff $w_{i}^b = \tilde{w}_{i}^{b'}$.
        \item If $(i,j)\cap (i'\!,j')=j$ and $b=b'$ then $\mathcal{V}$ accepts iff $w_j^b = \tilde{w}_j^{b'}$.
    \end{itemize}
    \label{prot:2prover}
\end{protocol}

In the 3-prover protocol (see Protocol~\ref{prot:3prover}), the distribution of challenges, denoted as $\mathcal{D}{\text{3RZKP}}$, follows the same pattern as $\mathcal{D}{\text{2RZKP}}$. However, in each round, the challenge given to the additional prover is randomly identical to either the first or the second prover's challenge. This approach is known to achieve soundness against quantum-entangled provers, as demonstrated in \cite{Crepeau_2019}.

\begin{protocol}[!h]
\caption{3-prover RZKP-3-COL (from \cite{Crepeau_2019})}
    Provers $\mathcal{P}_{A},\mathcal{P}_{B}$, and $\mathcal{P}_{C}$ pre-agree on a random 3-coloring of $G$: $\left\{ (i,c_{i}) | c_i \in \mathbb{F}_3\right\}_{i\in V_G}$ such that $({i},{j}) \in E_G \implies c_{{j}} \neq c_{{i}}$, and two labellings $w_i^0,w_i^1\in_R \mathbb{F}_3$ per vertex that should sum up to a three-coloring, namely, ${w_i^0+w_i^1= c_i}$. 
    
    {\bf Commit phase:}
    \begin{itemize}
    \item Verifier $\mathcal{V}$ picks $(((i,j),b), ((i'\!,j'), b'), ((i''\!, j''),b''))\in_{\mathcal{D}_{\text{3RZKP}}} \left(E_G\times \mathbb{F}_3\right)^3$, where we obtain $\mathcal{D}_{\text{3RZKP}}$ from $\mathcal{D}_{\text{2RZKP}}$ and set $((i''\!,j''),b'') = ((i,j),b)$ and $((i''\!,j''),b'') = ((i'\!,j'),b')$ with probability $\frac{1}{2}$ each. $\mathcal{V}$ then sends $((i,j),b)$ to $\mathcal{P}_A$, sends $((i'\!,j'),b')$ to $\mathcal{P}_B$, and sends 
     $((i''\!,j''), b'')$ to $\mathcal{P}_C$.
    \item If $(i,j)\in E_G$  then $\mathcal{P}_A$ replies $w_i^b$ and $w_{j}^b$.
    \item If $(i'\!,j')\in E_G$ then $\mathcal{P}_B$ replies $\tilde{w}_{i'}^{b'}$ and $\tilde{w}_{j'}^{b'}$.
    \item If $(i''\!,j'')\in E_G$ then $\mathcal{P}_C$ replies $\hat{w}_{i''}^{b''}$ and $\hat{w}_{j''}^{b''}$.
    \end{itemize}
    
    {\bf Check phase:}
    \begin{itemize}
    \item[] {\bf Consistency Test:}
    \item  If $((i''\!,j''),b'') = ((i,j),b)$, then $\mathcal{V}$ checks if $(w_{i}^b,w_{j}^b) = (\hat{w}_{i''}^{b''},\hat{w}_{j''}^{b''})$.
    \item  If $((i''\!,j''),b'') = ((i'\!,j'),b')$, then $\mathcal{V}$ checks if $(\tilde{w}_{i'}^{b'},\tilde{w}_{j'}^{b'}) = (\hat{w}_{i''}^{b''},\hat{w}_{j''}^{b''})$.
    \item[] {\bf Edge-Verification Test:}
    \item If $(i,j) = (i'\!,j')$ and $b\neq b'$ then $\mathcal{V}$
     checks if $w_{i}^{b}+\tilde{w}_{i'}^{b'} = w_{j}^{b}+\tilde{w}_{j'}^{b'}$.
    \item[] {\bf Well-Definition Test:}
    \item If $(i,j)= (i'\!,j')$ and $b=b'$ then $\mathcal{V}$ checks iff $(w_{i}^b = \tilde{w}_{i}^{b'})	\land (w_j^b = \tilde{w}_j^{b'})$.
    \item If $(i,j)\cap (i'\!,j')=i$ and $b=b'$ then $\mathcal{V}$ checks if $w_{i}^b = \tilde{w}_{i'}^{b'}$.
    \item If $(i,j)\cap (i'\!,j')=j$ and $b=b'$ then $\mathcal{V}$ checks if $w_j^b = \tilde{w}_{j'}^{b'}$.
    \item[] {\bf Accept Decision:}
    \item $\mathcal{V}$ accepts if all relevant checks passes.
    \end{itemize}
    \label{prot:3prover}
\end{protocol}

\subsection{Graph Generation}
\label{ssec:graph_gen}
In this section, we describe methods for efficiently generating large 3-colorable graph instances that are sufficiently challenging to solve.  
We draw on the concept introduced in \cite{MN07}. In essence, the authors propose a constructive algorithm for generating 3-colorable graph instances using a successive embedding method. Specifically, an instance is chosen uniformly at random from a specific set of seven 4-critical graphs to serve as a subgraph, and these are embedded one by one by replacing an edge of the first graph with the second one. It is important to note that each n4c-free 4-critical graph in the set is not 3-colorable on its own, but becomes 3-colorable by deleting any edge. This construction method ensures that the resulting graph instances are always 3-colorable. The authors further demonstrate through experimental results that the computational cost to solve the generated instances is exponential in the number of vertices.

In practice, we require the graph generator to keep track of the vertex coloring and ensure consistency throughout the generation process. In our implementation, we achieve this by using 
the 3-colorable instance generator specified in \cite{MN07}. Simultaneously, we perform graph coloring using the DSatur algorithm \cite{bre79} on the chosen subgraph (including the initial graph) before the embedding operation, and record the edge that is to be deleted. To ensure efficient tracking, we stipulate that the new edge to be deleted from the embedded graph should be selected from one of the three edges created during the embedding operation, and at least one vertex of this edge should have a degree of 3. We maintain the coloring of the first graph and permute the colors of the second graph to match after selecting the next edge to be deleted. This approach allows the coloring of the embedded graph to be easily adjusted to satisfy 3-colorability. Finally, we perform a rejection 3-coloring test on the final generated graph. Note that our implementation excludes MUG10 from the set of 4-critical graphs to achieve the above mechanism. Nevertheless, the overall difficulty of solving the 3-coloring problem on the generated graph instances remains intact.

\begin{figure}[!ht]
    \centering
    \includegraphics[width=0.7\linewidth]{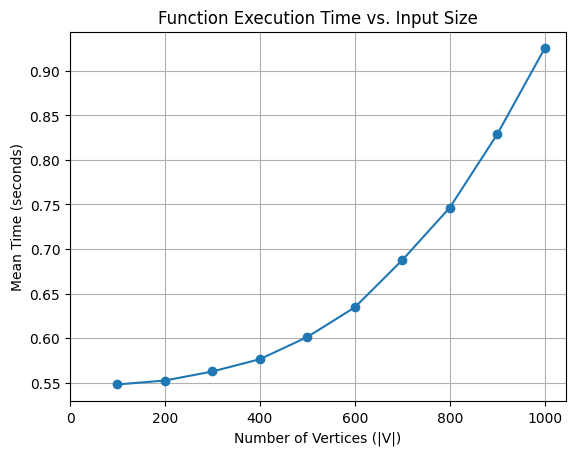}
    \caption{The generation rate of a 3-colorable graph with different size.}
    \label{fig:graph_gen}
\end{figure}

Regarding the operational complexity of the graph generation process, each graph expansion requires either \(O(|E_G|)\) or \(O(|V_G|^2)\) additional operations. This %complexity 
graph generation algorithm remains efficient enough to generate graph instances of the size we utilize in the protocol. As a reference, we present the CPU time required for generating 3-colorable graph instances in our implementation in Figure~\ref{fig:graph_gen}, along with their corresponding sizes in Table~\ref{table:graph_cons}. 

\begin{table}[!ht]
    \centering
    \begin{tblr}{
      cell{1}{2} = {c=7}{c},
      hlines,
      vlines,
    }
    3COL Graph (G)   & Size &     &      &      &      &      &      \\
    Vertices ($|V_G|$) & 200  & 400 & 600  & 800  & 1000 & 5000 & 6000 \\
    Edges ($|E_G|$, on average)    & 381  & 754 & 1128 & 1502 & 1874 & 9352 & 11218
    \end{tblr}
    \caption{The Size of 3-Colorable Graph Instances Construction in average}
    \label{table:graph_cons}
\end{table}

\subsection{Hardware}
\label{ssec:hardware}
In our implementation, each verifier and prover is equipped with an Arista 7130 series device, which includes an onboard Xilinx FPGA featuring the Virtex UltraScale+ VU9P, with a total memory capacity of 75.9 MB of block RAM. In addition to the Arista devices for each verifier, we have a host server with a Xeon Silver 4310 processor, which is responsible for monitoring the protocol execution and performing the final verification. Within each verifier-prover pair, challenges and responses are transmitted via a communication channel with a throughput of 10 Gbit/s, interfaced to the FPGA on both ends using \emph{small form-factor pluggable} (SFP) connectors. For two-prover experiments, we use a 1-meter \emph{direct attach copper} (DAC) cable, while for three-prover experiments, we use a 30-meter single-mode fiber.

To synchronize the verifiers at different locations, each verifier is equipped with a SyncServer S650 Enterprise NTP Time Server, which is equipped with a rubidium atomic clock. We further utilize a tunable PPS signal to trigger each verifier's FPGA, which is sourced from the time server via a direct \emph{SubMiniature version A} (SMA) connection. By combining both NTP/PTP and PPS, we achieve a very high precision of synchronization, ensuring that the verifiers can initiate execution almost simultaneously (within 10 ns).

\subsection{Data Processing and Transmission}
\label{ssec:proc_trans}
In this section, we introduce the architecture of our design, which we deploy in the implementation of the 2-prover and 3-prover RZKP protocols. By formalizing the hardware-software interactions and data processing into three main phases—\emph{Preparation}, \emph{Transmission}, and \emph{Verification}—we explain how messages are processed and transmitted in our implementation. Most importantly, we discuss how these messages can be compressed to enhance computational and communication efficiency while maintaining correctness.

Once a graph instance is generated as described earlier, its topology is made public, while the graph's 3-coloring remains a secret known only to the honest provers. Additionally, further randomness and alignment tags are required to properly encode the challenge and response messages for implementation purposes. Our implementation primarily follows the specifications from \cite{Alikhani_2021}, with modifications to coding techniques to improve efficiency and scalability

\subsubsection{Challenge Preparation}
\label{sssec:chall_pre}
For each round of execution, the verifiers encode the challenge message with the following information: the round index, the labeling index, and the edge to be challenged. To ensure that the challenge messages consistently fall into either an edge-verification test (i.e., the same edge with different labeling indices) or a well-definition test (i.e., edges that share a common vertex with the same labeling index) with distribution probabilities of $\frac{1}{5}$ and $\frac{4}{5}$, respectively (See Protocol \ref{prot:2prover} for further details), we implement an AES-based CTR-DRBG (Deterministic Random Bit Generator) as specified in \cite{BK15} on each verifier with unique fresh seeds. This setup determines: 1. The index of the label (0 or 1); 2. A random edge from 1 to $|E|$ according to the questions. This design is intended to meet the robustness and standard security goals for PRNGs under investigation \cite{cryptoeprint:2020/619}.

In our design, the verifiers generate the challenge messages within their respective servers and pre-load them onto their FPGAs. The messages are encoded in Ethernet frames and transmitted on the OSI data link layer. We denote the time taken to prepare the challenge messages before sending them to the provers as $t_{\text{prep\_v}}$.

\subsubsection{Response preparation} 
\label{sssec:resp_pre}
For each round, The response includes two vertex labels within an edge, with each label represented by a \emph{trit}. On the prover's side, preparation involves pre-sharing a randomized vector of \emph{coloring trits} based on the coloring vector, expanding its size from \(2 \times |V|\) to \(2 \times (|V| + \lceil 2 \log_3 |V| \rceil)\) in binary string form among the provers. Additionally, each round requires the provers to agree on a randomized color permutation (one of six permutations based on the previously generated graph coloring) and another shared randomized \emph{round trit} vector of size \(2 \times \lceil 2 \log_3 |V| \rceil + 1\). Generating labels in response involves specific modular arithmetic operations on each prover's FPGA, where trit vectors are directly loaded from block RAM using round indices as memory addresses. This operation is part of the execution logic and relates to relativistic constraints, which we discuss further in the next section.

In our implementation, the randomized round trit vectors and coloring permutations for each round are derived from the same DRBG instantiated on the verifier side, using fresh random seeds. We denote the time for each prover to prepare messages before receiving challenges as \(t_{\text{prep\_p}}\). Like the verifiers, the messages are encoded in Ethernet frames and transmitted on the OSI data link layer.

\subsubsection{Transmission} 
\label{sssec:trans}
In this section, we detail the transmission of challenges and responses between each verifier-prover pair, which is essential for determining spatial separation distance in a relativistic context. Our implementation utilizes a low-latency FPGA module to process messages, and we optimize communication and computation efficiency as well as time calibration.

With the 3-coloring pre-loaded from the provers' server onto the FPGA, responses are generated on the provers' FPGA upon receiving challenge messages from the verifiers. For each round, the provers compute each label by performing a scalar modular product between the round trit vector of size \(2 \times \lceil 2 \log_3 |V| \rceil + 1\) and the coloring trit vector, starting from the entry corresponding to the the first or second vertex of the edge specified in the challenge messages. By summing all trits in each produced vector, two vertex labels corresponding to the challenge edge are obtained for each round.

In our implementation, we optimize the above process from two perspectives: First, we utilize LUTs to simplify scalar product operations. Second, we employ an adder-tree construction to sum all trits. The encoding processes for generating the two labels are executed in parallel. Compared to \cite{Alikhani_2021}, our computation takes a single clock cycle with only 3.105 ns. Finally, each response between the FPGAs of a verifier-prover pair can be encoded into 2 data frames\footnote{Aligned with the AXI4-Stream protocol} with a 32-bit width per frame, and this can be accomplished within 3 clock cycles of the FPGA reference clock.

For each execution round in the 2-prover setup, we outline the execution and latency time parameters for each device in Table \ref{table:round_cons}, emphasizing the communication between each verifier-prover pair, which is pertinent to the relativistic setting. Each verifier-prover pair is housed within a single rack, using DAC cables for inter-communication. To measure the timing practically, a timer is initialized within each verifier's FPGA to track the duration when it synchronously issues the challenge and receives the response from the corresponding prover.

\begin{table}[!ht]
\centering
\begin{tblr}{
  cell{4}{3} = {r=6}{},
  vlines,
  hline{1-4,10-12,15} = {-}{},
  hline{5-9,13-15} = {1-2}{},
}
Components                                                 & Latency(ns) & Entity       \\
\textbf{---}                                               & \textbf{---}& Verifier     \\
DAC Cable (1 meter)                                        & 7.83        & Transmission \\
Front Panel                                                & 3           & Prover       \\
Rx Transceiver                                             & 20.31       &              \\
Rx CDC                                                     & 24.82       &              \\
Execution Logic (3 clock cycles)                           & 9.31        &              \\
Tx Transceiver                                             & 20.55       &              \\
Front Panel                                                & 3           &              \\
DAC Cable (1 meter)                                        & 7.83        & Transmission \\
\textbf{---}                                               & \textbf{---}& Verifier     \\
Ideal Latency (datasheet)                                  & 96.65       & Total        \\
Measured Latency (max)                                     & 98.98       &              \\
$t_{\text{trans}}$ (implementation)                        & 100         &              
\end{tblr}
\caption{Latency times for components related to relativistic constraints in our implementation.}
\label{table:round_cons}
\end{table}

To compute the minimal distance between locations that satisfies the relativistic constraints, we measure and obtain \(t_{\text{trans}}\), the upper-bounded time window between when each prover receives a challenge message and when the respective verifier receives the response in our implementation. We statistically measure and set a bound for the actual execution time of each round at 100 ns, and hence compute a secure lower bound for the minimal distance \(\text{Dist}_{\text{min}}\) as follows:

\[
\text{Dist}_{\text{min}} = c \times t_{\text{trans}} \approx 30~\text{m},
\]

where \(c\) denotes the speed of light. Note that we do not account for the time each challenge (or response) takes to pass through the Tx Transceiver (or Rx Transceiver) and the Front Panel of the verifier's FPGA, as these processes occur within the verifier's devices. However, these times are still part of the overall protocol challenge time \(t_{\text{chall}}\). 
% On the other hand, the responses received on the verifiers' FPGAs are immediately transmitted back to their respective host servers after wrapping-up, this will not be taken into account for the relativistic constraint as well. 

For the three-prover implementation, the transmission process is the same as described above, except that each intercommunication between the verifier and prover is conducted via a 30-meter single-mode fiber. By designing the implementation differently in Section~\ref{ssec:3pexp}, we make additional assumptions to ensure the relativistic constraints are met. 

\subsubsection{Verification} 
\label{sssec:verf}
Once each verifier's FPGA receives a response, it wraps the labels with a timestamp and forwards the message to the corresponding server. Finally, the verifiers perform a joint verification of the overall responses for each round with two separate checks: 

On one hand, the verifiers consolidate their responses to a single verifier, and check whether the provers respond within a specified time frame in every round. This is determined using the timestamp field that each verifier adds when it receives the prover's response. 

On the other hand, to check the validity of labels responded by the prover, we use the same DRBG and seed to simulate and compute the questions on the fly, rather than storing all of them in memory. Let \(t_{\text{vrfy}}\) denote the CPU time required for checking the responses by the verifiers. 

\vspace{4ex}

Finally, we modularize the execution of the relativistic ZKP protocol into the following procedures:
\begin{enumerate}
    \item \textbf{Preparation Phase}: This phase initiates the verifiers and provers, with the challenge messages and shared random information for responses generated and pre-loaded onto their respective FPGAs. Since the preparation of verifiers and provers can be done in parallel, the actual preparation time \(t_{\text{prep}}\) in our implementation will be the greater of \(t_{\text{prep\_v}}\) and \(t_{\text{prep\_p}}\), i.e., \(t_{\text{prep}} = \max\{t_{\text{prep\_v}}, t_{\text{prep\_p}}\}\).
    \item \textbf{Transmission Phase} This phase, with a total time \(t_{\text{chall}}\), involves each verifier issuing challenges and receiving responses sequentially. 
    \item \textbf{Verification Phase} In this phase, we check the validity of responses from the provers as described above, with a verification time \(t_{\text{vrfy}}\). 
    The phase also includes the time required for gathering the responses via network, and network packet processing. 
\end{enumerate}
By statistically demonstrating the time consumption for each step we have modularized through our experiments, we aim to provide a comprehensive overview of the software/hardware implementation with current devices and its performance at each step with different parameters. This analysis also aids in gaining a deeper understanding of the limitations of our current system and how it can be further optimized.

\subsection{Two-Prover RZKP Security}
\label{ssec:2proversound}
\subsubsection{Security Definition}

The security of RZKP is characterized by the soundness and zero-knowledge properties. 
Soundness examines the scenario where dishonest entangled provers attempt to pass the verifier's checks without knowledge of the secret (coloring).
It is typically related to the quantum value, where a sequential repetition of $m$ rounds gives a soundness of $\varepsilon_{sou}=\omega_q^m$.
More formally, the quantum value $\omega_q$ can be defined by
\begin{definition}[Quantum Value of Alt-RZKP-3-COL]
    Consider a graph $G$ that is not 3-colorable. Then, the Alt-RZKP-3-COL protocol has quantum value $\omega_q$ if for any entangled adversary, the winning probability of a single round of the Alt-RZKP-3-COL for extended graph $G'$ is $p_{win}<\omega_q$.
\end{definition}

Zero-knowledge involves a dishonest verifier which aims to gain knowledge of the secret (in our case, the coloring), and can be defined by
\begin{definition}[Perfect Zero-knowledge of Alt-RZKP-3-COL]
    The Alt-RZKP-3-COL protocol is perfect zero-knowledge if for any dishonest quantum verifier, there exists a simulator such that for any 3-colorable graph $G$, it can simulate the transcript of the communication between provers and the verifier. %\wy{Is this right?}  \ym{and should it be only one verifier in the MIP proof system?}
\end{definition}

\subsubsection{Soundness}

As presented earlier, the soundness can be demonstrated by the construction of a classical strategy for vertex-3-COL.
This can be summarized by
\begin{restatable}[]{theorem}{chainofthm}
% \begin{theorem}
\label{thm:Alt-RZKP-3-COL_to_Vertex-3-COL_simplifed}
    If there exists an $\varepsilon$-perfect strategy for Alt-RZKP-3-COL for graph $G'$, %with query challenge distribution $\mathcal{D}_{iji'j'b'}=\frac{1}{2\abs{E}}(\frac{\delta_{ii'}}{2\abs{\neigh(i)}}+\frac{\delta_{jj'}}{2\abs{\neigh(j)}})$, 
    then there exists a classical strategy for the vertex-3-COL game for graph $G$ which always passes the well-definition tests, $p_{win}^{WD}=1$, and passes the edge verification tests with probability 
    \begin{equation*}
    \begin{split}
        p_{win}^{EV}\geq&1-\varepsilon^{\frac{1}{4}}\left(\frac{9}{2}\abs{V_G}(\abs{V_G}-1)-8\abs{E_G}\right)\times\left[\frac{36+24\sqrt{6}+24\sqrt{3}}{3\abs{V_G}-3-2\max_{i\in V_G}\abs{\neigh(i)_G}}\right.\\
        &\left.+\frac{216\sqrt{3}(19+\sqrt{2})}{\abs{E_G}}\max_{i\in V_G}\abs{\neigh(i)_G}+\frac{(9+4\sqrt{2})(3+\sqrt{3})}{\abs{E_G}}\right],
    \end{split}
    \end{equation*}
    where $\abs{\neigh(i)_G}$ refers to the number of neighbors of vertex $i$.
% \end{theorem}
\end{restatable}
The proof of this construction can be broken up into a chain of theorems as follows:
% We can summarize the overall security proof as a proof chain below, of which the negation would give us the desired soundness guarantee.
\begin{equation*}
\label{eqn:Soundness_Main_Eqn}
\begin{gathered}
    \exists\,\varepsilon\text{-perfect strategy for Alt-RZKP-3-COL with } G'\\
    \Downarrow \text{Thm.~\ref{thm:Alt-RZKP-3-COL_to_Alt-Edge-3-COL}}\\
    \exists\,O(\sqrt{\varepsilon})\text{-perfect strategy for Alt-Edge-3-COL with } G'\\
    \Downarrow \text{Thm.~\ref{thm:Alt-Edge-3-COL_to_VCCC-BCS-3-COL}}\\
    \exists\,O(\sqrt{\varepsilon})\text{-EV VCCC strategy for BCS-3-COL with } G'\\
    \Downarrow\text{Thm.~\ref{thm:BCS-3-COL_to_Almost-Satisfying-Assignment}}\\
    \exists\,O(\varepsilon^{\frac{1}{4}})\text{-almost satisfying assignment for } G'\\
    \Downarrow \text{Thm.~\ref{thm:almost_qsa-to-fully_commuting_qsa}}\\
    \exists\,O(\varepsilon^{\frac{1}{4}})\text{-almost fully commuting satisfying assignment for } G\\
    \Downarrow \text{Thm.~\ref{thm:BCS-3-COL_to_Vertex-3-COL}}\\
    \exists\,O(\varepsilon^{\frac{1}{4}})\text{-EV classical strategy for vertex-3-COL with } G
\end{gathered}
\end{equation*}
where VCCC is an acronym for vertex-complete color-commuting, and $\varepsilon$-EV refers to winning the edge verification test with probability $p_{win}^{EV}\geq1-\varepsilon$.

The first three theorems forms the first phase where the $\varepsilon$-perfect strategy is reduced to an almost satisfying assignment.
Thm.~\ref{thm:Alt-RZKP-3-COL_to_Alt-Edge-3-COL} constructs a strategy for an non-local game where prover A is required to respond with the colors of an edge while prover B is required to respond with one of the vertices corresponding to the chosen edge.
This is achieved by having the provers perform two consecutive measurements each, allowing them to recover vertex colors $c_i=w_i^0+w_i^1$, and incurring a $O(\sqrt{\varepsilon})$ penalty from gentle measurement lemma~\cite{Wilde2017}.
Thm.~\ref{thm:Alt-Edge-3-COL_to_VCCC-BCS-3-COL} constructs a strategy that has commutative and completeness properties that can specifically pass the edge-verification tests of the BCS-3-COL game with high probability.
BCS-3-COL game (presented as Protocol~\ref{prot:Alt-Edge-3-COL} in Appendix~\ref{app:3-COL-Variants}) has prover A responding to either the color of a vertex or whether two neighboring vertices are of some color $\alpha$ while prover B responds with whether a vertex (any of the vertices asked to prover A) are of color $\alpha$.
This construction is straightforward, with the provers converting the colors measured to a YES/NO response to whether vertices are of color $\alpha$.
prover B's measurement operators for this strategy can form an almost satisfying assignment in Thm.~\ref{thm:BCS-3-COL_to_Almost-Satisfying-Assignment}, which relies on the almost commuting results from \cite{Paddock2024}.

In \cite{Ji_2013}, it was shown that if all projectors of neighboring vertices of the commutative gadget commute, then projectors of all vertices in the commutative gadget commute.
In Thm.~\ref{thm:almost_qsa-to-fully_commuting_qsa}, we use the properties of the almost satisfying assignment, and applied it to the proof presented in Lemmas 3 and 4 of \cite{Ji_2013}, without assuming perfect commutation and edge coloring.
This allows us to demonstrate in particular that the vertices $a$ and $e$ in the commutative gadget, which are also vertices in the original graph $G$ which are non-neighboring, also almost commute, i.e., $[B^a_{\alpha},B^e_{\beta}]\ket{\psi}$ has small trace.
Since almost satisfying assignment gives that neighboring vertices are almost commuting, the assignment for vertices corresponding to graph $G$ is almost commuting.

Finally, the almost commuting and tracial property of the projectors allows us to build a well-defined classical strategy for vertex-3-COL in Thm.~\ref{thm:BCS-3-COL_to_Vertex-3-COL}.
The construction involves the provers making consecutive measurements with the projectors in order of vertex label, $B^1_{c_1},\cdots,B^{m}_{c_{m}}$, for $m=\abs{V_G}$, on the pre-shared state $\ket{\psi}$ prior to the start of the protocol.
The provers then utilize the colors $c_1,\cdots,c_m$ as their responses to the BCS game, which we note immediately results in the provers always passing well-definition test.
We show that utilizing the tracial and almost commuting properties of the projectors, the probability of passing edge verification test is close to the value of the edge coloring property of the assignment.
This leads to a lower bound on the probability of winning edge verification tests for a vertex-3-COL game of $G$, as required.

With the chain of theorems providing a lower bound for the winning probability of a particular strategy constructed from $\varepsilon$-perfect Alt-RZKP-3-COL strategy, we now look at the upper bound of the winning probability for any strategy.
Presented more formally in Thm.~\ref{thm:Vertex-3-COL_classical_value}, it is clear that for a graph $G$ that is not 3-colorable, a well-defined strategy necessarily results in at least one edge failing the edge verification test, i.e. $p_{win}^{EV}\leq1-\frac{1}{\abs{E_G}}$.
Combining with Thm.~\ref{thm:Alt-RZKP-3-COL_to_Vertex-3-COL_simplifed}, with $p_{win}^{EV}\geq 1-\varphi\varepsilon^{\frac{1}{4}}$ (note that $\varphi\sim\abs{E_G}$ is the coefficient of $\varepsilon^{\frac{1}{4}}$ in Thm.~\ref{thm:Alt-RZKP-3-COL_to_Vertex-3-COL_simplifed}), we observe that a contradiction occurs when $\varphi\varepsilon^{\frac{1}{4}}<\frac{1}{\abs{E_G}}$.
Therefore, we necessarily require $\varepsilon\geq (\varphi\abs{E_G})^{-4}$, and the best any adversary can achieve is the quantum value $\omega_q=1-(\varphi\abs{E_G})^{-4}$, i.e. $\omega_q\sim 1-\frac{1}{\abs{E_G}^8}$ shown in Thm.~\ref{thm:soundness_scaling}.

\subsubsection{Zero-Knowledge}
\label{sec:Alt-RZKP-3-COL_ZK_Proof}

We can show that the proposed protocol is perfect zero-knowledge (see Thm.~\ref{thm:Alt-RZKP-3-COL_ZK}).
\begin{proof}[Proof of Thm.~\ref{thm:Alt-RZKP-3-COL_ZK}]
In Alt-RZKP-3-COL, any choice of edges $(i,j)\in E_{G'}$ and $(i^\prime, j^\prime)\in E_{G'}$ will always reveal the colors $c_i$ and $c_j$ (i.e. \emph{implicit unveiling}).
% it is trivial to observe that no matter how a dishonest verifier $\tilde{V}$ selects the edges $(i,j)$ and $(i^\prime, j^\prime)$ in $E$, the colors $c_i$ and $c_j$ are always unveiled (i.e. \emph{implicit unveiling}). 
Nevertheless, the responses received from prover B $\tilde{w}_{i'}^{b'}$ and $\tilde{w}_{j'}^{b'}$ is insufficient for a dishonest verifiers to gain any information on the colors for other vertices.
Indeed, all possible combinations of challenges a dishonest verifier may issue would be part of the scenarios shown in Figure 3 of the 3-prover-RZKP-3-COL in \cite{Crepeau_2019}: If $(i,j)\cap (i^\prime,j^\prime)=\emptyset$, the dishonest verifier does not learn $c_{i^\prime}$ and $c_{j^\prime}$ from $\tilde{w}_{i'}^{b'}$ and $\tilde{w}_{j'}^{b'}$ due to the \emph{forever hiding} property (Case 5).
If $(i,j)= (i^\prime,j^\prime)$, it performs the \emph{consistency testing}.
If $(i,j)\cap (i^\prime,j^\prime)=h$, the verifier either perform consistency testing on vertex $h$, or further learn the colors of a triangle in the graph (Case 6 or 8). 
Therefore, the dishonest verifier's view is also captured in the 3-prover-RZKP-3-COL case. 
Since 3-prover-RZKP-3-COL is shown to be perfect zero-knowledge~\cite{Crepeau_2019}, the Alt-RZKP-3-COL is also perfect zero-knowledge using the same simulator construction.
% In the alternative 2-prover interactive proof, it is trivial to observe that no matter how a dishonest verifier $\tilde{V}$ selects the edges $(i,j)$ and $(i^\prime, j^\prime)$ in $E$, the colors $c_i$ and $c_j$ are always unveiled (i.e. \emph{implicit unveiling}). Nevertheless, $\tilde{w}_{i'}^{b'}$ and $\tilde{w}_{j'}^{b'}$ received on the side is not sufficient for $\tilde{V}$ to gain colors of two end-points that do not form an edge. Indeed, all situations fall into the cases that shown in Figure 3 of the 3-prover interactive proof $\Pi_{qnl}^{(3)}$ in \cite{Crepeau_2019}: If $(i,j)\cap (i^\prime,j^\prime)=\emptyset$, $\tilde{V}$ does not learn $c_{i^\prime}$ and $c_{j^\prime}$ from $\tilde{w}_{i'}^{b'}$ and $\tilde{w}_{j'}^{b'}$ due to the \emph{forever hiding} property (i.e., Case 5); If $(i,j)= (i^\prime,j^\prime)$, it performs the \emph{consistency testing} (i.e., Case 7); And $(i,j)\cap (i^\prime,j^\prime)=h$ will either allows $\tilde{V}$ perform consistency testing on vertex $h$, or further learn the colors of a triangle in the graph (i.e., Case 6 or 8). That is to say, everything $\tilde{V}$ sees here is also captured in $\Pi_{qnl}^{(3)}$. Since $\Pi_{qnl}^{(3)}$ is proven perfect zero-knowledge, the alternative 2-prover interactive proof is also perfect zero-knowledge against quantum verifier with the same construction of simulator. This completes the proof. 
\end{proof}

\section*{Disclaimer}
This paper was prepared for informational purposes by the Global Technology Applied Research center of JPMorgan Chase \& Co. This paper is not a product of the Research Department of JPMorgan Chase \& Co. or its affiliates. Neither JPMorgan Chase \& Co. nor any of its affiliates makes any explicit or implied representation or warranty and none of them accept any liability in connection with this paper, including, without limitation, with respect to the completeness, accuracy, or reliability of the information contained herein and the potential legal, compliance, tax, or accounting effects thereof. This document is not intended as investment research or investment advice, or as a recommendation, offer, or solicitation for the purchase or sale of any security, financial instrument, financial product or service, or to be used in any way for evaluating the merits of participating in any transaction.

\bibliographystyle{alpha}
\bibliography{ref}

\newpage
\appendix

\section{Preliminaries}

\subsection{3-Coloring Game Variants}
\label{app:3-COL-Variants}

Let us consider a non-local game $G$, where provers A and B have to respond to challenges independently to win the game, i.e., obtain a score $V(x,y,a,b)=1$.
% \begin{definition}[Non-Local Game]
%     A non-local game $G$ is a game between two separate parties Alice and Bob, where each party independently receives challenge $x$ and $y$, and provides responses $a$ and $b$ respectively, and obtains a score $V(x,y,a,b)\in\{0,1\}$. Alice and Bob wins when they provide responses with a score $V(x,y,a,b)=1$.
% \end{definition}
Any strategy to win the non-local game $G$ can be defined by a pre-shared quantum system $\ket{\psi}$, and projective measurement operators of $\{A^x_a\}_{xa}$ for prover A and $\{B^y_b\}_{yb}$ for prover B\footnote{In general, the pre-shared quantum system, which includes any pre-shared classical information and randomness, can be a density matrix, while prover A and B's measurements can be POVMs. However, by purification argument and Neimark dilation theorem, we can always raise the dimension of the quantum system to obtain a pure pre-shared quantum state and projective measurement operators with the same winning probability}, with winning probability
\begin{equation}
    p_{win}=\sum_{xy}p_{xy}V(x,y,a,b)\mel{\psi}{A^x_a\otimes B^y_b}{\psi}.
\end{equation}
A strategy is defined as $\varepsilon$-perfect for a query challenge distribution $\mathcal{D}_{xy}$ if the strategy provides a winning probability of $p_{win}=1-\varepsilon$.
When defining the distribution $\mathcal{D}_{xy}$, we introduce the Dirac delta function $\delta_{ij}$, which takes on value 1 only if $i=j$ and the indicator function $I[\nu]$ which takes on value 1 only if $\nu$ is true.

A graph can be defined by a set of vertices and a set of edges connecting these vertices, $G=(V_G,E_G)$.
A 3-coloring of a graph is a set of labels $c_i\in\{0,1,2\}$ for $i\in V$, where no two adjacent vertices are of the same color, i.e. $c_i\neq c_j$ if $(i,j)\in E_G$.
We note that there are multiple variants of 3-coloring games which form valid non-local games.
In our analysis, there are a few variants of the 3-coloring game that are of importance: (1) alternative zero-knowledge 3-coloring game (Alt-RZKP-3-COL), (2) alternative edge 3-coloring game (Alt-Edge-3-COL), (3) BCS 3-coloring game (BCS-3-COL) and (4) vertex 3-coloring game (vertex-3-COL).
We note that for any edge challenge $(i,j)$, the challenge is always provided with $i<j$, and the corresponding measurement operators are always selected.
For notational simplicity, we take operators $A^{i,j}$ and $A^{j,i}$ to the the same operator, noting that provers should always take the operator matching $i<j$.

The Alt-Edge-3-COL game is similar to the protocol implemented in \cite{Alikhani_2021}, except additional questions are posed to the first prover, who has to respond with $w_i^0$, $w_i^1$, $w_j^0$, and $w_j^1$.%, allowing us to recover the permuted colors $c_i=w_i^0+w_i^1$ and $c_j=w_j^0+w_j^1$.
The protocol is presented as Protocol~\ref{prot:Alt-ZK-3-COL} in the main text.

% We can reduce the Alt-RZKP-3-COL-game to a simpler game which we term 
Alt-Edge-3-COL is a simpler 3-coloring game, where prover A receives an edge and prover B receives a vertex, and both are required to respond with the colors. 
This is presented as Protocol \ref{prot:Alt-Edge-3-COL} below.
\begin{protocol}[!h]
\caption{Alt-Edge-3-COL Game}
    Provers $\mathcal{P}_{A}$ and $\mathcal{P}_{B}$ pre-agree on a random 3-coloring of $G$: $\left\{ (i,c_{i}) | c_i \in \mathbb{F}_3\right\}_{i\in V_G}$ 
    such that $({i},{j}) \in E_G \implies c_{{j}} \neq c_{{i}}$.
    
    {\bf Commit phase:}
    \begin{itemize}
        \item Verifier $\mathcal{V}$ picks $((i,j), i')\in_{\mathcal{D}_{iji'}} \left(E_G\times V_G\right)$, where $\mathcal{D}_{iji'}=\frac{\delta_{ii'}+\delta_{ji'}}{2\abs{E_G}}$. $\mathcal{V}$ then sends $(i,j)$ to $\mathcal{P}_A$ and $i'$ to $\mathcal{P}_B$.
        \item If $(i,j) \in E_G$ then $\mathcal{P}_A$ replies $c_i$ and $c_j$.
        \item If $i' \in V_G$ then $\mathcal{P}_B$ replies $\tilde{c}_{i'}$.
    \end{itemize}
    
    {\bf Check phase:}
    \begin{itemize}
        \item[] {\bf Edge-Verification Test:}
        \item $\mathcal{V}$ checks if $c_i\neq c_j$.
        \item[] {\bf Well-Definition Test:}
        \item If $i'=i$  then $\mathcal{V}$
        checks if $c_i=\tilde{c}_{i'}$.
        \item If $i'=j$ then $\mathcal{V}$ checks if $c_j=\tilde{c}_{i'}$.
        \item[] {\bf Accept Decision:}
        \item $\mathcal{V}$ accepts if all relevant checks passes.
    \end{itemize}
    % \begin{itemize}
    %     \item If $i'=i$ then $\mathcal{V}$
    %      accepts iff $c_i\neq c_j$ and $c_i=\tilde{c}_{i'}$.
    %     \item If $i'=j$ then $\mathcal{V}$
    %      accepts iff $c_i\neq c_j$ and $c_j=\tilde{c}_{i'}$.
    % \end{itemize}
\label{prot:Alt-Edge-3-COL}
\end{protocol}

To link the 3-coloring games to the commutative gadget~\cite{Ji_2013}, we have to define another variant that forms a binary constraint system (BCS) game.
In this game, instead of requesting for the colors of a vertex, $c_i$, what is requested is a YES/NO response (i.e. an indicator) to whether the vertex $i$ has a color $\alpha$, i.e., responses are $i_{\alpha}\in\{0,1\}$.
Prover A would be asked the vertex color indicators for a constraint, while prover B would be asked a single vertex color indicator.
The set of constraints are presented in \cite{Ji_2013}, and we label the vertex color indicator challenge as a set $\mathcal{X}_{\mathcal{C}}=\{(i,0,1,2)\}_{i\in V_G}\cup\{(i,j,\alpha)\}_{(i,j)\in E_G,\alpha\in\mathbb{F}_3}$.
The full protocol is presented as Protocol \ref{prot:BCS-3-COL}.
\begin{protocol}[!h]
\caption{BCS-3-COL Game}
    Provers $\mathcal{P}_{A}$ and $\mathcal{P}_{B}$ pre-agree on an assignment of variables $\{i_{\alpha}\}_{i\in V_G,\alpha\in\mathbb{F}_3}$ of $G$ 
    such that $(i,j) \in E_G \implies i_{\alpha}j_{\alpha}=0$, $\forall \alpha\in\mathbb{F}_3$.
    
    {\bf Commit phase:}
    \begin{itemize}
        \item Verifier $\mathcal{V}$ picks $(x,k,\beta)\in_{\mathcal{D}_{xk\beta}} (\mathcal{X}_{\mathcal{C}}\times V_G\times\mathbb{F}_3)$, where $\mathcal{D}_{xk\beta}$ is formed from mixing $\mathcal{D}_{ijk\beta}=\frac{\delta_{ik}+\delta_{jk}}{6\abs{E_G}}I[(i,j)\in E_G]$ (corresponding to edge verification constraint $i_{\alpha}j_{\alpha}=0$) and $\mathcal{D}_{ik\beta}=\frac{\delta_{ik}}{3\abs{V_G}}$ (corresponding to consistency constraint $i_0+i_1+i_2=1$). $\mathcal{V}$ then sends $x$ to $\mathcal{P}_A$ and $(k,\alpha)$ to $\mathcal{P}_B$.
        \item If $x =(i,0,1,2)$ then $\mathcal{P}_A$ replies $i_0$, $i_1$ and $i_2$. If $x=(i,j,\alpha)$ then $\mathcal{P}_A$ replies $i_{\alpha}$ and $j_{\alpha}$
        \item If $k \in V_G$ then $\mathcal{P}_B$ replies $\tilde{k}_{\beta}$.
    \end{itemize}
    
    {\bf Check phase:}
    \begin{itemize}
        \item[] {\bf Constraint Satisfaction Test:}
        \item If $x=(i,0,1,2)$ then $\mathcal{V}$ checks if $i_0+i_1+i_2=1$.
        \item If $x=(i,j,\alpha)$ then $\mathcal{V}$ checks if $i_{\alpha}j_{\alpha}=0$.
        \item[] {\bf Well-Definition Test:}
        \item If $k,\beta\in x$ then $\mathcal{V}$
        checks if $k_{\beta}=\tilde{k}_{\beta}$.
        \item[] {\bf Accept Decision:}
        \item $\mathcal{V}$ accepts if all relevant checks passes.
    \end{itemize}
    % {\bf Check phase:}
    % \begin{itemize}
    %     \item If $x=(i,0,1,2)$ and $i=k$ then $\mathcal{V}$
    %      accepts if $i_0+i_1+i_2=1$ and $i_{\beta}=\tilde{k}_{\beta}$.
    %     \item If $x=(i,j,\alpha)$, $\alpha=\beta$ and $k=i$ then $\mathcal{V}$ accepts if $i_{\alpha}j_{\alpha}=0$ and $i_{\alpha}=\tilde{k}_{\beta}$.
    %      \item If $x=(i,j,\alpha)$, $\alpha=\beta$ and $k=j$ then $\mathcal{V}$ accepts if $i_{\alpha}j_{\alpha}=0$ and $j_{\alpha}=\tilde{k}_{\beta}$.
    % \end{itemize}
% \begin{enumerate}[leftmargin=*,topsep=0pt]
    % \item The challenges $x\in\mathcal{V}_{\mathcal{C}}$ and $y\in\{(k,\alpha)\}_{k,\beta}$ are sampled according to some distribution $\mathcal{D}_{xy}$.
    % \item Prover A responds with the corresponding vertex color indicators, either (1) $(i_0,i_1,i_2)$ or (2) $(i_{\alpha},j_{\alpha})$, depending on the challenge, and prover B responds with the bit value of $i_{\alpha}$.
    % \item Provers win if they respond with:
    % \begin{enumerate}[leftmargin=*,nosep]
    %     \item $i_0+i_1+i_2=1$ and $i_{\beta}=\tilde{i}_{\beta}$ when $x=(i,0,1,2)$ and $i=k$.
    %     \item $i_{\alpha}j_{\alpha}=0$ and $k_{\beta}=\tilde{k}_{\beta}$ when $x=(i,j,\alpha)$, $k\in\{i,j\}$ and $\alpha=\beta$.
        %\item $i_{\alpha}^2=i_{\alpha}$ and $i_{\alpha}=\tilde{k}_{\beta}$ when $x=(i,\alpha)$, $i=k$ and $\alpha=\beta$.
    % \end{enumerate}
% \end{enumerate}
\label{prot:BCS-3-COL}
\end{protocol}

The strategy of the BCS game would be utilized finally to build a classical strategy for the standard 3-coloring game, which we present as the vertex-3-COL game in Protocol~\ref{prot:Vertex-3-COL}.
\begin{protocol}[!h]
\caption{Vertex-3-COL Game}
    Provers $\mathcal{P}_{A}$ and $\mathcal{P}_{B}$ pre-agree on a random 3-coloring of $G$: $\left\{ (i,c_{i}) | c_i \in \mathbb{F}_3\right\}_{i\in V_{ij}}$ 
    such that $({i},{j}) \in E_G \implies c_{{j}} \neq c_{{i}}$.
    
    {\bf Commit phase:}
    \begin{itemize}
        \item Verifier $\mathcal{V}$ picks $(i,j)\in_{\mathcal{D}_{ij}} V_G^2$, where $\mathcal{D}_{ij}$ is a mixture of $\mathcal{D}_{ij}=\frac{\delta_{ij}}{\abs{V_G}}$ (well-definition test) and $\mathcal{D}_{ij}=\frac{I[(i,j)\in E_G]}{\abs{E_G}}$ (edge verification test). $\mathcal{V}$ then sends $i$ to $\mathcal{P}_A$ and $j$ to $\mathcal{P}_B$.
        \item If $i \in V_G$ then $\mathcal{P}_A$ replies $c_i$.
        \item If $j \in V_G$ then $\mathcal{P}_B$ replies $\tilde{c}_j$.
    \end{itemize}
    
    {\bf Check phase:}
    \begin{itemize}
        \item[] {\bf Edge Verification Test:}
        \item If $(i,j)\in E_G$ then $\mathcal{V}$
         checks if $c_i\neq \tilde{c}_j$.
        \item[] {\bf Well-Definition Test:}
        \item If $i=j$ then $\mathcal{V}$
         checks if $c_i=\tilde{c}_j$.
        \item[] {\bf Accept Decision:}
        \item $\mathcal{V}$ accepts if all relevant checks passes.
    \end{itemize}
    % {\bf Check phase:}
    % \begin{itemize}
    %     \item If $(i,j)\in E_G$ then $\mathcal{V}$
    %      accepts if $c_i\neq \tilde{c}_j$.
    %     \item If $i=j$ then $\mathcal{V}$
    %      accepts iff $c_i=\tilde{c}_j$.
    % \end{itemize}
% \begin{enumerate}[leftmargin=*,topsep=0pt]
%     \item The challenges $x=i$ and $y=j$ are sampled with $i,j\in V$ according to some distribution $\mathcal{D}_{ij}$.
%     \item Prover A responds with $c_i$ and prover B responds with $\tilde{c}_j$
%     \item Provers win if they respond with $c_i=\tilde{c}_j$ when $i=j$ (well-definition) and $c_i=\tilde{c}_j$ when $(i,j)\in E$ (edge-verification).
% \end{enumerate}
\label{prot:Vertex-3-COL}
\end{protocol}

\subsection{Quantum Assignments and Commutativity Gadget}

The BCS-3-COL game is a non-local BCS game, defined as
\begin{definition}[Non-local BCS Game~\cite{Ji_2013,Cleve_2014}]
    A binary constraint system is a collection of Boolean constraints $\nu_1,\cdots,\nu_m$ over binary variables $x_1,\cdots,x_n$. A non-local BCS game is a game where a randomly chosen constraint $\nu_i$ is sent to prover A and a randomly chosen variable $x_j\in\nu_i$ is sent to prover B. Prover A responds with the values of the binary variables in $\nu_i$ while prover B responds with the value of his binary variable. The provers win when prover A's responses satisfies constraint $\nu_i$ and both provers' response for $x_j$ matches.
\end{definition}
An important property of non-local BCS game was provided by \cite{Ji_2013,Cleve_2014}, which states that BCS games with perfect strategies (winning probability 1) has a quantum satisfying assignment to the BCS, i.e., there is a method of assigning Hermitian operators to each binary variable such that
\begin{enumerate}
    \item The polynomial constraints of BCS are satisfied with $X_j$ replacing $x_j$ and $\mathbb{I}$ replacing $1$.
    \item All $X_j$ operators are projectors, $X_j^2=X_j$.
    \item All operator pairs appearing in the same BCS constraint are commuting, i.e. $x_i,x_j\in\nu_k\implies[X_i,X_j]=0$.
\end{enumerate}

With commuting operators in neighboring nodes, it allows for the utilization of the commutativity gadget introduced by Ji~\cite{Ji_2013}.
For any pair of vertices $(i,j)$, one can introduce a triangular prism-shaped commutative gadget shown in Figure~\ref{fig:commutative_gadget}, by having vertex $i$ in position $a$ and vertex $j$ in position $e$.
More generally, for any graph $G$, we can form a graph $G'$ as an extension of $G$, by adding the commutative gadget between all non-neighboring vertices.
For a graph $G=(V_G,E_G)$, the extension introduces $\frac{\abs{V_G}(\abs{V_G}-1)}{2}-\abs{E_G}$ commutative gadgets, which results in a larger extended graph $G'=(V_{G'},E_{G'})$ with $\abs{V_{G'}}=2\abs{V_G}^2-\abs{V_G}-4\abs{E_G}$ and $\abs{E_{G'}}=\frac{9}{2}\abs{V_G}(\abs{V_G}-1)-8\abs{E_G}$.
Inserting this gadget allows vertices $i$ and $j$'s vertex color indicators to commute by symmetry of the gadget, of which importantly the indicators of $i$ and $j$ commute.
\begin{theorem}[Lemma 4 of \cite{Ji_2013}]
\label{thm:Commutative_Gadget}
    Suppose the commutative gadget is a sub-graph of a graph $G$ that admits a quantum satisfying assignment, then any pairwise assignment operators in the sub-graph commute, $[X_{u,\alpha},X_{v,\beta}]=0$ for any $u,v\in\{a,b,c,d,e,f\}$ and $\alpha,\beta\in\{0,1,2\}$.
\end{theorem}
The commutative nature of the assignment operators allows a classical strategy to be developed, which is capable of winning a simple vertex-3-COL game.

\subsection{Frobenius and Operator Norm Properties}

We note that when we bound the soundness, we consider an $\varepsilon$-perfect strategy instead, where the winning probability $p_{win}\geq 1-\varepsilon$.
As such, our analysis would not involve a quantum satisfying assignment, but an almost satisfying assignment instead, as described in \cite{Paddock2024}.
Here, we introduce some preliminaries that would aid us in that analysis.%, which will be introduced in the later sections.\\

The Frobenius norm is defined as $\norm{A}_F=\sqrt{\Tr[A^{\dagger}A]}=\sqrt{\sum_{ij}\abs{A_{ij}}^2}$, and the operator norm is defined as $\norm{A}_{op}=\min_{\norm{x}_2=1}\norm{Ax}_2$. 
Properties of the norm that are useful include:
\begin{enumerate}
    \item $\norm{A}_F\leq\norm{A}_{op}$.
    \item $\norm{A^{\dagger}}_F=\norm{A}_F$.
    \item $\norm{AB}_F\leq\norm{A}_{op}\norm{B}_F$, $\norm{AB}_F\leq\norm{B}_{op}\norm{A}_F$.
    \item Frobenius norm is unitarily invariant, i.e., for any unitary $U$, $\norm{UA}_F=\norm{AU}_F=\norm{A}_F$.
    \item Frobenius norm satisfies triangle inequality, $\norm{A+B}_F\leq\norm{A}_F+\norm{B}_F$, which when combined with 3, $\norm{A+A^{\dagger}}_F\leq 2\norm{A}_F$.
\end{enumerate}
%\wy{Is it true that $\Tr[A\rho]\leq\norm{A}_{op}\Tr[\rho]$ for any $\rho\geq0$?}
\begin{theorem}
\label{thm:Normal_Op_Frobenius_Norm}
    Let $\rho$ be a quantum state, i.e. $\rho\geq0$, $\Tr[\rho]=1$. Then, for any normal $A$ ($A^{\dagger}A=AA^{\dagger}$), $\abs{\Tr[A\rho^{\frac{1}{2}}]}\leq\norm{A}_F$.
\end{theorem}
\begin{proof}
Since $A$ is normal, it can be diagonalized as $A=UD_AU^{\dagger}$.
Similarly, we can express the square root of the quantum state by $\rho^{\frac{1}{2}}=V\Lambda^{\frac{1}{2}}V^{\dagger}$, where $\Lambda$ is the diagonal matrix containing the eignevalues of $\rho$, i.e. $\sum_i\lambda_i=1$ and $\lambda_i\geq0$.
\begin{equation*}
\begin{split}
    \abs{\Tr[A\rho^{\frac{1}{2}}]}=&\abs{\Tr[UD_AU^{\dagger}V\Lambda^{\frac{1}{2}}V^{\dagger}]}\\
    =&\abs{\Tr[V^{\dagger}UD_AU^{\dagger}V\Lambda^{\frac{1}{2}}]}.
\end{split}
\end{equation*}
Let us define a matrix $X=V^{\dagger}UD_AU^{\dagger}V$, then
\begin{equation*}
\begin{split}
    \abs{\Tr[A\rho^{\frac{1}{2}}]}=&\abs{\Tr[\sum_{ijk}X_{ij}\dyad{i}{j}\times\sqrt{\lambda_k}\dyad{k}]}\\
    =&\abs{\sum_iX_{ii}\sqrt{\lambda_i}}\\
    \leq&\sum_i\abs{X_{ii}}\sqrt{\lambda_i}
\end{split}
\end{equation*}
We can define a vector with entries $\abs{X_{ii}}$ and a vector with entries $\sqrt{\lambda_i}$.
As such, using the Cauchy-Schwarz inequality, we can bound
\begin{equation*}
\begin{split}
    \abs{\Tr[A\rho^{\frac{1}{2}}]}\leq&\sqrt{\sum_i\abs{X_{ii}}^2}\times\sqrt{\sum_i\lambda_i}\\
    \leq&\sqrt{\sum_{ij}\abs{X_{ij}}^2}\\
    =&\norm{V^{\dagger}UD_AU^{\dagger}V}_F\\
    =&\norm{A}_F.
\end{split}
\end{equation*}
\end{proof}
We also consider the operator norm of a chain of pinching map and projective operators,
\begin{theorem}
\label{thm:Chain_Pinching_Projector}
Consider projective operators $P_{1,i_1},\cdots,P_{n,i_n}$, with $\sum_{i_u}P_{u,i_u}=\mathbb{I}$.
Then, 
\begin{equation*}
    \norm{\sum_{i_1\cdots i_{n-1}\setminus i_u}P_{1,i_1}\cdots P_{n,i_n}\cdots P_{1,i_1}}_{op}\leq 1
\end{equation*}
\end{theorem}
\begin{proof}
We first begin with the fact that for any projector, $P_{n,i_n}\leq\mathbb{I}$.
For any operator $P\leq\mathbb{I}$, applying a pinching map gives
\begin{equation}
    \sum_iP_iPP_i\leq\sum_iP_i\mathbb{I}P_i\leq\sum_iP_i=\mathbb{I}.
\end{equation}
The same is true for any projective operation,
\begin{equation}
    P_iPP_i\leq P_i\mathbb{I}P_i=P_i\leq\mathbb{I}.
\end{equation}
As such, $\sum_{i_1\cdots i_{n-1}\setminus i_u}P_{1,i_1}\cdots P_{n,i_n}\cdots P_{1,i_1}\leq\mathbb{I}$ and its operator norm is bounded by 1.
\end{proof}

\section{Soundness of Alt-RZKP-3-COL}

As discussed in the main text, the soundness of the Alt-RZKP-3-COL protocol depends can be demonstrated by first proving a chain of theorems, culminating in the construction of a classical strategy of the vertex-3-COL game from an $\varepsilon$-perfect strategy for Alt-RZKP-3-COL, as in Thm.~\ref{thm:Alt-RZKP-3-COL_to_Vertex-3-COL_simplifed}.
Sec.~\ref{app:Alt-RZKP-3-COL_to_BCS-3-COL} and Sec.~\ref{app:BCS-3-COL_to_ASA} details the first phase, reducing a strategy for Alt-RZKP-3-COL to an almost satisfying assignment, Sec.~\ref{app:BCS-3-COL_to_FC-ASA} details the second phase where an almost fully commuting satisfying assignment is generated, and Sec.~\ref{app:FC-ASA_to_Classical} details the formation of the classical strategy.
The remaining sections summarizes the soundness argument and examines numerically the requirements necessary to provide security against entangled adversaries.

\subsection{Alt-RZKP-3-COL to BCS-3-COL}
\label{app:Alt-RZKP-3-COL_to_BCS-3-COL}

%To provide concrete bounds on the winning probability of Alt-RZKP-3-COL, 
We begin by proving a reduction to the Alt-Edge-3-COL game for $\varepsilon$-perfect strategies of the Alt-RZKP-3-COL game
% We begin with the reduction from Alt-RZKP-3-COL to Alt-Edge-3-COL
, which involves a consecutive measurement and the use of gentle measurement lemma~\cite{Wilde2017}. 
% The complete proof is provided in Appendix~\ref{app:Alt-RZKP-3-COL_to_Alt-Edge-3-COL}.
\begin{theorem}
\label{thm:Alt-RZKP-3-COL_to_Alt-Edge-3-COL}
    If there exists an $\varepsilon$-perfect strategy for Alt-RZKP-3-COL, %with query challenge distribution $\mathcal{D}_{iji'j'b}=\frac{1}{2\abs{E}}(\frac{\delta_{ii'}}{2\abs{\neigh(i)}}+\frac{\delta_{jj'}}{2\abs{\neigh(j)}})$, 
    then there exists an $(\varepsilon+2\sqrt{\varepsilon})$-perfect strategy for Alt-Edge-3-COL.% with query challenge distribution $\mathcal{D}_{iji'}=\frac{\delta_{ii'}+\delta_{ji'}}{2\abs{E}}$.
\end{theorem}
\begin{proof}
The query challenge distribution $\mathcal{D}_{iji'j'b}=\frac{1}{2\abs{E_G}}(\frac{\delta_{ii'}}{2\abs{\neigh(i)}}+\frac{\delta_{jj'}}{2\abs{\neigh(j)}})$ is one where the verifiers (1) randomly select edge $(i,j)\in E_G$, (2) randomly select $b\in\{0,1\}$, (3) randomly select vertex $i$ or $j$, (4) randomly select a neighbor of the selected vertex to form an edge $(i',j')$.
An $\varepsilon$-perfect strategy with the above challenge distribution includes a quantum state $\ket{\psi}$ shared between the provers, prover A's measurement operators $\{A^{i,j}_{w_i^0,w_j^0,w_i^1,w_j^1}\}_{(i,j)\in E_G}$ and prover B's measurement operators $\{B^{i,j,b}_{\tilde{w}_i^b,\tilde{w}_j^b}\}_{(i,j)\in E_G,b\in\{0,1\}}$, with
\begin{equation}
\label{eqn:Alt-RZKP_win_strategy}
\begin{split}
    &\frac{1}{2\abs{E_G}}\sum_{(i,j)\in E_G}\left[\frac{1}{2\abs{\neigh(i)}}\sum_{b,j'\in \neigh(i)\setminus j}\sum_{\substack{w_i^0+w_i^1\neq w_j^0+w_j^1\\ w_i^b=\tilde{w}_i^b}}\mel{\psi}{A^{i,j}_{w_i^0,w_j^0,w_i^1,w_j^1}\otimes B^{i,j',b}_{\tilde{w}_i^b,\tilde{w}_{j'}^b}}{\psi}\right.\\
    &+\left(\frac{1}{2\abs{\neigh(i)}}+\frac{1}{2\abs{\neigh(j)}}\right)\sum_b\sum_{\substack{w_i^0+w_i^1\neq w_j^0+w_j^1\\ w_i^b=\tilde{w}_i^b,w_j^b=\tilde{w}_j^b}}\mel{\psi}{A^{i,j}_{w_i^0,w_j^0,w_i^1,w_j^1}\otimes B^{i,j,b}_{\tilde{w}_i^b,\tilde{w}_j^b}}{\psi}\\
    &\left.+\frac{1}{2\abs{\neigh(j)}}\sum_{b,i'\in \neigh(j)\setminus i}\sum_{\substack{w_i^0+w_i^1\neq w_j^0+w_j^1\\ w_j^b=\tilde{w}_j^b}}\mel{\psi}{A^{i,j}_{w_i^0,w_j^0,w_i^1,w_j^1}\otimes B^{i',j,b}_{\tilde{w}_{i'}^b,\tilde{w}_{j}^b}}{\psi}\right]=1-\varepsilon,
\end{split}
\end{equation}
where we note a relabeling of $i'$ and $j'$ and swapping the index labels $B^{i,j,b}_{\tilde{w}_i^b,\tilde{w}_j^b}=B^{j,i,b}_{\tilde{w}_j^b,\tilde{w}_i^b}$ such that $i'=i$ in the first term while $j'=j$ in the last term.
Note that WLOG, the measurement operators can be chosen to be projective.
Consider the following strategy for the Alt-Edge-3-COL game:
\begin{enumerate}
    \item The provers pre-share the quantum state $\ket{\psi}$.
    \item When prover A receives her challenge $x=(i,j)$, she performs projective measurement $\{A^{i,j}_{w_i^0,w_j^0,w_i^1,w_j^1}\}$ and sets $c_i=w_i^0+w_i^1$ and $c_j=w_j^0+w_j^1$.
    \item When prover B receives his challenge $y=i'$, he randomly selects a neighboring edge $j'$ and bit value $b\in\{0,1\}$ and constructs the projective measurements $B^{i',b}_{\tilde{w}_{i'}^b}=\sum_{w_{j'}^b}B^{i',j',b}_{\tilde{w}_{i'}^b,\tilde{w}_{j'}^b}$. He performs projective measurement $\{B^{i',b}_{\tilde{w}_{i'}^b}\}$, followed by projective measurement $\{B^{i',\bar{b}}_{\tilde{w}_{i'}^{\bar{b}}}\}$ on the post-measured state, and sets $\tilde{c}_{i'}=\tilde{w}_{i'}^b+\tilde{w}_{i'}^{\bar{b}}$.
\end{enumerate}
We note that $B^{i',b}_{\tilde{w}_{i'}^b}$ remains orthogonal projectors since $(B^{i',b}_{\tilde{w}_{i'}^b})^2=B^{i',b}_{\tilde{w}_{i'}^b}$, and the components for different $\tilde{w}_{i'}^b$ are orthogonal.
What remains is to compute is the winning probability of the strategy.\\

Let us consider the winning probability of the Alt-Edge-3-COL game with uniform challenge distribution for $i'\in\{i,j\}$, $\mathcal{D}_{iji'}=\frac{\delta_{ii'}+\delta_{ji'}}{2\abs{E_G}}$.
Since prover B randomly selects the neighboring edge with probability $\frac{1}{\abs{\neigh(i')}}$, the winning probability can be given as
\begin{equation}
\begin{split}
    p_{win}=&\frac{1}{2\abs{E_G}}\sum_{(i,j)\in E_G}\frac{1}{2\abs{\neigh(i)}}\sum_{b,j'\in \neigh(i)}\sum_{\substack{w_i^0,w_j^0,w_i^1,w_j^1,\tilde{w}_i^0,\tilde{w}_i^1:\\ c_i\neq c_j,c_i=\tilde{c}_i}}f(\vec{w}_{ij},\vec{w}_i')\\
    &+\frac{1}{2\abs{E_G}}\sum_{(i,j)\in E_G}\frac{1}{2\abs{\neigh(j)}}\sum_{b,j'\in \neigh(j)}\sum_{\substack{w_i^0,w_j^0,w_i^1,w_j^1,\tilde{w}_j^0,\tilde{w}_j^1:\\ c_i\neq c_j,c_j=\tilde{c}_j}}f(\vec{w}_{ij},\vec{w}_j'),
\end{split}
\end{equation}
where 
\begin{equation}
\begin{gathered}
    f(\vec{w}_{ij},\vec{w}_{i'}')=f(c_i,c_j|w_i^0,w_i^1,w_j^0,w_j^1)f_B(\tilde{c}_{i'}|w_{i'}^0,w_{i'}^1)g(\vec{w}_{ij},\vec{w}_j')\\
    g(\vec{w}_{ij},\vec{w}_{i'}')= \mel{\psi}{A^{i,j}_{w_i^0,w_j^0,w_i^1,w_j^1}\otimes B^{i',b}_{\tilde{w}_{i'}^b}B^{i',\bar{b}}_{\tilde{w}_{i'}^{\bar{b}}}B^{i',b}_{\tilde{w}_{i'}^b}}{\psi}.
\end{gathered}
\end{equation}
We can lower bound the winning probability by a smaller set included in the sum, with constraints $w_i^0+w_i^1\neq w_j^0+w_j^1$, $w_{i'}^0=\tilde{w}_{i'}^0$ and $w_{i'}^1=\tilde{w}_{i'}^1$ (which implies $c_i\neq c_j$, $c_{i'}=\tilde{c}_{i'}$),
\begin{equation}
\begin{split}
    p_{win}\geq&\frac{1}{2\abs{E_G}}\sum_{(i,j)\in E_G}\frac{1}{2\abs{\neigh(i)}}\sum_{b,j'\in \neigh(i)}\sum_{\substack{w_i^0,w_j^0,w_i^1,w_j^1,\tilde{w}_i^0,\tilde{w}_i^1:\\ w_i^0+w_i^1\neq w_j^0+w_j^1\\ w_i^0=\tilde{w}_i^0,w_i^1=\tilde{w}_i^1}}g(\vec{w}_{ij},\vec{w}_i')\\
    &+\frac{1}{2\abs{E_G}}\sum_{(i,j)\in E_G}\frac{1}{2\abs{\neigh(j)}}\sum_{b,j'\in \neigh(j)}\sum_{\substack{w_i^0,w_j^0,w_i^1,w_j^1,\tilde{w}_i^0,\tilde{w}_i^1:\\ w_i^0+w_i^1\neq w_j^0+w_j^1\\ w_j^0=\tilde{w}_j^0,w_j^1=\tilde{w}_j^1}}g(\vec{w}_{ij},\vec{w}_j').
\end{split}
\end{equation}
Let us define $p(\vec{w}_{ij})=\mel{\psi}{A^{i,j}_{w_i^0,w_j^0,w_i^1,w_j^1}\otimes\mathbb{I}}{\psi}$, and $\rho^{\vec{w}_{ij}}$ as the quantum state on prover B's system conditioned on measurement outcome $(w_i^0,w_j^0,w_i^1,w_j^1)$.
Therefore, we can express
\begin{equation}
    \sum_{\substack{\tilde{w}_i^0,\tilde{w}_i^1:\\ w_{i'}^0=\tilde{w}_{i'}^0,w_{i'}^1=\tilde{w}_{i'}^1}}g(\vec{w}_{ij},\vec{w}_{i'}')=p(\vec{w}_{ij})\Tr[B^{i',\bar{b}}_{w_{i'}^{\bar{b}}}B^{i',b}_{w_{i'}^b}\rho^{\vec{w}_{ij}}B^{i',b}_{w_{i'}^b}B^{i',\bar{b}}_{w_{i'}^{\bar{b}}}].
\end{equation}
Let $\Tr[B^{i',b}_{w_{i'}^b}\rho^{\vec{w}_{ij}}]=1-\varepsilon_{\vec{w}_{ij}i'b}$, and we expect $p(\vec{w}_{ij})\varepsilon_{\vec{w}_{ij}i'b}$ to be generally small.
Therefore, by the gentle measurement lemma~\cite{Wilde2017}, we can bound
\begin{equation}
    \norm{B^{i',b}_{w_{i'}^b}\rho^{\vec{w}_{ij}}B^{i',b}_{w_{i'}^b}-\rho^{\vec{w}_{ij}}}_1\leq 2\sqrt{\varepsilon_{\vec{w}_{ij}i'b}}.
\end{equation}
Using the triangle inequality, and noting $\Tr[P\rho P]=\norm{P\rho P}_1$ for any projector $P$ and quantum state $\rho$, we can show
\begin{equation}
\begin{split}
    &\Tr[B^{i',\bar{b}}_{w_{i'}^{\bar{b}}}B^{i',b}_{w_{i'}^b}\rho^{\vec{w}_{ij}}B^{i',b}_{w_{i'}^b}B^{i',\bar{b}}_{w_{i'}^{\bar{b}}}]\\
    \geq &\Tr[B^{i',\bar{b}}_{w_{i'}^{\bar{b}}}\rho^{\vec{w}_{ij}}]-\norm{B^{i',\bar{b}}_{w_{i'}^{\bar{b}}}B^{i',b}_{w_{i'}^b}\rho^{\vec{w}_{ij}}B^{i',b}_{w_{i'}^b}B^{i',\bar{b}}_{w_{i'}^{\bar{b}}}-B^{i',\bar{b}}_{w_{i'}^{\bar{b}}}\rho^{\vec{w}_{ij}}B^{i',\bar{b}}_{w_{i'}^{\bar{b}}}}_1\\
    \geq& 1-\varepsilon_{\vec{w}_{ij}i'\bar{b}}-\norm{B^{i',b}_{w_{i'}^b}\rho^{\vec{w}_{ij}}B^{i',b}_{w_{i'}^b}-\rho^{\vec{w}_{ij}}}_1\\
    \geq& 1-\varepsilon_{\vec{w}_{ij}i'\bar{b}}-2\sqrt{\varepsilon_{\vec{w}_{ij}i'b}},
\end{split}
\end{equation}
where the second inequality uses the fact that completely-positive non-trace-increasing maps do not increase the trace norm.
As such, we can write
\begin{equation}
\begin{split}
    p_{win}\geq&\frac{1}{2\abs{E_G}}\sum_{(i,j)\in E_G}\frac{1}{2\abs{\neigh(i)}}\sum_{b,j'\in \neigh(i)}\sum_{\substack{w_i^0,w_j^0,w_i^1,w_j^1:\\ w_i^0+w_i^1\neq w_j^0+w_j^1}}p(\vec{w}_{ij})[1-\varepsilon_{\vec{w}_{ij}i\bar{b}}-2\sqrt{\varepsilon_{\vec{w}_{ij}ib}}]\\
    &+\frac{1}{2\abs{E_G}}\sum_{(i,j)\in E_G}\frac{1}{2\abs{\neigh(j)}}\sum_{b,j'\in \neigh(j)}\sum_{\substack{w_i^0,w_j^0,w_i^1,w_j^1:\\ w_i^0+w_i^1\neq w_j^0+w_j^1}}p(\vec{w}_{ij})[1-\varepsilon_{\vec{w}_{ij}j\bar{b}}-2\sqrt{\varepsilon_{\vec{w}_{ij}jb}}].
\end{split}
\end{equation}
Let us break the winning probability lower bound into two components, one with the summation over $(1-\varepsilon_{\vec{w}_{ij}i'\bar{b}})$ terms and one with the summation over $2\sqrt{\varepsilon_{\vec{w}_{ij}i'b}}$.
The first component, $p_{win,1}$, can be simplified by matching it to the Alt-RZKP-3-COL game, and noting that the checks there are more stringent.
We first re-express
\begin{equation}
\begin{split}
    p(\vec{w}_{ij})[1-\varepsilon_{\vec{w}_{ij}i'\bar{b}}]=&\Tr[B^{i',\bar{b}}_{w_{i'}^{\bar{b}}}p(\vec{w}_{ij})\rho^{\vec{w}_{ij}}]\\
    =&\mel{\psi}{A^{i,j}_{w_i^0,w_j^0,w_i^1,w_j^1}\otimes B^{i',\bar{b}}_{w_{i'}^{\bar{b}}}}{\psi}\\
    =&\sum_{\tilde{w}_{i'}^{\bar{b}}\tilde{w}_{j'}^{\bar{b}}:\tilde{w}_{i'}^{\bar{b}}=w_{i'}^{\bar{b}}}\mel{\psi}{A^{i,j}_{w_i^0,w_j^0,w_i^1,w_j^1}\otimes B^{i',j',\bar{b}}_{\tilde{w}_{i'}^{\bar{b}},\tilde{w}_{j'}^{\bar{b}}}}{\psi},
\end{split}
\end{equation}
allowing us to express the winning probability's first component as
\begin{equation}
\begin{split}
    p_{win,1}\geq&\frac{1}{2\abs{E_G}}\sum_{(i,j)\in E_G}\left[\frac{1}{2\abs{\neigh(i)}}\sum_{b,j'\in \neigh(i)\setminus j}\sum_{\substack{w_i^0+w_i^1\neq w_j^0+w_j^1\\ \tilde{w}_i^{\bar{b}}=w_i^{\bar{b}}}}\mel{\psi}{A^{i,j}_{w_i^0,w_j^0,w_i^1,w_j^1}\otimes B^{i,j',\bar{b}}_{\tilde{w}_{i}^{\bar{b}},\tilde{w}_{j'}^{\bar{b}}}}{\psi}\right.\\
    &+\frac{1}{2\abs{\neigh(i)}}\sum_{b}\sum_{\substack{w_i^0+w_i^1\neq w_j^0+w_j^1\\ \tilde{w}_i^{\bar{b}}=w_i^{\bar{b}}}}\mel{\psi}{A^{i,j}_{w_i^0,w_j^0,w_i^1,w_j^1}\otimes B^{i,j,\bar{b}}_{\tilde{w}_{i}^{\bar{b}},\tilde{w}_{j}^{\bar{b}}}}{\psi}\\
    &+\frac{1}{2\abs{\neigh(j)}}\sum_{b}\sum_{\substack{w_i^0+w_i^1\neq w_j^0+w_j^1\\ \tilde{w}_j^{\bar{b}}=w_j^{\bar{b}}}}\mel{\psi}{A^{i,j}_{w_i^0,w_j^0,w_i^1,w_j^1}\otimes B^{i,j,\bar{b}}_{\tilde{w}_{i}^{\bar{b}},\tilde{w}_{j}^{\bar{b}}}}{\psi}\\
    &\left.+\frac{1}{2\abs{\neigh(j)}}\sum_{b,j'\in \neigh(j)\setminus i}\sum_{\substack{w_i^0+w_i^1\neq w_j^0+w_j^1\\ \tilde{w}_j^{\bar{b}}=w_j^{\bar{b}}}}\mel{\psi}{A^{i,j}_{w_i^0,w_j^0,w_i^1,w_j^1}\otimes B^{j',j,\bar{b}}_{\tilde{w}_{j'}^{\bar{b}},\tilde{w}_{j}^{\bar{b}}}}{\psi}\right]\\
    \geq&1-\varepsilon.
\end{split}
\end{equation}
where we explicitly isolate the cases where prover B's choice of $(i',j')$ matches the edge $(i,j)$ in the first line. In the second line, we lower bound these $(i',j')=(i,j)$ cases by a smaller set where both vertex values are checked, which leads to the same form as Eqn.~\eqref{eqn:Alt-RZKP_win_strategy}. 
We can re-express $p_{win,1}$ as 
\begin{equation}
    p_{win,1}=\sum_{(i,j)\in E_G,b,i',j'\in\neigh(i'),\vec{w}_{ij}}p_{iji'j'b\vec{w}_{ij}}\left\{I[\vec{w}_{ij}:w_i^0+w_i^1\neq w_j^0+w_j^1](1-\varepsilon_{\vec{w}_{ij}i'b})\right\},
\end{equation}
where $p_{iji'j'b\vec{w}_{ij}}=\frac{p(\vec{w}_{ij})}{4\abs{E_G}\abs{\neigh(i')}}$ and $I[\vec{w}_{ij}:w_i^0+w_i^1\neq w_j^0+w_j^1]$ is an indicator function for $\vec{w}_{ij}$ satisfying $w_i^0+w_i^1\neq w_j^0+w_j^1$.
We can re-formulate this to get a bound on the expectation value of $\varepsilon_{\vec{w}_{ij}i'b}$,
\begin{equation}
    \sum_{(i,j)\in E_G,b,i',j'\in\neigh(i'),\vec{w}_{ij}}p_{iji'j'b\vec{w}_{ij}}\left\{I[\vec{w}_{ij}:w_i^0+w_i^1\neq w_j^0+w_j^1]\varepsilon_{\vec{w}_{ij}i'b}\right\}\leq \varepsilon-\delta,
\end{equation}
where $\delta=1-\sum_{(i,j)\in E_G,b,i',j'\in\neigh(i'),\vec{w}_{ij}}p_{iji'j'b\vec{w}_{ij}}I[\vec{w}_{ij}:w_i^0+w_i^1\neq w_j^0+w_j^1]\geq 0$, and $\varepsilon-\delta\geq0$ since $\varepsilon_{\vec{w}_{ij}i'b}\geq 0$.
The second component can be expressed as an expectation value of a different variable and can be simplified
\begin{equation}
\begin{split}
    p_{win,2}=&\sum_{(i,j)\in E_G,b,i',j'\in\neigh(i'),\vec{w}_{ij}}p_{iji'j'b\vec{w}_{ij}}(-2\sqrt{\varepsilon_{iji'\vec{w}_{ij}}I[\vec{w}_{ij}:w_i^0+w_i^1\neq w_j^0+w_j^1]})\\
    \geq&-2\sqrt{\sum_{(i,j)\in E_G,b,i',j'\in\neigh(i'),\vec{w}_{ij}}\varepsilon_{iji'\vec{w}_{ij}}I[\vec{w}_{ij}:w_i^0+w_i^1\neq w_j^0+w_j^1]}\\
    \geq&-2\sqrt{\varepsilon-\delta}\\
    \geq&-2\sqrt{\varepsilon},
\end{split}
\end{equation}
where we use Jensen's inequality and the fact that $-\sqrt{x}$ is a concave function.
Therefore, the overall winning probability is $p_{win,2}\geq 1-\varepsilon-2\sqrt{\varepsilon}$.
% We can consider a strategy for the Alt-Edge-3-COL game using the state and measurements of the Alt-RZKP-3-COL game. 
% In this strategy, Alice behaves in the same manner, while Bob would perform a measurement with a random neighboring vertex of $i'$ for both $b=0$ and $b=1$ (consecutive measurements).
% Both parties then compute the colors for the response using $c_i=w_i^0+w_i^1$.\\

% We observe that if the colors in the responses satisfy $c_i\neq c_j$ and $c_{i'}=\tilde{c}_{i'}$, it implies that the Alt-RZKP-3-COL responses $w_i^b$ satisfy the corresponding conditions.
% Therefore, we can simply consider the larger set of accepting responses on $w_i^b$.
% This includes requiring $w_{i'}^b=\tilde{w}_{i'}^b$ and $w_{i'}^{\bar{b}}=\tilde{w}_{i'}^{\bar{b}}$, i.e. both consecutive measurements should match the corresponding output from Alice.
% Since the winning probability is high, for a fixed outcome Alice receives, Bob's measurement outcome $\tilde{w}_{i'}^b=w_{i'}^b$ with high probability $\Tr[B^{i',b}_{\tilde{w}_{i'}^b}\rho^{w_{i'}^b}]\geq1-\varepsilon_{i',b}$.
% Therefore, by the gentle measurement lemma~\cite{Wilde2017}, the state remains $O(\sqrt{\varepsilon_{i',b}})$-close to the original, which means the second measurement is correct with high probability.
% Detailed computation can thus show that the strategy described is $O(\sqrt{\varepsilon})$-perfect. 

\end{proof}

We can next consider a reduction from Alt-Edge-3-COL to BCS-3-COL.
The construction of the strategy for BCS-3-COL provides certain useful properties of the operators that is useful to formulate the classical strategy.
As such, we prove the existence of a ``vertex-complete color-commuting" strategy, which wins the edge verification challenge with $\geq1-\varepsilon$ winning probability.
Note that the vertex challenge can be ignored since the vertex constraints are restricted by the vertex-completeness property, where prover A must always assign a color to the challenge vertex.
% The full proof is provided in Appendix~\ref{app:Alt-Edge-3-COL_to_VCCC-BCS-3-COL}.
\begin{theorem}
\label{thm:Alt-Edge-3-COL_to_VCCC-BCS-3-COL}
    If there exists an $\varepsilon$-perfect strategy for Alt-Egde-3-COL, %with challenge distribution $\mathcal{D}_{iji'}=\frac{\delta_{ii'}+\delta_{ji'}}{2\abs{E}}$, 
    then there exists a vertex-complete color-commuting strategy that can win the edge verification challenge ($i_{\alpha}j_{\alpha}=0$) of BCS-3-COL %with a distribution of $\mathcal{D}_{ijk\alpha}=\frac{\delta_{ik}+\delta_{jk}}{6\abs{E}}$ 
    with $p_{win}^{EV}\geq 1-\varepsilon$. Specifically, the vertex-complete color-commuting strategy has the following properties:
    \begin{enumerate}
        \item $\sum_{\alpha=0}^2\tilde{B}^{k\alpha}_1=\mathbb{I}$, $\forall k\in V$
        \item $[\tilde{B}^{k\alpha}_{b_{k\alpha}},\tilde{B}^{k\beta}_{b_{k\beta}}]=0$, $\forall k\in V$, $\alpha,\beta\in\{0,1,2\}$
        \item $[\tilde{A}^{ij\alpha}_{b_{i\alpha}b_{j\alpha}},\tilde{A}^{ij\beta}_{b_{i\beta}b_{j\beta}}]=0$, $\forall(i,j)\in E$, $\alpha,\beta\in\{0,1,2\}$
    \end{enumerate}
\end{theorem}
\begin{proof}
Consider an $\varepsilon$-perfect strategy for Alt-Edge-3-COL, with state $\ket{\psi}_{AB}$, prover A's measurement operators $\{A^{ij}_{c_ic_j}\}$ and prover B's measurement operators $\{B^k_{c_k}\}$.
WLOG, we can consider projective operators that satisfy completeness $\sum_{c_ic_j}A^{ij}_{c_ic_j}=\mathbb{I}$ and $\sum_{c_k}B^{k}_{c_k}=\mathbb{I}$.\footnote{From a general strategy, we can arrive at a projective strategy using Neimark's dilation theorem. A complete strategy can also be constructed from a general strategy by first noting that the adversary must always produce a response. As such, $\mel{\psi}{\mathbb{I}-\sum_{c_k}B^i_{c_k}}{\psi}=0$, and the operator $\mathbb{I}-\sum_{c_k}B^i_{c_k}$ is orthogonal to the state $\ket{\psi}$. This allows us to add $\mathbb{I}-\sum_{c_k}B^i_{c_k}$ to any of the measurement operators to give a complete strategy without change to the winning probability. The same can be done with prover A's operators.}
Since the strategy is $\varepsilon$-perfect, the strategy satisfy a winning probability
\begin{equation}
    \sum_{ij,i'\in\{i,j\}}\frac{1}{2\abs{E_G}}\sum_{c_i\neq c_j,c_{i'}=\tilde{c}_{i'}}\mel{\psi}{A^{ij}_{c_ic_j}\otimes B^{i'}_{\tilde{c}_{i'}}}{\psi}\geq1-\varepsilon.
\end{equation}

Consider a strategy for BCS-3-COL:
\begin{enumerate}
    \item The provers pre-share quantum state $\ket{\psi}$.
    \item After receiving his challenge $y=(k,\beta)$, prover B performs measurement $\{B^k_{c_k}\}$ on his part of the state, and outputs $\tilde{b}_{k\alpha}=\begin{cases} 1 & c_k=\alpha\\ 0 & c_k\neq\alpha \end{cases}$.
    \item Depending on the challenge received, prover A performs one of two actions:
    \begin{enumerate}
        \item If challenge $x=1$, prover A randomly selects a neighbor $j$ of vertex $i$ and measure the quantum state with $\{A^{ij}_{c_ic_j}\}$. She then outputs $b_{i\alpha}=\begin{cases} 1 & c_i=\alpha\\ 0 & c_i\neq\alpha \end{cases}$ for $\alpha=0,1,2$.
        \item If challenge $x=(i,j,\beta)$, prover A measures the quantum state with the challenge edge $\{A^{ij}_{c_ic_j}\}$ and responds with $b_{i\alpha}=\begin{cases} 1 & c_i=\alpha\\ 0 & c_i\neq\alpha \end{cases}$ and $b_{j\alpha}=\begin{cases} 1 & c_j=\alpha\\ 0 & c_j\neq\alpha \end{cases}$.
    \end{enumerate}  
\end{enumerate}
We first show that the resulting strategy is vertex-projective color-commuting.
We can define the measurement operators corresponding to prover B's actions by $\tilde{B}^{k\alpha}_1=B^k_{\alpha}$, $\tilde{B}^{k\alpha}_0=\mathbb{I}-B^k_{\alpha}$.
Since the original strategy is complete, the new strategy is vertex-complete,
\begin{equation}
    \sum_{\alpha=0}^2\tilde{B}^{k\alpha}_1=\sum_{\alpha=0}^2B^k_{\alpha}=\mathbb{I}.
\end{equation}
Since prover B's measurement operators are projective, $(B^i_{\alpha})^2=B^i_{\alpha}$, $B^i_{\alpha}B^i_{\beta}=0$, the new operators commute, $[\tilde{B}^{k\alpha}_{b_{k\alpha}},\tilde{B}^{k\beta}_{b_{k\beta}}]=0$.
We can also define the measurement operators corresponding to prover A's actions in case (b) by $\tilde{A}^{ij\alpha}_{b_{i\alpha}b_{j\alpha}}$.
Notably, since prover A's actual measurement $A^{ij}_{c_ic_j}$ is independent on the challenge color $\alpha$, she can provide responses to $b_{i\alpha}$ and $b_{j\alpha}$ for all three colors $\alpha$ with a single measurement.
As such, the corresponding measurement operators commute, $[\tilde{A}^{ij\alpha}_{b_{i\alpha}b_{j\alpha}},\tilde{A}^{ij\beta}_{b_{i\beta}b_{j\beta}}]=0$ since the second measurement would provide the same colors $c_ic_j$ as the first measurement.\\

We can compute the winning probability of the edge verification challenges,
\begin{equation}
\begin{split}
    p_{win}^{EV}=&\sum_{ij\alpha,k\in\{i,j\}}\frac{1}{6\abs{E_G}}\sum_{\substack{b_{k\alpha}=\tilde{b}_{k\alpha}\\b_{i\alpha}b_{j\alpha}=0}}p_A(b_{i\alpha}b_{j\alpha}|c_ic_j)p_B(\tilde{b}_{k\alpha}|\tilde{c}_k)\mel{\psi}{A^{ij}_{c_ic_j}\otimes B^{k}_{\tilde{c}_k}}{\psi}\\
    \geq&\sum_{ij\alpha,k\in\{i,j\}}\frac{1}{6\abs{E_G}}\sum_{\substack{c_{k}=\tilde{c}_{k}\\c_{i}\neq c_j}}\mel{\psi}{A^{ij}_{c_ic_j}\otimes B^{k}_{\tilde{c}_k}}{\psi}\\
    =&\frac{1}{2\abs{E_G}}\sum_{ij,i'\in\{i,j\}}\mel{\psi}{A^{ij}_{c_ic_j}\otimes B^{i'}_{\tilde{c}_{i'}}}{\psi}\\
    \geq&1-\varepsilon,
\end{split}
\end{equation}
where the second line uses the fact that $c_k=\tilde{c}_k\implies b_{k\alpha}=\tilde{b}_{k\alpha}$ and $c_i\neq c_j\implies b_{i\alpha}b_{j\alpha}=0$ for all $\alpha=0,1,2$.
% The BCS-3-COL game is similar to the Alt-Edge-3-COL, with the difference being that the colors that Alice and Bob has to reply with is replaced by a YES/NO response to whether a vertex has a particular color.
% Therefore, it is clear that the Alt-Edge-3-COL strategy can be utilized directly for BCS-3-COL, with the extra step of converting the color to a YES/NO response.
% Therefore, the corresponding measurement operators $\tilde{B}^{k\alpha}_{b_{k\alpha}}$ and $\tilde{A}^{ij\alpha}_{b_{i\alpha}b_{j\alpha}}$ for the same vertex or edge will commute since the quantum measurement step is identical.
% Notably the construction gives $\tilde{B}^{k\alpha}_1=B^k_{\alpha}$, which implies completeness of $\tilde{B}^{k\alpha}_1$ since $B^k_{\alpha}$ is a complete set of measurement.
% As such, the strategy developed in vertex-complete color-commuting.\\

% The winning probability of the edge verification challenge is lower bounded by a larger set of acceptance condition on the original Alt-Edge-3-COL measurements, where $c_k=\tilde{c}_k$ and $c_i\neq c_j$.
% Therefore, the winning probability for the edge verification challenge is lower bounded by $1-\varepsilon$.
\end{proof}

\subsection{Almost Satisfying Assignment from BCS-3-COL}
\label{app:BCS-3-COL_to_ASA}

% We also require a form similar to the triangle inequality for Frobenius norm,
% \begin{theorem}
% \label{thm:Frobenius_Norm_Triangle_Ineq}
%     $\norm{A+A^{\dagger}}_F\leq 2\norm{A}_F$.
% \end{theorem}
% \begin{proof}
% We can simplify by
% \begin{equation}
% \begin{split}
%     \norm{A+A^{\dagger}}_F^2=&\sum_{ij}\abs{A_{ij}+A^*_{ji}}^2\\
%     =&\sum_{ij}\abs{A_{ij}}^2+\abs{A_{ji}^*}^2+2\Re[A_{ij}A_{ji}]\\
%     \leq&\sum_{ij}\abs{A_{ij}}^2+\abs{A_{ji}}^2+2\abs{A_{ij}A_{ji}}\\
%     \leq&2\sum_{ij}\abs{A_{ij}}^2+2\sum_{ij}\abs{A_{ij}}\abs{A_{ji}}\\
%     \leq&2\norm{A}_F^2+2\sqrt{(\sum_{ij}\abs{A_{ij}}^2)(\sum_{ij}\abs{A_{ji}}^2)}\\
%     =&4\norm{A}_F^2,
% \end{split}
% \end{equation}
% which implies that $\norm{A+A^{\dagger}}_F\leq 2\norm{A}_F$.
% We note that the sixth line utilizes the Cauchy-Schwarz inequality.
% \end{proof}

To study the representation formed from prover B's measurement operators, we first define observables with eigenvalues $\pm 1$ from the measurement operators.
Let prover A's observable be
\begin{equation}
    Y^{ij\alpha i'}=\sum_{b_{i\alpha}b_{j\alpha}}(-1)^{b_{i'\alpha}+1}\tilde{A}^{ij\alpha}_{b_{i\alpha}b_{j\alpha}},
\end{equation}
where $i'\in\{i,j\}$, and the observable is dependent on only the color for a single vertex.
prover B's observable can be similarly defined,
\begin{equation}
    X^{k\alpha}=\sum_{\tilde{b}_{k\alpha}}(-1)^{\tilde{b}_{k\alpha}+1}\tilde{B}^{k\alpha}_{\tilde{b}_{k\alpha}}=2B^k_{\alpha}-\mathbb{I}.
\end{equation}
Some important properties of these observables include:
\begin{enumerate}
    \item Observables are Hermitian, $(Y^{ij\alpha i'})^{\dagger}=Y^{ij\alpha i'}$, $(X^{k\alpha})^{\dagger}=X^{k\alpha}$, stemming from the fact that the projectors are Hermitian.
    \item The observables are unitary, since the projectors are complete and corresponds to $\pm 1$ eigenvalues.
    \item The observables for $Y$ corresponding the same edge commutes, $[Y^{ij\alpha i},Y^{ij\beta i'}]=0$, since the projectors themselves commute.
\end{enumerate}
For each challenge choice, we can compute an expectation value of the observables, which is close to the winning probability.
We can define the probability of passing the edge verification test for a challenge choice $(i,j,\alpha,i')$ as $\sum_{\substack{b_{i\alpha}b_{j\alpha}=0\\ b_{i'\alpha}=\tilde{b}_{i'\alpha}}}\mel{\psi}{\tilde{A}^{ij\alpha}_{b_{i\alpha}b_{j\alpha}}\otimes\tilde{B}^{i'\alpha}_{\tilde{b}_{i'\alpha}}}{\psi}=1-\varepsilon_{ij\alpha i'}$.
\begin{theorem}
\label{thm:Pass_EV_Implies_Good_ExpVal}
    If the probability of passing the edge verification test (for BCS-3-COL) is $1-\varepsilon_{ij\alpha i'}$ for any challenge choice $(i,j,\alpha,i')$, the expectation value of the corresponding observable is $\mel{\psi}{Y^{ij\alpha i'}\otimes X^{i'\alpha}}{\psi}\geq 1-2\varepsilon_{ij\alpha i'}$.
\end{theorem}
\begin{proof}
We can expand
\begin{equation}
\begin{split}
    &\mel{\psi}{Y^{ij\alpha i'}\otimes X^{i'\alpha}}{\psi}\\
    =&\sum_{b_{i\alpha}b_{j\alpha}\tilde{b}_{i'\alpha}}(-1)^{b_{i'\alpha}+\tilde{b}_{i'\alpha}}\mel{\psi}{\tilde{A}^{ij\alpha}_{b_{i\alpha}b_{j\alpha}}\otimes\tilde{B}^{i'\alpha}_{\tilde{b}_{i'\alpha}}}{\psi}\\
    =&\sum_{\substack{b_{i\alpha}b_{j\alpha}=0\\ b_{i'\alpha}=\tilde{b}_{i'\alpha}}}\mel{\psi}{\tilde{A}^{ij\alpha}_{b_{i\alpha}b_{j\alpha}}\otimes\tilde{B}^{i'\alpha}_{\tilde{b}_{i'\alpha}}}{\psi}-\sum_{\substack{b_{i\alpha}b_{j\alpha}=0\\ b_{i'\alpha}=\tilde{b}_{i'\alpha}}}\mel{\psi}{\tilde{A}^{ij\alpha}_{b_{i\alpha}b_{j\alpha}}\otimes\tilde{B}^{i'\alpha}_{\tilde{b}_{i'\alpha}}}{\psi}\\
    &+\sum_{\substack{b_{i\alpha}b_{j\alpha}=1\\ \tilde{b}_{i'\alpha}}}(-1)^{b_{i'\alpha}+\tilde{b}_{i'\alpha}}\mel{\psi}{\tilde{A}^{ij\alpha}_{b_{i\alpha}b_{j\alpha}}\otimes\tilde{B}^{i'\alpha}_{\tilde{b}_{i'\alpha}}}{\psi}\\
    \geq&1-\varepsilon-\sum_{\substack{b_{i\alpha}b_{j\alpha}=0\\ b_{i'\alpha}=\tilde{b}_{i'\alpha}}}\mel{\psi}{\tilde{A}^{ij\alpha}_{b_{i\alpha}b_{j\alpha}}\otimes\tilde{B}^{i'\alpha}_{\tilde{b}_{i'\alpha}}}{\psi}-\sum_{\substack{b_{i\alpha}b_{j\alpha}=1\\ \tilde{b}_{i'\alpha}}}\mel{\psi}{\tilde{A}^{ij\alpha}_{b_{i\alpha}b_{j\alpha}}\otimes\tilde{B}^{i'\alpha}_{\tilde{b}_{i'\alpha}}}{\psi}\\
    =&1-\varepsilon-\left(1-\sum_{\substack{b_{i\alpha}b_{j\alpha}=0\\ b_{i'\alpha}=\tilde{b}_{i'\alpha}}}\mel{\psi}{\tilde{A}^{ij\alpha}_{b_{i\alpha}b_{j\alpha}}\otimes\tilde{B}^{i'\alpha}_{\tilde{b}_{i'\alpha}}}{\psi}\right)\\
    \geq &1-2\varepsilon_{ij\alpha i'}.
\end{split}
\end{equation}
\end{proof}

Due to the Hermitian and unitary nature of the observables, one can derive the tracial property that is critical to the classical strategy formulation.
% We provide a proof of this in Appendix~\ref{app:Tracial_Property_BCS}.
\begin{theorem}[Lemma 4.7 of \cite{Paddock2024}]
\label{thm:Tracial_Property_BCS}
    Let $X$ and $Y$ be Hermitian and unitary operators on a finite-dimensional Hilbert space, and $\ket{\psi}\in\mathcal{H}_A\otimes\mathcal{H}_B$, with $\rho=\Tr_A[\dyad{\psi}]$. If $\mel{\psi}{Y\otimes X}{\psi}\geq 1-\varepsilon$, then $\norm{X\rho^{\frac{1}{2}}-\rho^{\frac{1}{2}}X}_F\leq\sqrt{2\varepsilon(2-\varepsilon)}$ and $\norm{X\rho^{\frac{1}{2}}-\rho^{\frac{1}{2}}Y^T}_F\leq\sqrt{2\varepsilon(2-\varepsilon)}$.
\end{theorem}
\begin{proof}
We follow closely the proof of Proposition 5.4 from \cite{Slofstra2018}.
WLOG, we can decompose the quantum state in the Schmidt basis, $\ket{\psi}=\sum_ic_i\ket{ii}$, with $c_i\in\mathbb{R}$ and consider the operators when restricted to the Schmidt basis.
Note that to make the two states identical and $c_i\in\mathbb{R}$, the unitary in the general Schmidt basis expansion $\ket{\psi}=\sum_ic_i\ket{i}\otimes U\ket{i}$ can be pushed to the measurement operators and observables, without affecting the properties of these operators (they remain projective, complete etc).
We can express the state $\ket{\psi}=\sum_i\ket{i}\otimes\sqrt{\rho}\ket{i}$,
\begin{equation}
    \ket{\psi}=\sum_i\ket{i}\sqrt{c_i^2}\ket{i}=\sum_i\ket{i}\otimes\sum_j\sqrt{c_j^2}\dyad{j}{j}\ket{i}=\sum_i\ket{i}\otimes\sqrt{\rho}\ket{i}.
\end{equation}
This simplification allows us to re-express $\mel{\psi}{Y\otimes X}{\psi}$ as
\begin{equation}
\label{eqn:psi_Y_X_simplification}
\begin{split}
    \mel{\psi}{Y\otimes X}{\psi}=&\sum_{ij}\mel{i}{Y}{j}\mel{i}{\rho^{\frac{1}{2}}X\rho^{\frac{1}{2}}}{j}\\
    =&\sum_j\mel{j}{Y^T(\sum_i\dyad{i})\rho^{\frac{1}{2}}X\rho^{\frac{1}{2}}}{j}\\
    =&\Tr[\rho^{\frac{1}{4}}Y^T\rho^{\frac{1}{4}}\rho^{\frac{1}{4}}X\rho^{\frac{1}{4}}]\\
    =&\sum_{ij}A_{ij}B_{ji},\quad A=\rho^{\frac{1}{4}}Y^T\rho^{\frac{1}{4}},\, B=\rho^{\frac{1}{4}}X\rho^{\frac{1}{4}}\\
    \leq&\abs{\sum_{ij}(A_{ji}^{\dagger})^*B_{ji}}\\
    \leq&\sqrt{\sum_{ij}\abs{A^{\dagger}_{ji}}^2}\sqrt{\sum_{ij}\abs{B_{ji}}^2}\\
    =&\norm{\rho^{\frac{1}{4}}Y^T\rho^{\frac{1}{4}}}_F\norm{\rho^{\frac{1}{4}}X\rho^{\frac{1}{4}}}_F\\
    \leq&\norm{\rho^{\frac{1}{4}}X\rho^{\frac{1}{4}}}_F,
\end{split}
\end{equation}
where the second line uses the transpose property, $\mel{i}{A}{j}=\mel{j}{A^T}{i}$, the fourth line notes $a\leq\abs{a}$ (note in our case, $a$ is real), the sixth line uses the Cauchy-Schwarz inequality with the inner product on a complex vector space.
The final simplification uses the property
\begin{equation}
    \norm{\rho^{\frac{1}{4}}Y^T\rho^{\frac{1}{4}}}_F^2=\Tr[Y^T\rho^{\frac{1}{2}}Y^*\rho^{\frac{1}{2}}]\leq\norm{Y^T\rho^{\frac{1}{2}}Y^*}_{F}=\norm{\rho^{\frac{1}{2}}}_F=1,
\end{equation}
where we used Thm.~\ref{thm:Normal_Op_Frobenius_Norm}, and the fact that $Y$ is unitary.
Therefore, we can deduce $\norm{\rho^{\frac{1}{4}}X\rho^{\frac{1}{4}}}_F\geq 1-\varepsilon$.\\

We can now prove the tracial property for $\varepsilon\in[0,1]$,
\begin{equation}
\begin{split}
    \norm{X\rho^{\frac{1}{2}}-\rho^{\frac{1}{2}}X}_F^2=&2-2\Tr[X^{\dagger}\rho^{\frac{1}{2}}X\rho^{\frac{1}{2}}]\\
    =&2-2\norm{\rho^{\frac{1}{4}}X\rho^{\frac{1}{4}}}_F^2\\
    \leq &2-2(1-\varepsilon)^2\\
    =&2\varepsilon(2-\varepsilon),
\end{split}
\end{equation}
which gives the result in the theorem.
We note that the inequality $\norm{\rho^{\frac{1}{4}}X\rho^{\frac{1}{4}}}_F^2\geq (1-\varepsilon)^2$ may not hold for $\varepsilon\in(1,2]$, when $1-\varepsilon<0$.
For $\varepsilon\in(1,2]$, we can define $\tilde{X}=-X$, which gives $\mel{\psi}{Y\otimes\tilde{X}}{\psi}=1-(2-\varepsilon)$, where $2-\varepsilon\in[0,1)$.
Applying the same analysis with $\tilde{X}$, we arrive at $\norm{\tilde{X}\rho^{\frac{1}{2}}-\rho^{\frac{1}{2}}\tilde{X}}_F\leq2\varepsilon(2-\varepsilon)$.
Since the Frobenius norm is invariant to a sign change, we arrive at the same bound.
We can perform a similar expansion for the second property, with
\begin{equation}
    \norm{X\rho^{\frac{1}{2}}-\rho^{\frac{1}{2}}Y^T}_F^2=2-2\Tr[Y^T\rho^{\frac{1}{2}}X\rho^{\frac{1}{2}}]\leq 2\varepsilon(2-\varepsilon),
\end{equation}
where we used the fact that $\Tr[\rho^{\frac{1}{4}}Y^T\rho^{\frac{1}{4}}\rho^{\frac{1}{4}}X\rho^{\frac{1}{4}}]=\mel{\psi}{Y\otimes X}{\psi}$.
\end{proof}
We can combine Thm.~\ref{thm:Pass_EV_Implies_Good_ExpVal} and Thm.~\ref{thm:Tracial_Property_BCS} as a corollary,
\begin{corollary}
\label{thm:Pass_EV_Observables_Tracial}
    If the probability of passing the edge verification test is $1-\varepsilon_{ij\alpha i'}$ for all challenge choices $(i,j,\alpha,i')$, then for $\rho=\Tr_A[\dyad{\psi}]$, the corresponding observables satisfy the tracial properties $\norm{X^{i'\alpha}\rho^{\frac{1}{2}}-\rho^{\frac{1}{2}}X^{i'\alpha}}_F\leq2\sqrt{2\varepsilon_{ij\alpha i'}(2-\varepsilon_{ij\alpha i'})}$ and  $\norm{X^{i'\alpha}\rho^{\frac{1}{2}}-\rho^{\frac{1}{2}}(Y^{ij\alpha i'})^T}_F\leq2\sqrt{2\varepsilon_{ij\alpha i'}(2-\varepsilon_{ij\alpha i'})}$.
\end{corollary}

The tracial property is important to prove the two desired property of almost commuting and edge coloring, which we present in Thm.~\ref{thm:Pass_EV_Observables_Commute} and Thm.~\ref{thm:Pass_EV_Edge_Coloring}.
The almost commuting property can be shown by linking the commutation of $X$ and the commutation of $Y$ via the tracial properties, and noting that $Y$ commutes by construction.
The edge coloring property is demonstrated by linking the $X$ and $Y$ operators via the tracial properties, wherein the edge coloring constraint for $Y$ reduces to $4\tilde{A}^{ij\alpha}_{11}$, for which corresponds to a wrong response to the BCS edge verification challenge.
Before we proceed, we prove a useful property.
\begin{theorem}
\label{thm:Doubly_Tracial_Property}
    If the probability of passing the edge verification test is $1-\varepsilon_{ij\alpha i'}$ for all challenge choices $(i,j,\alpha,i')$, then for any edge $(i,j)\in E_G$, and any $\alpha,\beta\in\{0,1,2\}$ the corresponding observables satisfy
    \begin{equation*}
        \norm{X^{i\alpha}X^{j\beta}\rho^{\frac{1}{2}}-\rho^{\frac{1}{2}}(Y^{ij\beta j})^T(Y^{ij\alpha i})^T}_F\leq 2\sqrt{2\varepsilon_{ij\alpha i}(2-\varepsilon_{ij\alpha i})}+2\sqrt{2\varepsilon_{ij\beta j}(2-\varepsilon_{ij\beta j})}.
    \end{equation*}
\end{theorem}
\begin{proof}
By noting that the $Y$ observables commute, $[Y^{ij\alpha i},Y^{ij\beta j}]=0$, we can express
\begin{equation}
\begin{split}
    &\norm{X^{i\alpha}X^{j\beta}\rho^{\frac{1}{2}}-\rho^{\frac{1}{2}}(Y^{ij\beta j})^T(Y^{ij\alpha i})^T}_F\\
    =&\norm{X^{i\alpha}X^{j\beta}\rho^{\frac{1}{2}}(Y^{ij\beta j})^*(Y^{ij\alpha i})^*-\rho^{\frac{1}{2}}}_F\\
    \leq&\norm{X^{i\alpha}X^{j\beta}\rho^{\frac{1}{2}}(Y^{ij\beta j})^*(Y^{ij\alpha i})^*-X^{i\alpha}\rho^{\frac{1}{2}}(Y^{ij\alpha i})^*}_F+\norm{X^{i\alpha}\rho^{\frac{1}{2}}(Y^{ij\alpha i})^*-\rho^{\frac{1}{2}}}_F\\
    =&\norm{X^{j\beta}\rho^{\frac{1}{2}}-\rho^{\frac{1}{2}}(Y^{ij\beta j})^T}_F+\norm{X^{i\alpha}\rho^{\frac{1}{2}}-\rho^{\frac{1}{2}}(Y^{ij\alpha i})^T}_F\\
    \leq&2\sqrt{2\varepsilon_{ij\beta j}(2-\varepsilon_{ij\beta j})}+2\sqrt{2\varepsilon_{ij\alpha i}(2-\varepsilon_{ij\alpha i})},
\end{split}
\end{equation}
where the first line uses the unitary property of $Y$, and that Frobenius norm is unitary-invariant.
The second line uses the triangle inequality, and the third line uses the unitary property of both $X$ and $Y$, while the final line uses Corollary~\ref{thm:Pass_EV_Observables_Tracial}.
\end{proof}

We can now proceed to prove the almost commutation property of neighboring vertices.
\begin{theorem}
\label{thm:Pass_EV_Observables_Commute}
    If the probability of passing the edge verification test is $1-\varepsilon_{ij\alpha i'}$ for all challenge choices $(i,j,\alpha,i')$, then for any edge $(i,j)$ and any $\alpha,\beta\in\{0,1,2\}$, 
    \begin{equation*}
        \norm{[X^{i\alpha},X^{j\beta}]\rho^{\frac{1}{2}}}_F\leq 4\sqrt{2\varepsilon_{ij\beta j}(2-\varepsilon_{ij\beta j})}+4\sqrt{2\varepsilon_{ij\alpha i}(2-\varepsilon_{ij\alpha i})}
    \end{equation*}
\end{theorem}
\begin{proof}
We can expand the commutation relation and use the triangle inequality to simplify,
\begin{equation}
\begin{split}
    \norm{[X^{i\alpha},X^{j\beta}]\rho^{\frac{1}{2}}}_F\leq&\norm{X^{i\alpha}X^{j\beta}\rho^{\frac{1}{2}}-\rho^{\frac{1}{2}}(Y^{ij\beta j})^T(Y^{ij\alpha i})^T}_F\\
    &+\norm{\rho^{\frac{1}{2}}(Y^{ij\beta j})^T(Y^{ij\alpha i})^T-X^{j\beta}X^{i\alpha}\rho^{\frac{1}{2}}}_F\\
    =&\norm{X^{i\alpha}X^{j\beta}\rho^{\frac{1}{2}}-\rho^{\frac{1}{2}}(Y^{ij\beta j})^T(Y^{ij\alpha i})^T}_F\\
    &+\norm{X^{j\beta}X^{i\alpha}\rho^{\frac{1}{2}}-\rho^{\frac{1}{2}}(Y^{ij\alpha i})^T(Y^{ij\beta j})^T}_F\\
    \leq&4\sqrt{2\varepsilon_{ij\beta j}(2-\varepsilon_{ij\beta j})}+4\sqrt{2\varepsilon_{ij\alpha i}(2-\varepsilon_{ij\alpha i})},
\end{split}
\end{equation}
where we note that the $Y$ observables commute, $[Y^{ij\alpha i},Y^{ij\beta j}]=0$.
\end{proof}

We can follow \cite{Paddock2024} to construct the edge verification property, where we expect $X^{i\alpha}+X^{j\alpha}+X^{i\alpha}X^{j\alpha}+\mathbb{I}$ to be small.
Since the event where $b_{i\alpha}=b_{j\alpha}=1$ results in a large value, the guarantee above restricts the probability of the event $b_{i\alpha}b_{j\alpha}=1$.
\begin{theorem}
\label{thm:Pass_EV_Edge_Coloring}
    If the probability of passing the edge verification test is $1-\varepsilon_{ij\alpha i'}$ for all challenge choices $(i,j,\alpha,i')$, then for any edge $(i,j)\in E_G$, and any $\alpha,\beta\in\{0,1,2\}$ the corresponding observables satisfy
    \begin{equation*}
    \begin{split}
        &\norm{(X^{i\alpha}+X^{j\alpha}+X^{i\alpha}X^{j\alpha}+\mathbb{I})\rho^{\frac{1}{2}}}_F\\
        \leq &4\sqrt{2\varepsilon_{ij\alpha i}(2-\varepsilon_{ij\alpha i})}+4\sqrt{2\varepsilon_{ij\alpha j}(2-\varepsilon_{ij\alpha j})}+2\sqrt{2(\varepsilon_{ij\alpha i}+\varepsilon_{ij\alpha j})}.
    \end{split}
    \end{equation*}
\end{theorem}
\begin{proof}
Let us consider the same expression for the $Y$ observables,
\begin{equation}
\begin{split}
    &Y^{ij\alpha i}+Y^{ij\alpha j}+Y^{ij\alpha i}Y^{ij\alpha j}+\mathbb{I}\\
    =&\sum_{b_{i\alpha}b_{j\alpha}}[(-1)^{b_{i\alpha}+1}+(-1)^{b_{j\alpha}+1}+(-1)^{b_{i\alpha}+b_{j\alpha}}]\tilde{A}^{ij\alpha}_{b_{i\alpha}b_{j\alpha}}+\mathbb{I}\\
    =&-\tilde{A}^{ij\alpha}_{00}-\tilde{A}^{ij\alpha}_{01}-\tilde{A}^{ij\alpha}_{10}+3\tilde{A}^{ij\alpha}_{11}+\mathbb{I}\\
    =&4\tilde{A}^{ij\alpha}_{11},
\end{split}
\end{equation}
where the event $b_{i\alpha}=b_{j\alpha}=1$ corresponding to the final operator is expected to have low probability.
This can be shown from the Frobenius norm on the measurement operator acting on the state,
\begin{equation}
\begin{split}
    \norm{\rho^{\frac{1}{2}}(\tilde{A}^{ij\alpha}_{11})^T}_F=&\sqrt{\Tr[(\tilde{A}^{ij\alpha}_{11})^T\rho^{\frac{1}{2}}\mathbb{I}\rho^{\frac{1}{2}}]}\\
    =&\sqrt{\mel{\psi}{\tilde{A}^{ij\alpha}_{11}\otimes \mathbb{I}}{\psi}}\\
    =&\sqrt{\frac{1}{2}\left(\sum_{\tilde{b}_{i\alpha}}\mel{\psi}{\tilde{A}^{ij\alpha}_{11}\otimes \tilde{B}^{i\alpha}_{\tilde{b}_{i\alpha}}}{\psi}+\sum_{\tilde{b}_{j\alpha}}\mel{\psi}{\tilde{A}^{ij\alpha}_{11}\otimes \tilde{B}^{j\alpha}_{\tilde{b}_{j\alpha}}}{\psi}\right)}\\
    \leq&\sqrt{\frac{1}{2}(\varepsilon_{ij\alpha i}+\varepsilon_{ij\alpha j})},
\end{split}
\end{equation}
where the first line follows the definition of the Frobenius norm, the second line uses the simplification in Eqn.~\ref{eqn:psi_Y_X_simplification}.
The third line expands the identity operator into prover B's measurement operators, with a randomly chosen vertex in $\{i,j\}$.
The final line notes that the probability given by the measurement outcome matches the case when $b_{i\alpha}b_{j\alpha}=1$, which is bounded by the probability of failing the edge verification test.\\

We can therefore compute the constraint
\begin{equation}
\begin{split}
    &\norm{(X^{i\alpha}+X^{j\alpha}+X^{i\alpha}X^{j\alpha}+\mathbb{I})\rho^{\frac{1}{2}}}_F\\
    =&\norm{(X^{i\alpha}+X^{j\alpha}+X^{i\alpha}X^{j\alpha})\rho^{\frac{1}{2}}+\rho^{\frac{1}{2}}(4\tilde{A}^{ij\alpha}_{11}-Y^{ij\alpha i}-Y^{ij\alpha j}-Y^{ij\alpha i}Y^{ij\alpha j})^T}_F\\
    \leq&\norm{X^{i\alpha}\rho^{\frac{1}{2}}-\rho^{\frac{1}{2}}(Y^{ij\alpha i})^T}_F+\norm{X^{j\alpha}\rho^{\frac{1}{2}}-\rho^{\frac{1}{2}}(Y^{ij\alpha j})^T}_F+\norm{4\rho^{\frac{1}{2}}\tilde{A}^{ij\alpha}_{11}}_F\\
    &+\norm{X^{i\alpha}X^{j\alpha}\rho^{\frac{1}{2}}-\rho^{\frac{1}{2}}(Y^{ij\alpha j})^T(Y^{ij\alpha i})^T}_F\\
    \leq&4\sqrt{2\varepsilon_{ij\alpha i}(2-\varepsilon_{ij\alpha i})}+4\sqrt{2\varepsilon_{ij\alpha j}(2-\varepsilon_{ij\alpha j})}+2\sqrt{2(\varepsilon_{ij\alpha i}+\varepsilon_{ij\alpha j})},
\end{split}
\end{equation}
where the first line expands the identity operator as described earlier, the second line uses the triangle inequality, and the final line uses the fact that probability of $b_{i\alpha}=b_{j\alpha}=1$ is small, the tracial property and Thm.~\ref{thm:Doubly_Tracial_Property}.
\end{proof}

As such, we can demonstrate that an almost satisfying assignment can be formed from prover B's operators,
\begin{theorem}
\label{thm:BCS-3-COL_to_Almost-Satisfying-Assignment}
    Let $1-\varepsilon_{ij\alpha k}$ be the probability of the provers passing the edge verification test for challenge $(i,j,\alpha)$ to prover A and $(k,\alpha)$ to prover B.
    If there exists a vertex-complete color-commuting strategy that can win the edge verification challenge %with a distribution of $\mathcal{D}_{ijk\alpha}=\frac{\delta_{ik}+\delta_{jk}}{6\abs{E}}$ 
    with $p_{win}^{EV}\geq 1-\varepsilon$, then prover B's measurement operators $B^i_{\alpha}=\tilde{B}^{i\alpha}_1$ forms an almost satisfying assignment, satisfy the following conditions for any edge $(i,j)\in E_G$:
    \begin{enumerate}
        \item Tracial property: $\norm{[B^i_{\alpha},\rho^{\frac{1}{2}}]}_F\leq\sqrt{2\varepsilon_{ij\alpha i}(2-\varepsilon_{ij\alpha i})}+\sqrt{2\varepsilon_{ij\alpha j}(2-\varepsilon_{ij\alpha j})}$
        \item Almost commuting: $\norm{[B^i_{\alpha},B^j_{\beta}]\rho^{\frac{1}{2}}}_F\leq\sqrt{2\varepsilon_{ij\alpha i}(2-\varepsilon_{ij\alpha i})}+\sqrt{2\varepsilon_{ij\beta j}(2-\varepsilon_{ij\beta j})}$
        \item Edge coloring: $\norm{B^i_{\alpha}B^j_{\alpha}\rho^{\frac{1}{2}}}_F\leq\sqrt{2\varepsilon_{ij\alpha i}(2-\varepsilon_{ij\alpha i})}+\sqrt{2\varepsilon_{ij\alpha j}(2-\varepsilon_{ij\alpha j})}+\sqrt{\frac{1}{2}(\varepsilon_{ij\alpha i}+\varepsilon_{ij\alpha j})}$
    \end{enumerate}
    and $\frac{1}{6\abs{E_G}}\sum_{ij\alpha k}\varepsilon_{ij\alpha k}\leq\varepsilon$. 
\end{theorem}
\begin{proof}
By definition, the overall probability of passing the edge verification test is given by
\begin{equation}
    \frac{1}{6\abs{E_G}}\sum_{ij\alpha k}(1-\varepsilon_{ij\alpha k})\geq 1-\varepsilon.
\end{equation}
A simple rearrangement would give the upper bound on the average of $\varepsilon_{ij\alpha k}$.\\

With the definition of edge verification test, we can construct observables $X^{k\alpha}=2B^k_{\alpha}-\mathbb{I}$, which from Corollary~\ref{thm:Pass_EV_Observables_Tracial}, \ref{thm:Pass_EV_Observables_Commute}, and \ref{thm:Pass_EV_Edge_Coloring}, satisfy
\begin{equation}
\begin{gathered}
    \norm{[X^i_{\alpha},\rho^{\frac{1}{2}}]}_F\leq2\sqrt{2\varepsilon_{ij\alpha i'}(2-\varepsilon_{ij\alpha i'})}\\
    \norm{[X^{i\alpha},X^{j\beta}]\rho^{\frac{1}{2}}}_F\leq 4\sqrt{2\varepsilon_{ij\beta j}(2-\varepsilon_{ij\beta j})}+4\sqrt{2\varepsilon_{ij\alpha i}(2-\varepsilon_{ij\alpha i})}\\
    \norm{(X^{i\alpha}+X^{j\alpha}+X^{i\alpha}X^{j\alpha}+\mathbb{I})\rho^{\frac{1}{2}}}_F\leq 4\sqrt{2\varepsilon_{ij\alpha i}(2-\varepsilon_{ij\alpha i})}\\
    +4\sqrt{2\varepsilon_{ij\alpha j}(2-\varepsilon_{ij\alpha j})}+2\sqrt{2(\varepsilon_{ij\alpha i}+\varepsilon_{ij\alpha j})}.
\end{gathered}
\end{equation}
We can expand the observables for each condition.
The tracial property can be expanded, noting $\mathbb{I}$ commutes with any operator,
\begin{equation}
    \norm{[X^i_{\alpha},\rho^{\frac{1}{2}}]}_F=2\norm{[B^i_{\alpha},\rho^{\frac{1}{2}}]}_F.
\end{equation}
The commutation relation can be similarly expanded,
\begin{equation}
    \norm{[X^{i\alpha},X^{j\beta}]\rho^{\frac{1}{2}}}_F=4\norm{[B^i_{\alpha},X^j_{\beta}]\rho^{\frac{1}{2}}}_F.
\end{equation}
The edge coloring constraint reduces to a familiar form, where the two operators of neighboring vertices corresponding to the same color is almost orthogonal (wrt to the state),
\begin{equation}
    \norm{(X^{i\alpha}+X^{j\alpha}+X^{i\alpha}X^{j\alpha}+\mathbb{I})\rho^{\frac{1}{2}}}_F=\norm{4B^i_{\alpha}B^j_{\alpha}\rho^{\frac{1}{2}}}_F.
\end{equation}
Rearranging the inequalities yields the conditions in the theorem.
\end{proof}

\subsection{Almost Fully Commuting Satisfying Assignment from BCS-3-COL}
\label{app:BCS-3-COL_to_FC-ASA}

While the theorem establishing an almost satisfying assignment from prover B's measurement operators is useful, it remains insufficient to craft a classical strategy without additional structure to the operators.
A desirable property would be that the almost commuting property hold for all operators.
One method of achieving this property for a graph $G$ would be to introduce the commutative gadget (shown in Figure~\ref{fig:commutative_gadget}) from \cite{Ji_2013} to form a new graph $G'$.
In his analysis, Ji demonstrated when the operators for neighboring vertices in $G'$ commute for a perfect strategy, all operators in the graph $G$ (also $G'$) commutes.\\

We can replicate a similar property for the almost 
satisfying assignment for $G'$ following similar ideas to the proof in \cite{Ji_2013}, forming an almost satisfying assignment for $G$ where all assignment operators are almost commuting.
% A complete proof is provided in Appendix~\ref{app:almost_qsa-to-fully_commuting_qsa}.
\begin{theorem}
\label{thm:almost_qsa-to-fully_commuting_qsa}
    Let $1-\varepsilon_{ij\alpha k}$ be the probability of the provers passing the edge verification test for challenge $(i,j,\alpha)$ to prover A and $(k,\alpha)$ to prover B, and $G'$ be a graph constructed by inserting the commutative gadget in $G$.
    If there exists a vertex-complete color-commuting strategy that can win the edge verification challenge of $G'$ %with a distribution of $\mathcal{D}_{ijk\alpha}=\frac{\delta_{ik}+\delta_{jk}}{6\abs{E_{G'}}}$ 
    with $p_{win}^{EV}\geq 1-\varepsilon$, then prover B's measurement operators $B^i_{\alpha}=\tilde{B}^{i\alpha}_1$ are almost commuting for any non-neighboring vertex pair $i,j\in V_G$, $(i,j)\notin E_G$, i.e. 
    \begin{equation*}
        \norm{[B^i_{\alpha},B^j_{\beta}]\rho^{\frac{1}{2}}}_F\leq\sum_{(u,v)\in E_{cg},\alpha}\left[4\sqrt{\frac{1}{2}(\varepsilon_{uv\alpha u}+\varepsilon_{uv\alpha v})}+19\sum_{u'\in\{u,v\}}\sqrt{2\varepsilon_{uv\alpha u'}(2-\varepsilon_{uv\alpha u'})}\right],
    \end{equation*}
    where $E_{cg}$ are the edges in the commutative gadget.
\end{theorem}
\begin{proof}%[Proof of Thm.~\ref{thm:almost_qsa-to-fully_commuting_qsa}]
Let $\delta_{ij\alpha\beta}=\sqrt{2\varepsilon_{ij\alpha i}(2-\varepsilon_{ij\alpha i})}+\sqrt{2\varepsilon_{ij\beta j}(2-\varepsilon_{ij\beta j})}$ and $\delta_{ij\alpha}'=\delta_{ij\alpha\alpha}+\sqrt{\frac{1}{2}(\varepsilon_{ij\alpha i}+\varepsilon_{ij\alpha j})}$.
From Thm.~\ref{thm:BCS-3-COL_to_Almost-Satisfying-Assignment}, prover B's operators satisfies the tracial, almost commuting, and edge coloring properties for neighboring vertices in $E_{G'}$.
Before proving the main theorem, we prove several useful properties relating to the Frobenius norm of three consecutive measurement operators.
For three vertices $(i,j,k)$ that are connected in a triangle, 
\begin{equation}
\begin{split}
    \norm{B^i_{\alpha}B^j_{\alpha}B^k_{\beta}\rho^{\frac{1}{2}}}_F\leq&\norm{B^i_{\alpha}B^j_{\alpha}[B^k_{\beta},\rho^{\frac{1}{2}}]}_F+\norm{B^i_{\alpha}B^j_{\alpha}\rho^{\frac{1}{2}}B^k_{\beta}}_F\\
    \leq&\norm{[B^k_{\beta},\rho^{\frac{1}{2}}]}_F+\norm{B^i_{\alpha}B^j_{\alpha}\rho^{\frac{1}{2}}}_F\\
    \leq&\delta_{jk\beta\beta}+\delta_{ij\alpha}',
\end{split}
\end{equation}
and
\begin{equation}
\begin{split}
    \norm{B^i_{\alpha}B^j_{\beta}B^k_{\alpha}\rho^{\frac{1}{2}}}_F\leq&\norm{B^i_{\alpha}[B^j_{\beta},B^k_{\alpha}]\rho^{\frac{1}{2}}}_F+\norm{B^i_{\alpha}B^k_{\alpha}B^j_{\beta}\rho^{\frac{1}{2}}}_F\\
    \leq&\delta_{jk\alpha\beta}+\delta_{jk\beta\beta}+\delta'_{ik\alpha}.
\end{split}
\end{equation}
For three vertices $(i,j,k)$ with edges $(i,j)$ and $(j,k)$, we have that
\begin{equation}
\begin{split}
    \norm{B^j_{\alpha}B^k_{\alpha}B^i_{\beta}\rho^{\frac{1}{2}}}_F\leq&\norm{B^j_{\alpha}B^k_{\alpha}[B^i_{\beta},\rho^{\frac{1}{2}}]}_F+\norm{B^j_{\alpha}B^k_{\alpha}\rho^{\frac{1}{2}}B^i_{\beta}}_F\\
    \leq&\norm{[B^i_{\beta},\rho^{\frac{1}{2}}]}_F+\norm{B^j_{\alpha}B^k_{\alpha}\rho^{\frac{1}{2}}}_F\\
    \leq&\delta_{ij\beta\beta}+\delta_{jk\alpha}'.
\end{split}
\end{equation}
For a difference of three operators, we have
\begin{equation}
\begin{split}
    &\norm{(B^i_{\alpha}B^j_{\beta}B^k_{\alpha}-B^k_{\alpha}B^j_{\beta}B^i_{\alpha})\rho^{\frac{1}{2}}}_F\\
    \leq&\norm{(B^i_{\alpha}B^k_{\alpha}B^j_{\beta}-B^k_{\alpha}B^i_{\alpha}B^j_{\beta})\rho^{\frac{1}{2}}}_F+\norm{B^i_{\alpha}[B^j_{\beta},B^k_{\alpha}]\rho^{\frac{1}{2}}}_F+\norm{B^k_{\alpha}[B^j_{\beta},B^i_{\alpha}]\rho^{\frac{1}{2}}}_F\\
    \leq&\norm{[B^i_{\alpha},B^k_{\alpha}]\rho^{\frac{1}{2}}B^j_{\beta}}_F+2\norm{[B^j_{\alpha},\rho^{\frac{1}{2}}]}_F +\delta_{jk\beta\alpha}+\delta_{ij\alpha\beta}\\
    \leq&\delta_{ik\alpha\alpha}+2\delta_{ij\alpha\alpha}+\delta_{jk\beta\alpha}+\delta_{ij\alpha\beta}.
\end{split}
\end{equation}

The proof of the main theorem can be split into three parts, corresponding to Lemma 4 and the two parts of Lemma 5 of \cite{Ji_2013}.
The first part proves that three vertices that are connected in a triangle must be of different colors, the second part uses this property to prove that operators corresponding to the same color commutes, and the third part extends this to operators corresponding to different colors.
For the first part, we consider $B^i_{\alpha}+B^j_{\alpha}+B^k_{\alpha}-\mathbb{I}$, which we expect to be small since one of the three vertices $i,j,k$ should give $\alpha$.
We can expand this expression,
\begin{equation}
\begin{split}
    &B^i_{\alpha}+B^j_{\alpha}+B^k_{\alpha}-\mathbb{I}\\
    =&-B^i_{\beta}-B^i_{\gamma}+B^j_{\alpha}+B^k_{\alpha}\\
    =&-(B^i_{\beta}+B^i_{\gamma})(B^j_{\alpha}+B^j_{\beta}+B^j_{\gamma})+(B^i_{\alpha}+B^i_{\beta}+B^i_{\gamma})B^j_{\alpha}+B^k_{\alpha}\\
    =&B^i_{\alpha}B^j_{\alpha}-B^i_{\beta}B^j_{\beta}-B^i_{\gamma}B^j_{\gamma}-(B^i_{\beta}B^j_{\gamma}+B^i_{\gamma}B^j_{\beta})(B^k_{\alpha}+B^k_{\beta}+B^k_{\gamma})\\
    &+(B^i_{\alpha}+B^i_{\beta}+B^i_{\gamma})(B^j_{\alpha}+B^j_{\beta}+B^j_{\gamma})B^k_{\alpha}\\
    =&B^i_{\alpha}B^j_{\alpha}-B^i_{\beta}B^j_{\beta}-B^i_{\gamma}B^j_{\gamma}-B^i_{\beta}B^j_{\gamma}B^k_{\beta}-B^i_{\beta}B^j_{\gamma}B^k_{\gamma}-B^i_{\gamma}B^j_{\beta}B^k_{\beta}-B^i_{\gamma}B^j_{\beta}B^k_{\gamma}\\
    &+B^i_{\alpha}B^k_{\alpha}+B^i_{\beta}B^j_{\alpha}B^k_{\alpha}+B^i_{\beta}B^j_{\beta}B^k_{\alpha}+B^i_{\gamma}B^j_{\alpha}B^k_{\alpha}+B^i_{\gamma}B^j_{\gamma}B^k_{\alpha},
\end{split}
\end{equation}
using the fact that prover B's operators are vertex-complete.
As such, we can compute the Frobenius norm
\begin{equation}
\begin{split}
    &\norm{(B^i_{\alpha}+B^j_{\alpha}+B^k_{\alpha}-\mathbb{I})\rho^{\frac{1}{2}}}_F\\
    \leq&\norm{B^i_{\alpha}B^j_{\alpha}\rho^{\frac{1}{2}}}_F+\norm{B^i_{\beta}B^j_{\beta}\rho^{\frac{1}{2}}}_F+\norm{B^i_{\gamma}B^j_{\gamma}\rho^{\frac{1}{2}}}_F+\norm{B^i_{\beta}B^j_{\gamma}B^k_{\beta}\rho^{\frac{1}{2}}}_F\\
    &+\norm{B^i_{\beta}B^j_{\gamma}B^k_{\gamma}\rho^{\frac{1}{2}}}_F+\norm{B^i_{\gamma}B^j_{\beta}B^k_{\beta}\rho^{\frac{1}{2}}}_F+\norm{B^i_{\gamma}B^j_{\beta}B^k_{\gamma}\rho^{\frac{1}{2}}}_F+\norm{B^i_{\alpha}B^k_{\alpha}\rho^{\frac{1}{2}}}_F\\
    &+\norm{B^i_{\beta}B^j_{\alpha}B^k_{\alpha}\rho^{\frac{1}{2}}}_F+\norm{B^i_{\beta}B^j_{\beta}B^k_{\alpha}\rho^{\frac{1}{2}}}_F+\norm{B^i_{\gamma}B^j_{\alpha}B^k_{\alpha}\rho^{\frac{1}{2}}}_F+\norm{B^i_{\gamma}B^j_{\gamma}B^k_{\alpha}\rho^{\frac{1}{2}}}_F\\
    \leq&\sum_{\alpha=0}^2\delta_{ij\alpha}'+\delta_{jk\beta}+\delta_{jk\gamma\beta}+\delta_{jk\gamma\gamma}+\delta_{ik\beta}'+\delta_{jk\gamma}'+\delta_{jk\beta}'+\delta_{jk\gamma\beta}+\delta_{jk\beta\beta}\\
    &+\delta_{ik\gamma}'+\delta_{ik\alpha}'+2\delta_{jk\alpha}'+\delta_{jk\alpha\alpha}+\delta_{ij\beta}'+\delta_{jk\alpha\alpha}+\delta_{ij\gamma}'.
\end{split}
\end{equation}
Noting that the expansion choice with vertex $i$ as focus is arbitrary, and that we can perform the expansion similarly with $j$ or $k$, we are able to symmetrize and simplify the result,
\begin{equation}
    \norm{(B^i_{\alpha}+B^j_{\alpha}+B^k_{\alpha}-\mathbb{I})\rho^{\frac{1}{2}}}_F\leq2\sum_{u\neq v\in\{i,j,k\},\alpha}\delta_{uv\alpha}'+\frac{2}{3}\sum_{u\neq v\in\{i,j,k\},\alpha,\beta}\delta_{uv\alpha\beta},
\end{equation}
noting that many terms are added to reach a symmetric result.\\

The second part of the proof examines the commutator between the operators corresponding to vertices in $G$, which are by construction the operators at positions $a$ and $e$ of the commutative gadget in $G'$.
As such, we simply need to show that the operators at these two positions almost commute.
We begin with the case where the two operators correspond to the same color,
\begin{equation}
\begin{split}
    &\norm{[B^a_{\alpha},B^e_{\alpha}]\rho^{\frac{1}{2}}}_F\\
    \leq&\norm{([B^a_{\alpha},B^e_{\alpha}]-B^c_{\alpha}B^f_{\alpha}B^a_{\alpha}+B^a_{\alpha}B^f_{\alpha}B^c_{\alpha})\rho^{\frac{1}{2}}}_F+\norm{B^c_{\alpha}B^f_{\alpha}B^a_{\alpha}\rho^{\frac{1}{2}}}_F+\norm{B^a_{\alpha}B^f_{\alpha}B^c_{\alpha}\rho^{\frac{1}{2}}}_F\\
    \leq&\norm{([B^a_{\alpha},B^e_{\alpha}]-(\mathbb{I}-B^a_{\alpha}-B^b_{\alpha})(\mathbb{I}-B^d_{\alpha}-B^e_{\alpha})B^a_{\alpha}+B^a_{\alpha}(\mathbb{I}-B^d_{\alpha}-B^e_{\alpha})(\mathbb{I}-B^a_{\alpha}-B^b_{\alpha}))\rho^{\frac{1}{2}}}_F\\
    &+\norm{(B^c_{\alpha}B^f_{\alpha}B^a_{\alpha}-(\mathbb{I}-B^a_{\alpha}-B^b_{\alpha})(\mathbb{I}-B^d_{\alpha}-B^e_{\alpha})B^a_{\alpha})\rho^{\frac{1}{2}}}_F\\
    &+\norm{(B^a_{\alpha}B^f_{\alpha}B^c_{\alpha}-B^a_{\alpha}(\mathbb{I}-B^d_{\alpha}-B^e_{\alpha})(\mathbb{I}-B^a_{\alpha}-B^b_{\alpha}))\rho^{\frac{1}{2}}}_F\\
    &+\delta_{ac\alpha\alpha}+2\delta_{cf\alpha}'.
\end{split}
\end{equation}
We can simplify the three Frobenius norm terms separately.
In the first term, we can expand
\begin{equation}
\begin{split}
    &[B^a_{\alpha},B^e_{\alpha}]-(\mathbb{I}-B^a_{\alpha}-B^b_{\alpha})(\mathbb{I}-B^d_{\alpha}-B^e_{\alpha})B^a_{\alpha}+B^a_{\alpha}(\mathbb{I}-B^d_{\alpha}-B^e_{\alpha})(\mathbb{I}-B^a_{\alpha}-B^b_{\alpha})\\
    =&B^b_{\alpha}B^a_{\alpha}-B^d_{\alpha}B^a_{\alpha}-B^b_{\alpha}B^e_{\alpha}B^a_{\alpha}-B^a_{\alpha}B^b_{\alpha}+B^a_{\alpha}B^d_{\alpha}+B^a_{\alpha}B^e_{\alpha}B^b_{\alpha}\\
    =&[B^b_{\alpha},B^a_{\alpha}]+[B^a_{\alpha},B^d_{\alpha}]+B^a_{\alpha}B^e_{\alpha}B^b_{\alpha}-B^b_{\alpha}B^e_{\alpha}B^a_{\alpha}.
\end{split}
\end{equation}
Therefore, by using the triangle inequality, we can bound the Frobenius norm by $\delta_{ab\alpha\alpha}+\delta_{ad\alpha\alpha}+\delta_{be\alpha}'+\delta_{ab\alpha\alpha}+\delta_{be\alpha}'$.
For the second Frobenius norm, we can introduce an intermediate state to simplify, and split the Frobenius norm into two.
The first component simplifies as
\begin{equation}
\begin{split}
    &\norm{(B^c_{\alpha}B^f_{\alpha}B^a_{\alpha}-(\mathbb{I}-B^a_{\alpha}-B^b_{\alpha})B^f_{\alpha}B^a_{\alpha})\rho^{\frac{1}{2}}}_F\\
    \leq&\norm{(B^a_{\alpha}+B^b_{\alpha}+B^c_{\alpha}+\mathbb{I})B^f_{\alpha}[B^a_{\alpha},\rho^{\frac{1}{2}}]}_F+\norm{(B^a_{\alpha}+B^b_{\alpha}+B^c_{\alpha}-\mathbb{I})B^f_{\alpha}\rho^{\frac{1}{2}}B^a_{\alpha}}_F\\
    \leq& 4\delta_{ab\alpha\alpha}+\norm{(B^a_{\alpha}+B^b_{\alpha}+B^c_{\alpha}+\mathbb{I})[B^f_{\alpha},\rho^{\frac{1}{2}}]}_F+\norm{(B^a_{\alpha}+B^b_{\alpha}+B^c_{\alpha}-\mathbb{I})\rho^{\frac{1}{2}}}_F\\
    \leq& 4\delta_{ab\alpha\alpha}+4\delta_{df\alpha\alpha}+2\sum_{u\neq v\in\{a,b,c\},\alpha}\delta_{uv\alpha}'+\frac{2}{3}\sum_{u\neq v\in\{a,b,c\},\alpha,\beta}\delta_{uv\alpha\beta}.
\end{split}
\end{equation}
The second component can be simplified as
\begin{equation}
\begin{split}
    &\norm{(\mathbb{I}-B^a_{\alpha}-B^b_{\alpha})(B^f_{\alpha}B^a_{\alpha}-(\mathbb{I}-B^d_{\alpha}-B^e_{\alpha})B^a_{\alpha})\rho^{\frac{1}{2}}}_F\\
    \leq&3\norm{(B^f_{\alpha}+B^d_{\alpha}+B^e_{\alpha}-\mathbb{I})B^a_{\alpha}\rho^{\frac{1}{2}}}_F\\
    \leq&3\norm{(B^f_{\alpha}+B^d_{\alpha}+B^e_{\alpha}-\mathbb{I})[B^a_{\alpha},\rho^{\frac{1}{2}}]}_F+3\norm{(B^f_{\alpha}+B^d_{\alpha}+B^e_{\alpha}-\mathbb{I})\rho^{\frac{1}{2}}B^a_{\alpha}}_F\\
    \leq&12\delta_{ad\alpha\alpha}+2\sum_{u\neq v\in\{d,e,f\},\alpha}\delta_{uv\alpha}'+\frac{2}{3}\sum_{u\neq v\in\{d,e,f\},\alpha,\beta}\delta_{uv\alpha\beta}.
\end{split}
\end{equation}
For the third Frobenius norm, we can similarly introduce an intermediate state and split the norm.
The first component simplifies as
\begin{equation}
\begin{split}
    &\norm{(B^a_{\alpha}B^f_{\alpha}B^c_{\alpha}-B^a_{\alpha}(\mathbb{I}-B^d_{\alpha}-B^e_{\alpha})B^c_{\alpha})\rho^{\frac{1}{2}}}_F\\
    \leq&\norm{(B^f_{\alpha}+B^d_{\alpha}+B^e_{\alpha}-\mathbb{I})[B^c_{\alpha},\rho^{\frac{1}{2}}]}_F+\norm{(B^f_{\alpha}+B^d_{\alpha}+B^e_{\alpha}-\mathbb{I})\rho^{\frac{1}{2}}B^c_{\alpha}}_F\\
    \leq&4\delta_{cf\alpha\alpha}+2\sum_{u\neq v\in\{d,e,f\},\alpha}\delta_{uv\alpha}'+\frac{2}{3}\sum_{u\neq v\in\{d,e,f\},\alpha,\beta}\delta_{uv\alpha\beta}.
\end{split}
\end{equation}
The second component simplifies as
\begin{equation}
\begin{split}
    &\norm{(B^a_{\alpha}(\mathbb{I}-B^d_{\alpha}-B^e_{\alpha})B^c_{\alpha}-B^a_{\alpha}(\mathbb{I}-B^d_{\alpha}-B^e_{\alpha})(\mathbb{I}-B^a_{\alpha}-B^b_{\alpha}))\rho^{\frac{1}{2}}}_F\\
    \leq&3\norm{(B^a_{\alpha}+B^b_{\alpha}+B^c_{\alpha}-\mathbb{I})\rho^{\frac{1}{2}}}_F\\
    \leq&2\sum_{u\neq v\in\{a,b,c\},\alpha}\delta_{uv\alpha}'+\frac{2}{3}\sum_{u\neq v\in\{a,b,c\},\alpha,\beta}\delta_{uv\alpha\beta}.
\end{split}
\end{equation}
Combining the results, and adding terms to provide symmetry, we can bound
\begin{equation}
    \norm{[B^a_{\alpha},B^e_{\alpha}]\rho^{\frac{1}{2}}}_F\leq2\sum_{(u,v)\in E_{cg},\alpha}\delta_{uv\alpha}'+12\sum_{(u,v)\in E_{cg},\alpha,\beta}\delta_{uv\alpha\beta}.
\end{equation}

The final part of the proof examines the commutator between operators for vertices in $G$ that corresponding to different colors.
The Frobenius norm can be split
\begin{equation}
\begin{split}
    &\norm{[B^a_{\alpha},B^e_{\beta}]\rho^{\frac{1}{2}}}_F\\
    \leq&\norm{([B^a_{\alpha},B^e_{\beta}]-B^e_{\gamma}B^d_{\gamma}B^a_{\alpha}+B^a_{\alpha}B^d_{\gamma}B^e_{\gamma})\rho^{\frac{1}{2}}}_F+\norm{B^e_{\gamma}B^d_{\gamma}B^a_{\alpha}\rho^{\frac{1}{2}}}_F+\norm{B^a_{\alpha}B^d_{\gamma}B^e_{\gamma}\rho^{\frac{1}{2}}}_F\\
    \leq&\norm{([B^a_{\alpha},B^e_{\beta}]-(\mathbb{I}-B^e_{\alpha}-B^e_{\beta})(\mathbb{I}-B^d_{\alpha}-B^d_{\beta})B^a_{\alpha}+B^a_{\alpha}(\mathbb{I}-B^d_{\alpha}-B^d_{\beta})(\mathbb{I}-B^e_{\alpha}-B^e_{\beta}))\rho^{\frac{1}{2}}}_F\\
    &+\norm{B^e_{\gamma}B^d_{\gamma}[B^a_{\alpha},\rho^{\frac{1}{2}}]}_F+\norm{B^e_{\gamma}B^d_{\gamma}\rho^{\frac{1}{2}}B^a_{\alpha}}_F+\norm{B^a_{\alpha}B^d_{\gamma}B^e_{\gamma}\rho^{\frac{1}{2}}}_F\\
    \leq&\norm{[B^e_{\alpha},B^a_{\alpha}]\rho^{\frac{1}{2}}}_F+\norm{[B^d_{\alpha},B^a_{\alpha}]\rho^{\frac{1}{2}}}_F+\norm{[B^d_{\beta},B^a_{\alpha}]\rho^{\frac{1}{2}}}_F+\norm{(B^a_{\alpha}B^d_{\beta}B^e_{\alpha}-B^e_{\alpha}B^d_{\beta}B^a_{\alpha})\rho^{\frac{1}{2}}}_F\\
    &+\norm{B^e_{\alpha}B^d_{\alpha}B^a_{\alpha}\rho^{\frac{1}{2}}}_F+\norm{B^e_{\beta}B^d_{\alpha}B^a_{\alpha}\rho^{\frac{1}{2}}}_F+\norm{B^e_{\beta}B^d_{\beta}B^a_{\alpha}\rho^{\frac{1}{2}}}_F+\norm{B^a_{\alpha}B^d_{\alpha}B^e_{\alpha}\rho^{\frac{1}{2}}}_F\\
    &+\norm{B^a_{\alpha}B^d_{\alpha}B^e_{\beta}\rho^{\frac{1}{2}}}_F+\norm{B^a_{\beta}B^d_{\beta}B^e_{\alpha}\rho^{\frac{1}{2}}}_F+\delta_{ab\alpha\alpha}+2\delta_{de\gamma}'\\
    \leq&2\sum_{(u,v)\in E_{cg},\alpha}\delta_{uv\alpha}'+12\sum_{(u,v)\in E_{cg},\alpha,\beta}\delta_{uv\alpha\beta}+\delta_{ad\alpha\alpha}+\delta_{ad\alpha\beta}+\delta_{ae\alpha\alpha}\\
    &+2\delta_{ad\alpha\alpha}+\delta_{de\beta\alpha}+\delta_{ad\alpha\beta}+2\delta_{ad\alpha}'+\delta_{ae\alpha\alpha}+\delta_{de\beta}'+\delta_{de\alpha}'+\delta_{ae\beta\beta}\\
    &+\delta_{ad\alpha}'+\delta_{ae\alpha\alpha}+\delta_{ad\beta}'+\delta_{ab\alpha\alpha}+2\delta_{de\gamma}'\\
    \leq&4\sum_{(u,v)\in E_{cg},\alpha}\delta_{uv\alpha}'+15\sum_{(u,v)\in E_{cg},\alpha,\beta}\delta_{uv\alpha\beta}.
\end{split}
\end{equation}
Therefore, we can express
\begin{equation}
\begin{split}
    &\norm{[B^a_{\alpha},B^e_{\beta}]\rho^{\frac{1}{2}}}_F\\
    \leq&19\sum_{(u,v)\in E_{cg},\alpha,\beta}\left[\sqrt{2\varepsilon_{uv\alpha u}(2-\varepsilon_{uv\alpha u})}+\sqrt{2\varepsilon_{uv\beta v}(2-\varepsilon_{uv\beta v})}\right]\\
    &+4\sum_{(u,v)\in E_{cg},\alpha}\sqrt{\frac{1}{2}(\varepsilon_{uv\alpha u}+\varepsilon_{uv\alpha v})}\\
    =&\sum_{(u,v)\in E_{cg},\alpha}\left[4\sqrt{\frac{1}{2}(\varepsilon_{uv\alpha u}+\varepsilon_{uv\alpha v})}+19\sum_{u'\in\{u,v\}}\sqrt{2\varepsilon_{uv\alpha u'}(2-\varepsilon_{uv\alpha u'})}\right]
\end{split}
\end{equation}
\end{proof}
% \begin{proof}[Sketch Proof]
% The proof is split into three parts corresponding to Lemma 4 and the two parts of Lemma 5 of Ref.~\cite{Ji_2013}, showing (1) three vertices connected in a triangle are of different colors, (2) operators of the commutativity gadget corresponding to the same colors commute, and (3) operators of the commutativity gadget corresponding to different colors commute.\\

% Part 1 shows $\norm{(B^i_{\alpha}+B^j_{\alpha}+B^k_{\alpha}-\mathbb{I})\rho^{\frac{1}{2}}}_F$ is bounded by expanding the operators into a set of $B^iB^jB^k$ operators using the property that Bob's operators are vertex-complete.
% The expanded operators in the Frobenius norm can then be reduced using the tracial, commutation and edge coloring conditions.\\

% Part 2 and 3 shows $\norm{[B^a_{\alpha},B^e_{\alpha}]\rho^{\frac{1}{2}}}_F$ and $\norm{[B^a_{\alpha},B^e_{\beta}]\rho^{\frac{1}{2}}}_F$ are bounded by performing similar expansions as in Ref.~\cite{Ji_2013}, where $(a,e)$ correspond to vertices on the commutative gadget that are part of graph $G$.
% A similar reduction of the expanded operators can be performed using the quantum assignment properties.
% \end{proof}

We summarize the results in the central theorem, showing that a vertex-complete color-commuting strategy winning the edge verification challenge for $G'$ with high probability implies prover B's operators form an almost commuting representation for $G$.
\begin{theorem}
\label{thm:BCS-3-COL_to_All-Commuting-Representation}
    Let $1-\varepsilon_{ij\alpha k}$ be the probability of the prover passing the edge verification test for challenge $(i,j,\alpha)$ to prover A and $(k,\alpha)$ to prover B.
    If there exists a vertex-complete color-commuting strategy that can win the edge verification challenge for BCS-3-COL %with a distribution of $\mathcal{D}_{ijk\alpha}=\frac{\delta_{ik}+\delta_{jk}}{6\abs{E_{G'}}}$ 
    with $p_{win}^{EV}\geq 1-\varepsilon$ for graph $G'$, then prover B's measurement operators $B^i_{\alpha}=\tilde{B}^{i\alpha}_1$ satisfy the following conditions:
    \begin{enumerate}
        \item Tracial property: For any edge $(i,j)\in E_{G'}$, $\norm{[B^i_{\alpha},\rho^{\frac{1}{2}}]}_F\leq\sqrt{2\varepsilon_{ij\alpha i}(2-\varepsilon_{ij\alpha i})}+\sqrt{2\varepsilon_{ij\alpha j}(2-\varepsilon_{ij\alpha j})}$
        \item Almost commuting in $G'$: For any edge $(i,j)\in E_{G'}$, $\norm{[B^i_{\alpha},B^j_{\beta}]\rho^{\frac{1}{2}}}_F\leq\sqrt{2\varepsilon_{ij\alpha i}(2-\varepsilon_{ij\alpha i})}+\sqrt{2\varepsilon_{ij\beta j}(2-\varepsilon_{ij\beta j})}$
        \item Almost fully commuting in $G$: For any non-neighboring vertex pairs in $G$, $i,j\in V_G$, $(i,j)\notin E_G$, 
        \begin{equation*}
            \norm{[B^i_{\alpha},B^j_{\beta}]\rho^{\frac{1}{2}}}_F\leq \sum_{(u,v)\in E_{cg}^{ij},\alpha}\left[4\sqrt{\frac{1}{2}(\varepsilon_{uv\alpha u}+\varepsilon_{uv\alpha v})}+19\sum_{u'\in\{u,v\}}\sqrt{2\varepsilon_{uv\alpha u'}(2-\varepsilon_{uv\alpha u'})}\right]
        \end{equation*}
        \item Edge coloring: For any edge $(i,j)\in E_{G'}$, $\norm{B^i_{\alpha}B^j_{\alpha}\rho^{\frac{1}{2}}}_F\leq\sqrt{2\varepsilon_{ij\alpha i}(2-\varepsilon_{ij\alpha i})}+\sqrt{2\varepsilon_{ij\alpha j}(2-\varepsilon_{ij\alpha j})}+\sqrt{\frac{1}{2}(\varepsilon_{ij\alpha i}+\varepsilon_{ij\alpha j})}$
    \end{enumerate}
    and $\frac{1}{6\abs{E_{G'}}}\sum_{ij\alpha k}\varepsilon_{ij\alpha k}\leq\varepsilon$. 
\end{theorem}

% We can also provide the representation of a BCS algebra that can be formed by an almost-perfect strategy.
% \begin{theorem}[Proposition 5.7 of Ref.~\cite{Paddock2024}]
% \label{thm:BCS_to_Rep}
%     Let $\rho$ be the reduced density matrix of the state $\ket{\psi}$. If $\ket{\psi}$ is the state and $\{X_{u,\alpha}\}_{u,\alpha}$ is Bob's measurement in a $\varepsilon$-perfect strategy of the BCS-3-COL game, then $\{X_{u,\alpha}\}_{u,\alpha}$ is a $(O(\varepsilon^{\frac{1}{2}}),\rho)$-representation of the BCS algebra that is also $O(\varepsilon^{\frac{1}{2}})$-tracial, i.e.
%     \begin{gather*}
%         \norm{[X_{u,\alpha},X_{v,\beta}]\rho^{\frac{1}{2}}}_F\leq \delta\\
%         \norm{(\mathbb{I}+X_{u,\alpha})(\mathbb{I}+X_{v,\alpha})\rho^{\frac{1}{2}}}_F\leq2\delta \\
%         \norm{[X_{u,\alpha},\rho^{\frac{1}{2}}]}_F\leq\delta
%     \end{gather*}
%     where $\delta=O(\varepsilon^{\frac{1}{2}})$.
% \end{theorem}
% We note that the Frobenius norm values can be converted to the projectors via $X_{u,\alpha}=P^u_{\alpha}-(\mathbb{I}-P^u_{\alpha})$, which gives rise to
% \begin{gather*}
%     \norm{[P^u_{\alpha},P^v_{\beta}]\rho^{\frac{1}{2}}}_F\leq \frac{\delta}{4}\\
%     \norm{P^u_{\alpha}P^v_{\alpha}\rho^{\frac{1}{2}}}_F\leq\frac{\delta}{2} \\
%     \norm{[P^u_{\alpha},\rho^{\frac{1}{2}}]}_F\leq\frac{\delta}{2},
% \end{gather*}
% which we can use to create a classical strategy for the vertex-3-COL game.

\subsection{Classical Strategy from Almost Fully Commuting Satisfying Assignment}
\label{app:FC-ASA_to_Classical}

Consider the following classical strategy formed using a series of consecutive measurements:
\begin{enumerate}
    \item Prover B prepares the quantum state $\rho=\Tr_A[\dyad{\psi}_{AB}]$.
    \item Prover B performs sequential measurement with projectors $\{B^i_{\alpha_i}\}_{\alpha_i=0,1,2}$, starting with $i=1$ to $i=\abs{V}$.
    \item Prover B determine $c_i=\alpha_i$ for each $i\in\{1,\cdots,\abs{V}\}$, and inform prover A of the colors $c_i$.
    \item When challenged with any vertex, the provers would respond with the colors respectively.
\end{enumerate}
We show that this strategy can win the vertex-3-COL game with a high probability.%, with the complete proof provided in Appendix~\ref{app:BCS-3-COL_to_Vertex-3-COL}.
\begin{theorem}
\label{thm:BCS-3-COL_to_Vertex-3-COL}
    If there exists a vertex-complete color-commuting strategy that can win the edge verification challenge of BCS-3-COL %with a distribution of $\mathcal{D}_{ijk\alpha}=\frac{\delta_{ik}+\delta_{jk}}{6\abs{E_{G'}}}$ 
    with $p_{win}^{EV}\geq 1-\varepsilon$ for graph $G'$, then there exists a classical strategy that can win the vertex-3-COL game with probability $p_{win}^{WD}=1$ for the well-definition tests and
    \begin{equation*}
    \begin{split}
        p_{win}^{EV}\geq&1-\frac{6\abs{E_{G'}}}{\min_{i\in V_G}\abs{\neigh(i)_{G'}}}(2\varepsilon+4(1+\sqrt{2})\sqrt{\varepsilon})\\
        &-\frac{216(19+\sqrt{2})\abs{E_{G'}}}{\abs{E_G}}\max_{i\in V_G}\abs{\neigh(i)_G}\sqrt{\varepsilon}-\frac{(9+4\sqrt{2})\abs{E_{G'}}}{2\abs{E_G}}(\varepsilon+\sqrt{\varepsilon})
    \end{split}
    \end{equation*}
    for the edge verification tests.
\end{theorem}
\begin{proof}%[Proof of Thm.~\ref{thm:BCS-3-COL_to_Vertex-3-COL}]
We note that if there exists an vertex-complete color-commuting strategy that can win the edge verification challenge of BCS-3-COL with a distribution of $\mathcal{D}_{ijk\alpha}=\frac{\delta_{ik}+\delta_{jk}}{6\abs{E_{G'}}}$ with $p_{win}^{EV}\geq 1-\varepsilon$ for graph $G'$, then by Thm.~\ref{thm:BCS-3-COL_to_All-Commuting-Representation}, prover B's measurement operator satisfies the set of conditions for any edge $(i,j)\in E_{G'}$,
\begin{gather*}
    \norm{[B^i_{\alpha},\rho^{\frac{1}{2}}]}_F\leq\sqrt{2\varepsilon_{ij\alpha i}(2-\varepsilon_{ij\alpha i})}+\sqrt{2\varepsilon_{ij\alpha j}(2-\varepsilon_{ij\alpha j})}\\
    \norm{[B^i_{\alpha},B^j_{\beta}]\rho^{\frac{1}{2}}}_F\leq\sqrt{2\varepsilon_{ij\alpha i}(2-\varepsilon_{ij\alpha i})}+\sqrt{2\varepsilon_{ij\beta j}(2-\varepsilon_{ij\beta j})}\\
    \norm{B^i_{\alpha}B^j_{\alpha}\rho^{\frac{1}{2}}}_F\leq\sqrt{2\varepsilon_{ij\alpha i}(2-\varepsilon_{ij\alpha i})}+\sqrt{2\varepsilon_{ij\alpha j}(2-\varepsilon_{ij\alpha j})}+\sqrt{\frac{1}{2}(\varepsilon_{ij\alpha i}+\varepsilon_{ij\alpha j})}
\end{gather*}
and for any pair of vertex $i,j\in V_G$ but do not form an edge $(i,j)\notin E_G$,
\begin{equation*}
    \norm{[B^i_{\alpha},B^j_{\beta}]\rho^{\frac{1}{2}}}_F\leq \sum_{(u,v)\in E_{cg}^{ij},\alpha}\left[4\sqrt{\frac{1}{2}(\varepsilon_{uv\alpha u}+\varepsilon_{uv\alpha v})}+19\sum_{u'\in\{u,v\}}\sqrt{2\varepsilon_{uv\alpha u'}(2-\varepsilon_{uv\alpha u'})}\right],
\end{equation*}
where $\varepsilon_{ij\alpha j}$ is the probability of winning the edge verification test when challenge $(i,j,\alpha)$ and $(i,\alpha)$ are issued in the BCS game, with $\frac{1}{6\abs{E_{G'}}}\sum_{ij\alpha k}\varepsilon_{ij\alpha k}\leq\varepsilon$. 
For simplicity, let us label the bounds for the conditions by $\delta_{i\alpha}^{tr}$, $\delta_{ij\alpha\beta}^{com}$, $\delta_{ij\alpha}^{ec}$, and $\delta_{ij\alpha}^{cg}$ respectively, and taking $\tilde{\delta}_{ij\alpha\beta}^{com}$ to be either $\delta_{ij\alpha\beta}^{com}$ or $\delta_{ij\alpha}^{cg}$ depending on whether $(i,j)\in E_G$.\footnote{Respectively, the abbreviation represent tracial, commutation, edge coloring, and commutative gadget.}
The probability of obtaining any combination of colors for the graph is given by
\begin{equation}
    P(\alpha_1,\cdots,\alpha_m)=\Tr[B^m_{\alpha_m}\cdots B^1_{\alpha_1}\rho B^1_{\alpha_1}\cdots B^m_{\alpha_m}],
\end{equation}
where we set $m=\abs{V_G}$.\\

Let us consider the vertex-3-COL game.
By construction, the classical strategy described always wins the well-definition test (when $i=j$) since both provers will always give the same response.
As such, we can focus on the edge verification test, where a uniform choice of $(i,j)\in E_G$ would be asked to both provers.
The failure probability for any given edge $(i,j)\in E_G$ can be computed by
\begin{equation}
    p_{fail}^{i,j}=\sum_{\substack{\alpha_1,\cdots,\alpha_m\\ \alpha_i=\alpha_j}}\Tr[B^m_{\alpha_m}\cdots B^1_{\alpha_1}\rho B^1_{\alpha_1}\cdots B^m_{\alpha_m}].
\end{equation}
WLOG, let us consider a $(i,j) \in E$ with $1<i<j<m$.
The goal would be to reduce the failure probability to match the Frobenius norm corresponding to the edge verification constraint, i.e. removing all projectors except $B^i_{c_i}$ and $B^j_{c_j}$.
We begin by removing projectors for vertices with index $u>j$,
\begin{equation}
\begin{split}
    p_{fail}^{i,j}=&\sum_{\substack{\alpha_1,\cdots,\alpha_{m-1}\\ \alpha_i=\alpha_j}}\Tr[(\sum_{\alpha_m}B^m_{\alpha_m})B^{m-1}_{\alpha_{m-1}}\cdots B^j_{\alpha_j}\cdots B^1_{\alpha_1}\rho B^1_{\alpha_1}\cdots B^j_{\alpha_j}\cdots B^{m-1}_{\alpha_{m-1}}]\\
    =&\sum_{\substack{\alpha_1,\cdots,\alpha_{m-1}\\ \alpha_i=\alpha_j}}\Tr[B^{m-1}_{\alpha_{m-1}}\cdots B^j_{\alpha_j}\cdots B^1_{\alpha_1}\rho B^1_{\alpha_1}\cdots B^j_{\alpha_j}\cdots B^{m-1}_{\alpha_{m-1}}]\\
    =&\cdots\\
    =&\sum_{\substack{\alpha_1,\cdots,\alpha_j\\ \alpha_i=\alpha_j}}\Tr[B^j_{\alpha_j}\cdots B^1_{\alpha_1}\rho B^1_{\alpha_1}\cdots B^j_{\alpha_j}],
\end{split}
\end{equation}
where the first line utilizes the cyclic property of trace and that $B^m_{\alpha_m}$ is projective, the second line utilizes the fact that $\sum_{\alpha_m}B^m_{\alpha_m}=\mathbb{I}$, and with repeated simplification, we arrive at the final form.
Our next target is to remove projectors corresponding to indices $u<i$.
We can use the commutator to expand,
\begin{equation}
\begin{split}
    p_{fail}^{i,j}=&\sum_{\substack{\alpha_1,\cdots,\alpha_j\\ \alpha_i=\alpha_j}}\Tr[B^j_{\alpha_j}\cdots B^2_{\alpha_2}B^1_{\alpha_1}\rho B^1_{\alpha_1}B^2_{\alpha_2}\cdots B^j_{\alpha_j}]\\
    =&\sum_{\substack{\alpha_1,\cdots,\alpha_j\\ \alpha_i=\alpha_j}}\Tr[B^j_{\alpha_j}\cdots B^2_{\alpha_2}([B^1_{\alpha_1},\rho^{\frac{1}{2}}]+\rho^{\frac{1}{2}}B^1_{\alpha_1})([\rho^{\frac{1}{2}},B^1_{\alpha_1}]+B^1_{\alpha_1}\rho^{\frac{1}{2}})B^2_{\alpha_2}\cdots B^j_{\alpha_j}]\\
    =&\sum_{\substack{\alpha_1,\alpha_i,\alpha_j\\ \alpha_i=\alpha_j}}\Tr[\sum_{\alpha_2,\cdots,\alpha_{j-1}\setminus \alpha_i}P^2_{\alpha_2}\cdots B^j_{\alpha_j}\cdots B^2_{\alpha_2}[P^1_{\alpha_1},\rho^{\frac{1}{2}}][\rho^{\frac{1}{2}},B^1_{\alpha_1}]]\\
    &+\sum_{\substack{\alpha_1,\cdots,\alpha_j\\ \alpha_i=\alpha_j}}\Tr[B^j_{\alpha_j}\cdots B^2_{\alpha_2}([B^1_{\alpha_1},\rho^{\frac{1}{2}}]B^1_{\alpha_1}\rho^{\frac{1}{2}}+\rho^{\frac{1}{2}}B^1_{\alpha_1}[\rho^{\frac{1}{2}},B^1_{\alpha_1}])B^2_{\alpha_2}\cdots B^j_{\alpha_j}]\\
    &+\sum_{\substack{\alpha_1,\cdots,\alpha_j\\ \alpha_i=\alpha_j}}\Tr[B^j_{\alpha_j}\cdots B^2_{\alpha_2}\rho^{\frac{1}{2}}B^1_{\alpha_1} \rho^{\frac{1}{2}}B^2_{\alpha_2}\cdots B^j_{\alpha_j}].
\end{split} 
\end{equation}
We examine each of the terms separately.
The first term can be simplified as
\begin{equation}
\begin{split}
    &\sum_{\substack{\alpha_1,\alpha_i,\alpha_j\\ \alpha_i=\alpha_j}}\Tr[\sum_{\alpha_2,\cdots,\alpha_{j-1}\setminus \alpha_i}B^2_{\alpha_2}\cdots B^j_{\alpha_j}\cdots B^2_{\alpha_2}[B^1_{\alpha_1},\rho^{\frac{1}{2}}][\rho^{\frac{1}{2}},B^1_{\alpha_1}]]\\
    \leq&\sum_{\substack{\alpha_1,\alpha_i,\alpha_j\\ \alpha_i=\alpha_j}}\norm{\sum_{\alpha_2,\cdots,\alpha_j\setminus \alpha_i}B^2_{\alpha_2}\cdots B^i_{\alpha_j}\cdots B^2_{\alpha_2}}_{op}\norm{[B^1_{\alpha_1},\rho^{\frac{1}{2}}]}_F^2\\
    \leq& 3\sum_{\alpha}(\delta_{1\alpha}^{tr})^2
\end{split}
\end{equation}
where the operator norm property is utilized, noting the fact that $[B_{\alpha_1}^1,\rho^{\frac{1}{2}}]=([\rho^{\frac{1}{2}},B_{\alpha_1}^1])^{\dagger}$ and thus $[B_{\alpha_1}^1,\rho^{\frac{1}{2}}][\rho^{\frac{1}{2}},B_{\alpha_1}^1]\geq 0$ in the first line.
In the final line, we use the tracial property and Thm.~\ref{thm:Chain_Pinching_Projector}.
The second term can be simplified by noting the chain of operators can be made into complex conjugate of each other,
\begin{equation}
\begin{split}
    &\sum_{\substack{\alpha_1,\cdots,\alpha_j\\ \alpha_i=\alpha_j}}\Tr[B^j_{\alpha_j}\cdots B^2_{\alpha_2}([B^1_{\alpha_1},\rho^{\frac{1}{2}}]B^1_{\alpha_1}\rho^{\frac{1}{2}}+\rho^{\frac{1}{2}}B^1_{\alpha_1}[\rho^{\frac{1}{2}},B^1_{\alpha_1}])B^2_{\alpha_2}\cdots B^j_{\alpha_j}]\\
    =&\sum_{\substack{\alpha_1,\alpha_i,\alpha_j\\ \alpha_i=\alpha_j}}\Tr[\left(\sum_{\alpha_2,\cdots,\alpha_j\setminus \alpha_i}B^2_{\alpha_2}\cdots B^j_{\alpha_j}\cdots B^2_{\alpha_2}\right)[B^1_{\alpha_1},\rho^{\frac{1}{2}}]B^1_{\alpha_1}\rho^{\frac{1}{2}}]\\
    &+\sum_{\substack{\alpha_1,\alpha_i,\alpha_j\\ \alpha_i=\alpha_j}}\Tr[B^1_{\alpha_1}[\rho^{\frac{1}{2}},B^1_{\alpha_1}]\left(\sum_{\alpha_2,\cdots,\alpha_j\setminus \alpha_i}B^2_{\alpha_2}\cdots B^j_{\alpha_j}\cdots B^2_{\alpha_2}\right)\rho^{\frac{1}{2}}]\\
    =&\sum_{\substack{\alpha_1,\alpha_i,\alpha_j\\ \alpha_i=\alpha_j}}\Tr[(X+X^{\dagger})\rho^{\frac{1}{2}}],\quad X=\left(\sum_{\alpha_2,\cdots,\alpha_j\setminus \alpha_i}B^2_{\alpha_2}\cdots B^j_{\alpha_j}\cdots B^2_{\alpha_2}\right)[B^1_{\alpha_1},\rho^{\frac{1}{2}}]B^1_{\alpha_1}\\
    \leq&\sum_{\substack{\alpha_1,\alpha_i,\alpha_j\\ \alpha_i=\alpha_j}}\norm{X+X^{\dagger}}_F\\
    \leq&2\sum_{\substack{\alpha_1,\alpha_i,\alpha_j\\ \alpha_i=\alpha_j}}\norm{\left(\sum_{\alpha_2,\cdots,\alpha_j\setminus \alpha_i}B^2_{\alpha_2}\cdots B^j_{\alpha_j}\cdots B^2_{\alpha_2}\right)[B^1_{\alpha_1},\rho^{\frac{1}{2}}]B^1_{\alpha_1}}_F\\
    \leq&2\sum_{\substack{\alpha_1,\alpha_i,\alpha_j\\ \alpha_i=\alpha_j}}\norm{\left(\sum_{\alpha_2,\cdots,\alpha_j\setminus \alpha_i}B^2_{\alpha_2}\cdots B^j_{\alpha_j}\cdots B^2_{\alpha_2}\right)}_{op}\norm{[B^1_{\alpha_1},\rho^{\frac{1}{2}}]B^1_{\alpha_1}}_F\\
    \leq&2\sum_{\substack{\alpha_1,\alpha_i,\alpha_j\\ \alpha_i=\alpha_j}}\norm{B^1_{\alpha_1}[B^1_{\alpha_1},\rho^{\frac{1}{2}}]}_F\\
    \leq&6\sum_{\alpha_1}\norm{B^1_{\alpha_1}}_{op}\norm{[B^1_{\alpha_1},\rho^{\frac{1}{2}}]}_F\\
    \leq&6\sum_{\alpha}\delta_{1\alpha}^{tr}
\end{split}
\end{equation}
where the first line uses the cyclic property of trace, the third line utilizes Thm.~\ref{thm:Normal_Op_Frobenius_Norm} since $(X+X^{\dagger})$ is normal. The fourth line utilizes the triangle inequality, while the fifth line uses the property of the Frobenius norm.
The sixth line notes the Frobenius norm is invariant to complex conjugation and the final line uses the tracial property.
The third term corresponds to one where $B_{\alpha_1}^1$ is removed,
\begin{equation}
    \sum_{\substack{\alpha_1,\cdots,\alpha_j\\ \alpha_i=\alpha_j}}\Tr[B^j_{\alpha_j}\cdots B^2_{\alpha_2}\rho^{\frac{1}{2}}B^1_{\alpha_1} \rho^{\frac{1}{2}}B^2_{\alpha_2}\cdots B^j_{\alpha_j}]=\sum_{\substack{\alpha_2,\cdots,\alpha_j\\ \alpha_i=\alpha_j}}\Tr[B^j_{\alpha_j}\cdots B^2_{\alpha_2} \rho B^2_{\alpha_2}\cdots B^j_{\alpha_j}].
\end{equation}
Therefore, we are able to remove the first operator $B^1_{\alpha_1}$ at a penalty of $3\sum_{\alpha}[(\delta_{1\alpha}^{tr})^2+2\delta_{1\alpha}^{tr}]$.
By repeating this removal process up to operator with index $u=i-1$, we have that
\begin{equation}
    p_{fail}^{i,j}=\sum_{\substack{\alpha_i,\cdots,\alpha_j\\ \alpha_i=\alpha_j}}\Tr[B^j_{\alpha_j}\cdots B^i_{\alpha_i} \rho B^i_{\alpha_i}\cdots B^j_{\alpha_j}]+3\sum_{u\in[1,i-1],\alpha}[(\delta_{u\alpha}^{tr})^2+2\delta_{u\alpha}^{tr}],
\end{equation}
simplifying the failing probability.\\

The final set of operators to remove are those between $B^i_{\alpha_i}$ and $B^j_{\alpha_j}$.
This is performed by first swapping the positions of $B^i_{\alpha_i}$ and $B^{i+1}_{\alpha_{i+1}}$, before utilizing the same analysis above to remove $B^{i+1}_{\alpha_{i+1}}$.
We again use the commutator to expand,
\begin{equation}
\begin{split}
    &\sum_{\substack{\alpha_i,\cdots,\alpha_j\\ \alpha_i=\alpha_j}}\Tr[B^j_{\alpha_j}\cdots B^i_{\alpha_i} \rho B^i_{\alpha_i}\cdots B^j_{\alpha_j}]\\
    =&\sum_{\substack{\alpha_i,\cdots,\alpha_j\\ \alpha_i=\alpha_j}}\Tr[B^j_{\alpha_j}\cdots B^{i+2}_{\alpha_{i+2}}([B^{i+1}_{\alpha_{i+1}},B^i_{\alpha_i}]+B^i_{\alpha_i}B^{i+1}_{\alpha_{i+1}}) \rho ([B^i_{\alpha_i},B^{i+1}_{\alpha_{i+1}}]+B^{i+1}_{\alpha_{i+1}}B^i_{\alpha_i})B^{i+2}_{\alpha_{i+2}}\cdots B^j_{\alpha_j}]\\
    =&\sum_{\substack{\alpha_i,\alpha_j,\alpha_{i+1}\\ \alpha_i=\alpha_j}}\Tr[\left(\sum_{\alpha_{i+2},\cdots,\alpha_{j-1}}B^{i+2}_{\alpha_{i+2}}\cdots B^j_{\alpha_j}\cdots B^{i+2}_{\alpha_{i+2}}\right)[B^{i+1}_{\alpha_{i+1}},B^i_{\alpha_i}] \rho [B^i_{\alpha_i},B^{i+1}_{\alpha_{i+1}}]]\\
    +&\sum_{\substack{\alpha_i,\cdots,\alpha_j\\ \alpha_i=\alpha_j}}\Tr[B^j_{\alpha_j}\cdots B^{i+2}_{\alpha_{i+2}}([B^{i+1}_{\alpha_{i+1}},B^i_{\alpha_i}]\rho B^{i+1}_{\alpha_{i+1}}B^i_{\alpha_i}+B^i_{\alpha_i}B^{i+1}_{\alpha_{i+1}} \rho [B^i_{\alpha_i},B^{i+1}_{\alpha_{i+1}}])B^{i+2}_{\alpha_{i+2}}\cdots B^j_{\alpha_j}]\\
    +&\sum_{\substack{\alpha_i,\cdots,\alpha_j\\ \alpha_i=\alpha_j}}\Tr[B^j_{\alpha_j}\cdots B^{i+2}_{\alpha_{i+2}}B^i_{\alpha_i}B^{i+1}_{\alpha_{i+1}} \rho B^{i+1}_{\alpha_{i+1}}B^i_{\alpha_i}B^{i+2}_{\alpha_{i+2}}\cdots B^j_{\alpha_j}],
\end{split}
\end{equation}
with three separate terms to simplify.
The first term can be simplified since $[B^{i+1}_{\alpha_{i+1}},B^i_{\alpha_i}] \rho [B^i_{\alpha_i},B^{i+1}_{\alpha_{i+1}}]\geq 0$,
\begin{equation}
\begin{split}
    &\sum_{\substack{\alpha_i,\alpha_j,\alpha_{i+1}\\ \alpha_i=\alpha_j}}\Tr[\left(\sum_{\alpha_{i+2},\cdots,\alpha_{j-1}}B^{i+2}_{\alpha_{i+2}}\cdots B^j_{\alpha_j}\cdots B^{i+2}_{\alpha_{i+2}}\right)[B^{i+1}_{\alpha_{i+1}},B^i_{\alpha_i}] \rho [B^i_{\alpha_i},B^{i+1}_{\alpha_{i+1}}]]\\
    \leq&\sum_{\substack{\alpha_i,\alpha_j,\alpha_{i+1}\\ \alpha_i=\alpha_j}}\norm{\sum_{\alpha_{i+2},\cdots,\alpha_{j-1}}B^{i+2}_{\alpha_{i+2}}\cdots B^j_{\alpha_j}\cdots B^{i+2}_{\alpha_{i+2}}}_{op}\norm{[B^{i+1}_{\alpha_{i+1}},B^i_{\alpha_i}] \rho^{\frac{1}{2}}}_F\\
    \leq&\sum_{\alpha\beta}\tilde{\delta}^{com}_{i+1,i\alpha\beta},
\end{split}
\end{equation}
where the last line utilize the representation property from the edge verification constraint.
The second term can be simplified as well,
\begin{equation}
\begin{split}
    &\sum_{\substack{\alpha_i,\cdots,\alpha_j\\ \alpha_i=\alpha_j}}\Tr[B^j_{\alpha_j}\cdots B^{i+2}_{\alpha_{i+2}}([B^{i+1}_{\alpha_{i+1}},B^i_{\alpha_i}]\rho B^{i+1}_{\alpha_{i+1}}B^i_{\alpha_i}+B^i_{\alpha_i}B^{i+1}_{\alpha_{i+1}} \rho [B^i_{\alpha_i},B^{i+1}_{\alpha_{i+1}}])B^{i+2}_{\alpha_{i+2}}\cdots B^j_{\alpha_j}]\\
    =&\sum_{\substack{\alpha_i,\alpha_j,\alpha_{i+1}\\ \alpha_i=\alpha_j}}\Tr[B^{i+1}_{\alpha_{i+1}}B^i_{\alpha_i}(\sum_{\alpha_{i+2},\cdots,\alpha_{j-1}}B^{i+2}_{\alpha_{i+2}}\cdots B^j_{\alpha_j}\cdots B^{i+2}_{\alpha_{i+2}})[B^{i+1}_{\alpha_{i+1}},B^i_{\alpha_i}]\rho^{\frac{1}{2}} \rho^{\frac{1}{2}}]\\
    &+\sum_{\substack{\alpha_i,\alpha_j,\alpha_{i+1}\\ \alpha_i=\alpha_j}}\Tr[\rho^{\frac{1}{2}}[B^i_{\alpha_i},B^{i+1}_{\alpha_{i+1}}](\sum_{\alpha_{i+2},\cdots,\alpha_{j-1}}B^{i+2}_{\alpha_{i+2}}\cdots B^j_{\alpha_j}\cdots B^{i+2}_{\alpha_{i+2}})B^i_{\alpha_i}B^{i+1}_{\alpha_{i+1}} \rho^{\frac{1}{2}}]\\
    =&\sum_{\substack{\alpha_i,\alpha_j,\alpha_{i+1}\\ \alpha_i=\alpha_j}}\Tr[(Y+Y^{\dagger}) \rho^{\frac{1}{2}}]\\
    \leq&\sum_{\substack{\alpha_i,\alpha_j,\alpha_{i+1}\\ \alpha_i=\alpha_j}}\norm{Y+Y^{\dagger}}_F\\
    \leq&2\sum_{\substack{\alpha_i,\alpha_j,\alpha_{i+1}\\ \alpha_i=\alpha_j}}\norm{B^{i+1}_{\alpha_{i+1}}B^i_{\alpha_i}(\sum_{\alpha_{i+2},\cdots,\alpha_{j-1}}B^{i+2}_{\alpha_{i+2}}\cdots B^j_{\alpha_j}\cdots B^{i+2}_{\alpha_{i+2}})[B^{i+1}_{\alpha_{i+1}},B^i_{\alpha_i}]\rho^{\frac{1}{2}}}_F\\
    \leq&2\sum_{\substack{\alpha_i,\alpha_j,\alpha_{i+1}\\ \alpha_i=\alpha_j}}\norm{B^{i+1}_{\alpha_{i+1}}}_{op}\norm{B^i_{\alpha_i}}_{op}\norm{\sum_{\alpha_{i+2},\cdots,\alpha_{j-1}}B^{i+2}_{\alpha_{i+2}}\cdots B^j_{\alpha_j}\cdots B^{i+2}_{\alpha_{i+2}}}_{op}\norm{[B^{i+1}_{\alpha_{i+1}},B^i_{\alpha_i}]\rho^{\frac{1}{2}}}_F\\
    \leq&2\sum_{\alpha\beta}\tilde{\delta}^{com}_{i+1,i\alpha\beta}
\end{split}
\end{equation}
where the first line uses the cyclic property of trace, the second line defines $Y=B^{i+1}_{\alpha_{i+1}}B^i_{\alpha_i}(\sum_{\alpha_{i+2},\cdots,\alpha_{j-1}}B^{i+2}_{\alpha_{i+2}}\cdots B^j_{\alpha_j}\cdots B^{i+2}_{\alpha_{i+2}})[B^{i+1}_{\alpha_{i+1}},B^i_{\alpha_i}]\rho^{\frac{1}{2}}$, the third line uses Thm.~\ref{thm:Normal_Op_Frobenius_Norm} and the fourth line uses the triangle inequality.
The fifth line uses the property of the Frobenius norm and the sixth line uses the approximate representation property.
The final term can be simplified using the same method as before since $B_{\alpha_{i+1}}^{i+1}$ is now no longer between $i$ and $j$.
Therefore,
\begin{equation}
\begin{split}
    &\sum_{\substack{\alpha_i,\cdots,\alpha_j\\ \alpha_i=\alpha_j}}\Tr[B^j_{\alpha_j}\cdots B^{i+2}_{\alpha_{i+2}}B^i_{\alpha_i}B^{i+1}_{\alpha_{i+1}} \rho B^{i+1}_{\alpha_{i+1}}B^i_{\alpha_i}B^{i+2}_{\alpha_{i+2}}\cdots B^j_{\alpha_j}]\\
    =&\sum_{\substack{\alpha_i,\alpha_{i+2},\cdots,\alpha_j\\ \alpha_i=\alpha_j}}\Tr[B^j_{\alpha_j}\cdots B^{i+2}_{\alpha_{i+2}}B^i_{\alpha_i} \rho B^i_{\alpha_i}B^{i+2}_{\alpha_{i+2}}\cdots B^j_{\alpha_j}]+3\sum_{\alpha}[(\delta_{i+1,\alpha}^{tr})^2+2\delta_{i+1,\alpha}^{tr}].
\end{split}
\end{equation}
Therefore, the $B^{i+1}_{\alpha_{i+1}}$ operator can be removed with penalty of $3\sum_{\alpha}[(\delta_{i+1,\alpha}^{tr})^2+2\delta_{i+1,\alpha}^{tr}]+3\sum_{\alpha\beta}\tilde{\delta}^{com}_{i+1,i\alpha\beta}$.
The process can be repeated for other operators between indices $i$ and $j$, which gives the overall bound
\begin{equation}
\begin{split}
    p_{fail}^{i,j}\leq &3\sum_{u\in[1,i-1]\cup[i+1,j-1],\alpha}[(\delta_{u,\alpha}^{tr})^2+2\delta_{u,\alpha}^{tr}]+3\sum_{v\in[i+1,j-1]\alpha\beta}\tilde{\delta}^{com}_{v,i\alpha\beta}\\
    &+\sum_{\alpha_i=\alpha_j}\Tr[B^j_{\alpha_j}B^i_{\alpha_i}\rho B^i_{\alpha_i}B^j_{\alpha_j}]\\
    \leq&3\sum_{u\in[1,i-1]\cup[i+1,j-1],\alpha}[(\delta_{u,\alpha}^{tr})^2+2\delta_{u,\alpha}^{tr}]+3\sum_{v\in[i+1,j-1]\alpha\beta}\tilde{\delta}^{com}_{v,i\alpha\beta}\\
    &+\sum_{\alpha}(\delta^{ec}_{ij\alpha})^2,
\end{split}
\end{equation}
where we use the definition of the Frobenius norm.\\

The overall failure probability for edge verification is an average of the failure probability for each edge $(i,j)\in E_G$.
As such, we can bound by looking at the average of the individual terms.
Before proceeding, let us note that by Jensen's inequality and the concave nature of square root,
\begin{equation}
    \frac{1}{6\abs{E_{G'}}}\sum_{uv\alpha k}\sqrt{\varepsilon_{uv\alpha k}}\leq\sqrt{\frac{1}{6\abs{E_{G'}}}\sum_{uv\alpha k}\varepsilon_{uv\alpha k}}\leq\sqrt{\varepsilon}.
\end{equation}
We also note that for $x,y\in[0,1]$,
\begin{equation}
\begin{split}
    (\sqrt{2x}+\sqrt{2y})^2=&2x+2y+2\sqrt{4xy}\\
    \leq&2x+2y+\sqrt{2}\sqrt{2x}+\sqrt{2}\sqrt{2y}
\end{split}
\end{equation}
The first term can be simplified by including the terms for all vertices.
Since we are free to choose the corresponding vertex $j$ in $\delta_{i,\alpha}$, we uniformly choose $j\in\neigh(i)_{G'}$, noting that there can be at most double counting for each edge once -- $i$ selecting $j$ and $j$ selecting $i$, which gives a factor of 2,
\begin{equation}
\begin{split}
    &\sum_{u\in[1,i-1]\cup[i+1,j-1],\alpha}[(\delta_{u,\alpha}^{tr})^2+2\delta_{u,\alpha}^{tr}]\\
    \leq&\sum_{u\in V_G,\alpha}[(\delta_{u,\alpha}^{tr})^2+2\delta_{u,\alpha}^{tr}]\\
    \leq&\sum_{u\in V_G,v\in\neigh(v)_{G'},\alpha,k}\frac{1}{\abs{\neigh(u)_{G'}}}f(\varepsilon_{uv\alpha k})\\
    \leq&\frac{1}{\min_{u\in V_G}\abs{\neigh(u)_{G'}}}\sum_{uv\alpha k}[2\varepsilon_{uv\alpha k}+(4+2\sqrt{2})\sqrt{2\varepsilon_{uv\alpha k}}]\\
    \leq&\frac{6\abs{E_{G'}}}{\min_{u\in V_G}\abs{\neigh(u)_{G'}}}(2\varepsilon+4(1+\sqrt{2})\sqrt{\varepsilon}).
\end{split}
\end{equation}
where $f(\varepsilon_{uv\alpha k})=g(\varepsilon_{uv\alpha k})^2+(2+\sqrt{2})g(\varepsilon_{uv\alpha k})$ and $g(\varepsilon_{uv\alpha k})=\sqrt{2\varepsilon_{uv\alpha k}(2-\varepsilon_{uv\alpha k})}$.
The fourth line uses the bound $2-\varepsilon_{uv\alpha k}\leq 2$.
Since this value is independent of $(i,j)$, the average would have the same upper bound.
We can average the second term,
\begin{equation}
\begin{split}
    &\frac{1}{\abs{E_G}}\sum_{(i,j)\in E_G}\sum_{v\in[i+1,j-1]\alpha\beta}\tilde{\delta}^{com}_{vi\alpha\beta}\\
    \leq&\frac{1}{\abs{E_G}}\max_{u\in V_G}\abs{\neigh(u)_G}\sum_{i\in V_G}\sum_{v\in V_G,\alpha\beta}\tilde{\delta}^{com}_{vi\alpha\beta}\\
    \leq&\frac{18}{\abs{E_G}}\max_{u\in V_G}\abs{\neigh(u)_G}\sum_{(i,v)\in E_{G'},\alpha\beta}\left[4\sqrt{\frac{1}{2}(\varepsilon_{iv\alpha i}+\varepsilon_{iv\alpha j})}+19\sum_{i'\in\{i,j\}}\sqrt{2\varepsilon_{iv\alpha i'}(2-\varepsilon_{iv\alpha i'})}\right]\\
    \leq&\frac{18}{\abs{E_G}}\max_{u\in V_G}\abs{\neigh(u)_G}(6\abs{E_{G'}})(2\sqrt{2}\sqrt{\varepsilon}+38\sqrt{\varepsilon})\\
    \leq&\frac{216(19+\sqrt{2})\abs{E_{G'}}}{\abs{E_G}}\max_{u\in V_G}\abs{\neigh(u)_G}\sqrt{\varepsilon},
\end{split}
\end{equation}
where in the second line, we note that the sum over all edges would at most result in calling each vertex $i\in V_G$ at most $\abs{\neigh(i)_G}$ times.
The third line notes that for any pair of $(i,v)$, they either take on $\delta_{vi\alpha\beta}^{com}$ if $(i,v)\in E_G$, while taking on $\delta_{vi\alpha}^{cg}$ if $(i,v)\notin E_G$.
Since the latter value is larger, and noting that each $\varepsilon_{iv\alpha i'}$ only appears in at most two $(i,v)$ pairs, we can upper bound by the sum over all edges of the graph.
In the fourth line, we utilize Jensen's inequality and note $2-\varepsilon_{iv\alpha k}\leq 2$.
Averaging of the final term is straightforward,
\begin{equation}
\begin{split}
    &\frac{1}{\abs{E_G}}\sum_{(i,j)\in V_G}\sum_{\alpha}(\delta_{ij\alpha}^{ec})^2\\
    \leq&\frac{1}{\abs{E_G}}\sum_{i,j,\alpha}\left[2\sqrt{\varepsilon_{ij\alpha i}}+2\sqrt{\varepsilon_{ij\alpha j}}+\sqrt{\frac{1}{2}(\varepsilon_{ij\alpha i}+\varepsilon_{ij\alpha j})}\right]^2\\
    \leq&\frac{1}{\abs{E_G}}\sum_{i,j,\alpha}(2+\frac{1}{\sqrt{2}})^2\left[\sqrt{\varepsilon_{ij\alpha i}}+\sqrt{\varepsilon_{ij\alpha j}}\right]^2\\
    \leq&\frac{1}{\abs{E_G}}\sum_{i,j,\alpha,i'}(\frac{9}{2}+2\sqrt{2})\left[\varepsilon_{ij\alpha i'}+\sqrt{\varepsilon_{ij\alpha i'}}\right]\\
    \leq&\frac{(9+4\sqrt{2})\abs{E_{G'}}}{2\abs{E_G}}(\varepsilon+\sqrt{\varepsilon}).
\end{split}
\end{equation}
Combining the results, we obtain
\begin{equation}
\begin{split}
    p_{fail}^{EV}\leq&\frac{6\abs{E_{G'}}}{\min_{i\in V_G}\abs{\neigh(i)_{G'}}}(2\varepsilon+4(1+\sqrt{2})\sqrt{\varepsilon})\\
    &+\frac{216(19+\sqrt{2})\abs{E_{G'}}}{\abs{E_G}}\max_{i\in V_G}\abs{\neigh(i)_G}\sqrt{\varepsilon}+\frac{(9+4\sqrt{2})\abs{E_{G'}}}{2\abs{E_G}}(\varepsilon+\sqrt{\varepsilon}).
\end{split}
\end{equation}
% Taking the worst case selection of indices $(u,v)$, the probability of failing the edge verification test is upper bounded by
% \begin{equation}
%     p_{fail}^{EV}\leq\frac{9(7\delta+\delta^2)(m-2)}{4}.
% \end{equation}
\end{proof}

We can further simplify the winning probability obtained with some relaxation, noting $\varepsilon\leq\sqrt{\varepsilon}$ when $\varepsilon\leq 1$.
We recall Thm.~\ref{thm:Alt-RZKP-3-COL_to_Vertex-3-COL_simplifed}:
\chainofthm*
% \begin{theorem}
% \label{thm:Alt-RZKP-3-COL_to_Vertex-3-COL_simplifed}
%     If there exists an $\varepsilon$-perfect strategy for Alt-RZKP-3-COL with query challenge distribution $\mathcal{D}_{iji'j'b'}=\frac{1}{2\abs{E}}(\frac{\delta_{ii'}}{2\abs{\neigh(i)}}+\frac{\delta_{jj'}}{2\abs{\neigh(j)}})$, then there exists a classical strategy that can win the vertex-3-COL game with $p_{win}^{WD}=1$ for the well-definition tests and 
%     \begin{equation*}
%     \begin{split}
%         p_{win}^{EV}\geq&1-\varepsilon^{\frac{1}{4}}\left(\frac{9}{2}\abs{V_G}(\abs{V_G}-1)-8\abs{E_G}\right)\times\left[\frac{36+24\sqrt{6}+24\sqrt{3}}{3\abs{V_G}-3-2\max_{i\in V_G}\abs{\neigh(i)_G}}\right.\\
%         &\left.+\frac{216\sqrt{3}(19+\sqrt{2})}{\abs{E_G}}\max_{i\in V_G}\abs{\neigh(i)_G}+\frac{(9+4\sqrt{2})(3+\sqrt{3})}{\abs{E_G}}\right].
%     \end{split}
%     \end{equation*}
%     for the edge verification tests.
% \end{theorem}
\begin{proof}
From Thm.~\ref{thm:Alt-RZKP-3-COL_to_Alt-Edge-3-COL}, we know that the existence of an $\varepsilon$-perfect strategy for Alt-RZKP-3-COL with query challenge distribution $\mathcal{D}_{iji'j'b'}=\frac{1}{2\abs{E}}(\frac{\delta_{ii'}}{2\abs{\neigh(i)}}+\frac{\delta_{jj'}}{2\abs{\neigh(j)}})$ implies the existence of a $3\sqrt{\varepsilon}$-perfect strategy for Alt-Edge-3-COL, noting $\sqrt{\varepsilon}\leq\varepsilon$.
From Thm.~\ref{thm:Alt-Edge-3-COL_to_VCCC-BCS-3-COL}, this in turn implies that there exists a vertex-complete color-commuting strategy that can win the edge verification challenge with a distribution of $\mathcal{D}_{ijk\alpha}=\frac{1}{6\abs{E}}$ with $p_{win}^{EV}\geq 1-3\sqrt{\varepsilon}$.
As a consequence, from Thm.~\ref{thm:BCS-3-COL_to_Vertex-3-COL}, the winning probability for the edge verification tests are
\begin{equation}
\begin{split}
    p_{win}^{EV}\geq&1-\frac{6\abs{E_{G'}}}{\min_{i\in V_G}\abs{\neigh(i)_{G'}}}[6\sqrt{\varepsilon}+4\sqrt{3}(1+\sqrt{2})\varepsilon^{\frac{1}{4}}]\\
    &-\frac{216\sqrt{3}(19+\sqrt{2})\abs{E_{G'}}}{\abs{E_G}}\max_{i\in V_G}\abs{\neigh(i)_G}\varepsilon^{\frac{1}{4}}-\frac{(9+4\sqrt{2})\abs{E_{G'}}}{\abs{E_G}}\left(3\sqrt{\varepsilon}+\sqrt{3}\varepsilon^{\frac{1}{4}}\right).
\end{split}
\end{equation}
Let us now expand some of the terms.
By construction, we know that $\abs{E_{G'}}=\frac{9}{2}\abs{V_G}(\abs{V_G}-1)-8\abs{E_G}$.
The number of neighbors for $i\in V_G$ can also be computed by construction,
\begin{equation}
    \abs{\neigh(i)_{G'}}=\abs{\neigh(i)_G}+3(\abs{V_G}-1-\abs{\neigh(i)_G})=3\abs{V_G}-3-2\abs{\neigh(i)_G}.
\end{equation}
Therefore,
\begin{equation}
    \min_{i\in V_G}\abs{\neigh(i)_{G'}}=3\abs{V_G}-3-2\max_{i\in V_G}\abs{\neigh(i)_G}.
\end{equation}
We can further bound $\sqrt{\varepsilon}\leq\varepsilon^{\frac{1}{4}}$, which simplifies the winning probability
\begin{equation}
\begin{split}
    p_{win}^{EV}\geq&1-\varepsilon^{\frac{1}{4}}\left(\frac{9}{2}\abs{V_G}(\abs{V_G}-1)-8\abs{E_G}\right)\times\left[\frac{36+24\sqrt{6}+24\sqrt{3}}{3\abs{V_G}-3-2\max_{i\in V_G}\abs{\neigh(i)_G}}\right.\\
    &\left.+\frac{216\sqrt{3}(19+\sqrt{2})}{\abs{E_G}}\max_{i\in V_G}\abs{\neigh(i)_G}+\frac{(9+4\sqrt{2})(3+\sqrt{3})}{\abs{E_G}}\right].
\end{split}
\end{equation}
\end{proof}

\subsection{Soundness}

We have demonstrated that an $\varepsilon$-perfect strategy for Alt-RZKP-3-COL allows us to construct a classical strategy for vertex-3-COL which always passes the well-definition test and has a high chance of passing the edge verification test, $p_{win}^{EV}\geq1-f(\varepsilon)$.
Here, we can argue that any classical strategy for vertex-3-COL that always passes the well-definition test cannot pass the edge verification test with arbitrarily high probability.
\begin{theorem}
\label{thm:Vertex-3-COL_classical_value}
    If the graph $G$ is not 3-colorable, then all classical strategies for vertex-3-COL for $G$ always passes the well-definition test passes the edge verification test with $p_{win}^{EV}\leq 1-\frac{1}{\abs{E_G}}$.
\end{theorem}
\begin{proof}
WLOG, we can consider a deterministic classical strategy, since any probabilistic strategy is a convex combination of deterministic strategies -- and it always benefits the adversary to pick the best deterministic strategy out of this combination.
For the adversary to pass the well-definition test, both provers must agree on a coloring of the graph $G$.
Since $G$ is not 3-colorable, there exists $(i,j)\in E_G$ where $c_i=c_j$, i.e., both provers will fail the edge verification test when challenged with this edge (out of $\abs{E_G}$ edge choices).
Therefore, the winning probability is bounded
\begin{equation}
    p_{win}^{EV}\leq 1-\frac{1}{\abs{E_G}}.
\end{equation}
\end{proof}

As such, there exists a contradiction which can be used to demonstrate the soundness by computing a bound on the quantum value,
\begin{theorem}
\label{thm:soundness_main_thm}
    Suppose $G$ is not 3-colorable. For any adversary, the winning probability for the Alt-RZKP-3-COL game for graph $G'$ % with query challenge distribution $\mathcal{D}_{iji'j'b'}=\frac{1}{2\abs{E}}(\frac{\delta_{ii'}}{2\abs{\neigh(i)}}+\frac{\delta_{jj'}}{2\abs{\neigh(j)}})$ 
    cannot exceed
    \begin{equation*}
    \begin{split}
        p_{win}\leq&1-\frac{1}{\abs{E_G}^4(\frac{9}{2}\abs{V_G}(\abs{V_G}-1)-8\abs{E_G})^4}\times\\
        &\left[\frac{(324+108\sqrt{2})\abs{E_G}}{3\abs{V_G}-3-2\max_{i\in V_G}\abs{\neigh(i)_G}}+20412\sqrt{2}\max_{i\in V_G}\abs{\neigh(i)_G}+117\right]^{-4},
    \end{split}
    \end{equation*}
    i.e., the protocol has quantum value 
    \begin{equation*}
    \begin{split}
        \omega_q=&1-\frac{1}{\abs{E_G}^4(\frac{9}{2}\abs{V_G}(\abs{V_G}-1)-8\abs{E_G})^4}\times\\
        &\left[\frac{(324+108\sqrt{2})\abs{E_G}}{3\abs{V_G}-3-2\max_{i\in V_G}\abs{\neigh(i)_G}}+20412\sqrt{2}\max_{i\in V_G}\abs{\neigh(i)_G}+117\right]^{-4},
    \end{split}
    \end{equation*}
\end{theorem}
\begin{proof}
From Thm.~\ref{thm:Alt-RZKP-3-COL_to_Vertex-3-COL_simplifed} and Thm.~\ref{thm:Vertex-3-COL_classical_value}, it is clear that there cannot exist any $\varepsilon$-perfect strategy for Alt-RZKP-3-COL with
\begin{equation}
\begin{gathered}
    \left[\frac{36+24\sqrt{6}+24\sqrt{3}}{3\abs{V_G}-3-2\max_{i\in V_G}\abs{\neigh(i)_G}}+\frac{216\sqrt{3}(19+\sqrt{2})}{\abs{E_G}}\max_{i\in V_G}\abs{\neigh(i)_G}\right.\\
    \left.+\frac{(9+4\sqrt{2})(3+\sqrt{3})}{\abs{E_G}}\right]\times\left(\frac{9}{2}\abs{V_G}(\abs{V_G}-1)-8\abs{E_G}\right)\varepsilon^{\frac{1}{4}}<\frac{1}{\abs{E_G}}
\end{gathered}
\end{equation}
since it triggers a contradiction.\\

Therefore, we require that
\begin{equation}
\begin{gathered}
    \varepsilon\geq\frac{1}{\abs{E_G}^4(\frac{9}{2}\abs{V_G}(\abs{V_G}-1)-8\abs{E_G})^4}\times\left[\frac{36+24\sqrt{6}+24\sqrt{3}}{3\abs{V_G}-3-2\max_{i\in V_G}\abs{\neigh(i)_G}}\right.\\
    \left.+\frac{216\sqrt{3}(19+\sqrt{2})}{\abs{E_G}}\max_{i\in V_G}\abs{\neigh(i)_G}+\frac{(9+4\sqrt{2})(3+\sqrt{3})}{\abs{E_G}}\right]^{-4},
\end{gathered}
\end{equation}
implying the winning probability
\begin{equation}
\begin{split}
    p_{win}\leq&1-\frac{1}{\abs{E_G}^4(\frac{9}{2}\abs{V_G}(\abs{V_G}-1)-8\abs{E_G})^4}\times\left[\frac{36+24\sqrt{6}+24\sqrt{3}}{3\abs{V_G}-3-2\max_{i\in V_G}\abs{\neigh(i)_G}}\right.\\
    &\left.+\frac{216\sqrt{3}(19+\sqrt{2})}{\abs{E_G}}\max_{i\in V_G}\abs{\neigh(i)_G}+\frac{(9+4\sqrt{2})(3+\sqrt{3})}{\abs{E_G}}\right]^{-4}
\end{split}
\end{equation}
\end{proof}

\subsection{Performance}
\label{app:perf}
Here, we provide some values for comparison.
Consider the target winning probability after $m$ repetition as $\varepsilon_{sou}=e^{-k}$.
Therefore, %the target is $(1-\delta)^m= e^{-k}$, which yields
we can estimate approximately $m\approx k(1-\omega_q)^{-1}$, i.e.
\begin{equation}
\begin{split}
    m\approx&k\abs{E_G}^4(\frac{9}{2}\abs{V_G}(\abs{V_G}-1)-8\abs{E_G})^4\times\left[\frac{36+24\sqrt{6}+24\sqrt{3}}{3\abs{V_G}-3-2\max_{i\in V_G}\abs{\neigh(i)_G}}\right.\\
    &\left.+\frac{216\sqrt{3}(19+\sqrt{2})}{\abs{E_G}}\max_{i\in V_G}\abs{\neigh(i)_G}+\frac{(9+4\sqrt{2})(3+\sqrt{3})}{\abs{E_G}}\right]^{4}.
\end{split}
\end{equation}
Assuming $k=100$~\cite{Alikhani_2021}, the number of rounds are shown in Table~\ref{table:num_rounds_security}.

\begin{table}[!h]
    \centering
    \begin{tabular}{|c|c|c|c|c|}
    \hline
    Nodes ($\abs{V_G}$) & Edges ($\abs{E_G}$) & Nodes ($\abs{V_{G'}}$) & Edges ($\abs{E_{G'}}$) & Rounds \\
    \hline
    $200$ & $380$ & $78280$ & $176060$ & $8.54\times 10^{40}$\\
    \hline
    $600$ & $1122$ & $714912$ & $1608324$ & $5.95\times 10^{44}$\\
    \hline
    $900$ & $1695$ & $1612320$ & $3627390$ & $1.54\times 10^{46}$\\
    \hline
    \end{tabular}
    \caption{Number of rounds required for $e^{-100}$ soundness for various graph choices, with maximum number of neighbors $\max_{i\in V_G}\abs{\neigh(i)_G}=4$.}
    \label{table:num_rounds_security}
\end{table}

A quick numerical analysis gives the scaling of the problem as approximately $\abs{E_G}^8$ if $\abs{V_G}$ and $\abs{E_G}$ are linearly related (or equivalently $\abs{V_{G'}}^4$).
We note that this is worse than the $\abs{E_G}^4$ scaling of the three-party protocol~\cite{Alikhani_2021}.
There are three main contribution to the scaling:
\begin{enumerate}
    \item Gentle measurement lemma for reducing Alt-RZKP-3-COL to Alt-Edge-3-COL gives an order of 2 scaling. The penalty paid here stems from the protocol design which provides zero-knowledge properties.
    \item Formulation of almost commuting operators from BCS non-local game strategy, where the order 2 scaling is contributed by an argument similar to gentle measurement lemma -- high probability of conclusive measurement outcomes result in tracial relations~\cite{Slofstra2018,Paddock2024}.
    \item A mixture of the effect of accumulation of $O(\abs{V_G})$ terms in the reduction from the almost fully commuting satisfying assignment to the classical strategy and the gap between original game with $G'$ and the security guarantee from $G$'s non-3-colorability. %The introduction of gadgets results in almost an order 2 scaling of the number of vertices and edges. 
    Consolidating errors over $E_{G'}$ challenges of Alt-Edge-3-COL into $E_G$ challenges in vertex-3-COL leads to a ratio $\frac{\abs{E_{G'}}}{\abs{E_G}}$ penalty which is significant when the graph has low connectivity. On the other hand, when the graph is highly connected, the $O(\abs{V_G})$ terms appears from the number of neighbors. %\wy{Check this again, it might instead be due to the same accumulation cause of trace distance, though the accumulation amount might be different.}
\end{enumerate}
% \wy{Do we want to consider an experiment/proof for $K_4$ graph to maybe see gap between SDP and theory? Probably can tighten a lot of the values at the same time since it's small enough to compute the exact form by hand.}

% \subsection{Perfect Zero-knowledge of Alternative Two-Prover Protocol}

% \section{Security Analysis Proofs}

% \wy{Note this will merge back.}
% \subsection{Alt-RZKP-3-COL to Alt-Edge-3-COL Reduction}
% \label{app:Alt-RZKP-3-COL_to_Alt-Edge-3-COL}

% \subsection{Alt-Edge-3-COL to BCS-3-COL}
% \label{app:Alt-Edge-3-COL_to_VCCC-BCS-3-COL}

% \subsection{Tracial Property}
% \label{app:Tracial_Property_BCS}

% \subsection{BCS-3-COL to Quantum Assignment}

% \subsection{Fully Commuting Almost Satisfying Assignment}
% \label{app:almost_qsa-to-fully_commuting_qsa}

% \subsection{Classical Strategy for Vertex-3-COL}
% \label{app:BCS-3-COL_to_Vertex-3-COL}

\end{document}